\documentclass[12pt]{article}

\usepackage{hyperref,amsfonts,amsmath,amssymb,bbm,epsf,fancybox}
\usepackage{graphics} 
\advance\textheight by \topskip
\newcounter{defthm}
 
\newtheorem{defthm}{\whattheorem}[section]

\newcommand\sect[1]{\section{#1}\setcounter{equation}0\setcounter{defthm}0}

\newcommand\be            {\begin{equation}}
\newcommand\bea           {\begin{equation}\begin{array}l\displaystyle}
\newcommand\bearll        {\begin{array}{ll}\displaystyle}
\newcommand\ee            {\end{equation}}
\newcommand\eE            {{\mathrm e}}
\newcommand\eear          {\end{array}}
\newcommand\enl           {\\[1em]\displaystyle}
\newcommand\enL           {\\[.4em]\displaystyle}
\newcommand\etb           {& \displaystyle}
\newcommand\erf[1]        {(\ref{#1})}
\newcommand\Frac[2]       {\mbox{\large$\frac{#1}{#2}$}}
\newcommand\id            {{\mathrm{id}}}
\newcommand\ii            {{\mathrm i}}
\newcommand\labl[1]       {\label{#1}\ee}
\newcommand\noo           {\mbox{\bf:}}

\newcommand\FB[1]         {\text{Bos}(#1)}
\newcommand\Hom           {\text{Hom}}

\newcommand\one           {{\bf1}}
\newcommand\ol            {\overline}
\newcommand\oor           {^{(r)}}
\newcommand\oos           {^{(s)}}
\newcommand\ors           {^{(rs)}}
\newcommand\osr           {^{(sr)}}
\newcommand\oti           {\,{\otimes}\,}
\newcommand\rmq           {{\mathrm q}}
\newcommand\RR            {\hat R}
\newcommand\Rsd           {R_\text{s.d.}}
\newcommand\su            {\mathrm{su}}
\newcommand\suh           {\widehat{\mathrm{su}}}
\newcommand\Ue            {{\mathrm u(1)}}
\newcommand\Ueh           {{\widehat{\mathrm u}(1)}}
\renewcommand\varTheta    {\theta}

\newcommand\Cb            {{\mathbbm{C}}}

\newcommand\Qb            {\mathbb{Q}}
\newcommand\Rb            {\mathbb{R}}
\newcommand\Zb            {\mathbb{Z}}

\newcommand\Cc            {\mathcal{C}}
\newcommand\Dc            {\mathcal{D}}
\newcommand\Hc            {\mathcal{H}}
\newcommand\Ic            {\mathcal{I}}

\newcommand\Uc            {\mathcal{U}}

\newcommand\Fs            {\mathsf{F}}
\newcommand\Rs            {\mathsf{R}}

\begin{document}
\thispagestyle{empty}
\def\thefootnote{\fnsymbol{footnote}}
\begin{flushright}
{\sf KCL-MTH-07-05}\\[1mm]
{\sf ZMP-HH/2007-06}\\[1mm]
{\sf Hamburger$\;$Beitr\"age$\;$zur$\;$Mathematik$\;$Nr.$\;$271}
\end{flushright}
\vskip 2.0em
\begin{center}\Large TOPOLOGICAL DEFECTS FOR THE FREE BOSON CFT
\end{center}\vskip 1.5em
\begin{center}
  J\"urgen Fuchs\,$^{a}$,
  ~Matthias R.~Gaberdiel\,$^{b}$,
  ~Ingo Runkel\,$^{c}$,
  ~Christoph Schweigert\,$^{d}$\footnote{\scriptsize 
  Emails: 
  jfuchs@fuchs.tekn.kau.se, gaberdiel@itp.phys.ethz.ch,
  ingo.runkel@kcl.ac.uk, schweigert@math.uni-hamburg.de}
\end{center}

\begin{center}\it$^a$
Teoretisk fysik, \ Karlstads Universitet\\
Universitetsgatan 5, \ S\,--\,651\,88\, Karlstad
\end{center}
\begin{center}\it$^b$
Institut f\"ur Theoretische Physik, ETH Z\"urich \\
8093 Z\"urich, Switzerland
\end{center}
\begin{center}\it$^c$
  Department of Mathematics, King's College London \\
  Strand, London WC2R 2LS, United Kingdom  
\end{center}
\begin{center}\it$^c$
  Organisationseinheit Mathematik, \ Universit\"at Hamburg\\
  Schwerpunkt Algebra und Zahlentheorie\\
  Bundesstra\ss e 55, \ D\,--\,20\,146\, Hamburg
\end{center}
\vskip 1.5em
\begin{center} May 2007 \end{center}
\vskip 2em
\begin{abstract}
Two different conformal field theories can be joined
together along a defect line. We study such defects for the case where
the conformal field theories on either side are single free bosons 
compactified on a circle. We concentrate on topological defects for
which the left- and right-moving Virasoro algebras are separately
preserved, but not necessarily any additional symmetries.
For the case where both radii are rational multiples of the self-dual
radius we classify these topological defects. We also show that the
isomorphism between two T-dual free boson conformal field theories can
be described by the action of a topological defect, and hence that
T-duality can be understood as a special type of order-disorder
duality. 
\end{abstract}

\setcounter{footnote}{0}
\def\thefootnote{\arabic{footnote}}
\newpage

\tableofcontents

\sect{Introduction}

In this paper we investigate how one can join two free boson conformal field
theories along a line in a conformally invariant manner. More specifically,
we are interested in interfaces which preserve the left- and right-moving 
conformal symmetries separately. Such interfaces are special types of conformal
defects that appear    naturally in conformal field theory; conformal defects   
have recently attracted some attention, see e.g.\ 
\cite{Petkova:2000ip,Petkova:2001ag,Chui:2001kw,Coquereaux:2001di,%
Bachas:2001vj,Quella:2002ct,Fuchs:2002cm,Graham:2003nc,Frohlich:2004ef,%
Bachas:2004sy,Frohlich:2006ch,Quella:2006de}.
As we shall review momentarily, the special (topological) defects we
consider in this paper have a number of interesting and useful properties. 

Let us consider the case that the world sheet is the complex plane and the 
interface runs along the real axis. We take the conformal field theory on the 
upper half plane to be the compactified free boson of radius $R_1$ and the 
theory on the lower half plane to be the free boson theory at radius $R_2$. 
Denoting the left- and 
right-moving stress tensors of the two theories by $T^{i}$ and $\bar{T}^{i}$,
$i\,{=}\,1,2$, respectively, general conformal defects are interfaces that obey 
  \be
  T^{1}(x) - \bar{T}^{1}(x) = T^{2}(x) - \bar{T}^{2}(x)
  \qquad \text{for all} ~~ x \,{\in}\, \Rb \,.
  \labl{eq:free-boson-interface}
Via the folding trick \cite{Wong:1994pa}, solutions to 
\erf{eq:free-boson-interface} correspond to
conformal boundary conditions for the product theory consisting of two 
free bosons compactified to $R_1$ and $R_2$, respectively. (From a target space
perspective we are thus looking for the conformal D-branes of the theory on a
rectangular torus with radii $R_1$ and $R_2$.) This method to investigate the
conformal interfaces was used in \cite{Bachas:2001vj}.  

While all conformal boundary conditions for the theory of a single
free boson on a circle are known
\cite{Friedan,Gaberdiel:2001xm,Gaberdiel:2001zq,Janik:2001hb}, 
the classification of all conformal boundaries of the $c=2$ theory in
question is presently out of reach. Thus we cannot hope to find all
interfaces obeying \erf{eq:free-boson-interface}. There is, however,
an interesting subclass of conformal interfaces for which a
classification can be achieved. These are the interfaces
for which the condition \erf{eq:free-boson-interface} is strengthened to 
  \be
  T^{1}(x) = T^{2}(x) ~~~\text{and}~~~ \bar{T}^{1}(x) = \bar{T}^{2}(x) 
  \qquad \text{for all} ~~ x \,{\in}\, \Rb \,.
  \labl{eq:free-boson-top-interface}
In other words, we require that the stress tensor is continuous across
the interface. In this case the interface commutes with the generators
of local conformal transformations, and the interface line can
be continuously deformed without affecting the value of correlators as
long it does not cross any field insertion points. These interfaces are 
therefore called {\em topological defects} \cite{Bachas:2004sy}
(or sometimes also totally transmissive defects). 
They have, at least, two nice properties: 
\def\leftmargini{1.1em}\begin{itemize}
\item 
Topological defects can be fused together (by letting the interfaces merge), 
and thus carry a multiplicative structure \cite{Petkova:2000ip,%
Petkova:2001ag,Chui:2001kw,Coquereaux:2001di,Frohlich:2006ch}. They therefore 
possess more structure than the corresponding boundary conditions. 
\item 
Topological defects contain information about symmetries of the conformal field  
theory (in the present case the free boson), as well as about order-disorder 
dualities \cite{Frohlich:2004ef,Frohlich:2006ch}. For the
the free boson we identify a topological defect of the latter type, which 
generates the T-duality symmetry.  
\end{itemize}
 
\noindent
In this paper we shall describe a large class of such topological defects for 
the compactified free boson, and give a classification for specific values of 
the radii. We should stress that even if the two radii are different, 
$R_1\,{\neq}\, R_2$, many topological defects exist -- this will become clear 
in section~\ref{sec:topdef-sum}; the precise form of the defects then depends 
on the arithmetic properties of $R_1$ and $R_2$. 
Furthermore, via the folding trick, these topological defects correspond 
to new conformal boundary conditions in the product of two free boson CFTs.

\smallskip

It is difficult to verify that a given collection of topological defects is
consistent. In principle one has to specify all correlators involving such 
defect lines and verify the relevant sewing conditions. (A complete list of 
sewing constraints for correlators involving defects has not been written down; 
it will be an extension of the constraints for correlators involving boundaries 
given in \cite{Cardy:1991tv,Lewellen:1991tb}). In this paper we take two 
approaches to this problem. The first is to use the TFT-formulation for 
constructing CFT correlators \cite{Fuchs:2002cm,Frohlich:2006ch,Fjelstad:2006aw},
which does give all collections of correlators (involving boundaries and 
topological defect lines) that are consistent with sewing.  However, this 
approach only applies to {\em rational\/} conformal field theories and to the 
defects which preserve a rational symmetry. This approach thus gives a 
collection of defects that are guaranteed to be consistent, but these will not 
be all. We can, however, give a criterion for a
deformation of topological defects by a defect field to be exactly
marginal, and in this way we will also gain some information about the 
neighbourhood of these `rational' topological defects.  

The second approach is to select some of the necessary consistency conditions 
which are relatively easy to analyse (at least for the free boson) and to try 
to classify all of their solutions. One example is the analogue of the Cardy
constraint for boundary conditions \cite{Cardy:1986gw}; it arises from
analysing torus partition functions with insertions of defect lines
\cite{Petkova:2000ip}. We will also use an additional condition which
is obtained by deforming defect lines in the presence of bulk
fields. This gives an upper bound on the complete list of consistent
defects. By comparing the results from the two approaches we can thus
propose a fairly convincing picture of what all the consistent defects
between free boson theories at $c\,{=}\,1$ are.

\smallskip

The paper is organised as follows. In section \ref{sec:top-def} we give a
brief introduction to topological defects; section~3 contains our
conventions for the free boson theory, and in section~4 we give a
non-technical summary of our results. The details of the first
approach are presented in section \ref{sec:rat-def},
and those of the second approach in section \ref{sec:topdef-gen}. Some
technical calculations have been collected in several appendices.


\sect{Topological defect lines}\label{sec:top-def}

A defect is a one-dimensional interface separating two conformal
field theories CFT$_1$ and CFT$_2$. The defect is called conformal iff
the stress-energy tensors of the two theories are related as in 
(\ref{eq:free-boson-interface}). In the following we shall concentrate
on topological defects for which the stronger condition 
$T^1 \,{=}\, T^2$ and $\overline T{}^1 \,{=}\, \overline T{}^2$ 
holds -- see  \erf{eq:free-boson-top-interface}. In other words, 
we shall require that the two independent components of the
stress-energy tensor are continuous across the defect line.  

Suppose $D$ is a topological defect joining the two conformal field theories 
CFT$_1$ and CFT$_2$, and denote by $\Hc_1$ and $\Hc_2$ the corresponding 
spaces of bulk states. The topological defect $D$ then gives rise to a linear 
operator $\hat{D}{:}\ \Hc_1 \,{\rightarrow}\, \Hc_2$ which is obtained
by taking 
$D$ to wind around the field insertion with the appropriate orientation.\,%
  \footnote{~Here, as in the following, we shall
  identify bulk fields $\phi(z)$ with the corresponding states $\phi$ in $\Hc$.}
The defect $\bar{D}$ with the opposite orientation then defines a map 
$\hat{\bar{D}}{:}\ \Hc_2 \,{\rightarrow}\, \Hc_1$. Pictorially,
  \be
  \raisebox{-11pt}{
  \begin{picture}(70,44)
   \put(20,0){\scalebox{.4}{\includegraphics{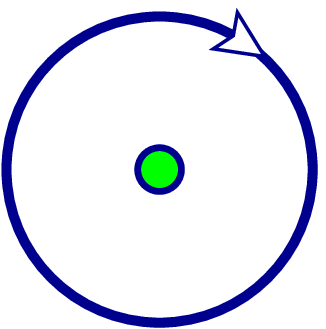}}}
    \put(20,0){
     \setlength{\unitlength}{.4pt}\put(0,0){
     \put(53,30)  {\scriptsize $ \phi_1 $}
     \put(89,70)  {\scriptsize $ D $}
     \put(-60, 0) {\footnotesize\shadowbox{$2$}}
     \put(-50,60) {\footnotesize\shadowbox{$1$}
               \put(0,0){\line(3,-1){30}}\put(0,0.2){\line(3,-1){30}}}
     }\setlength{\unitlength}{1pt}}
  \end{picture}}
   =:~ (\hat{D} \phi_1)(z) 
  \qquad \text{and} \qquad~
  \raisebox{-16pt}{
  \begin{picture}(70,44)
    \put(20,0){\scalebox{.4}{\includegraphics{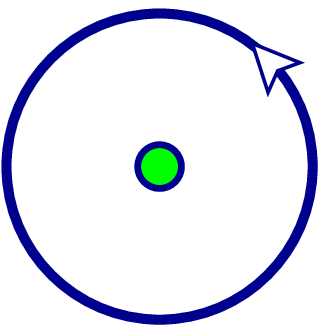}}}
    \put(20,0){
     \setlength{\unitlength}{.4pt}\put(0,0){
     \put(53,30)  {\scriptsize $ \phi_2 $}
     \put(89,70)  {\scriptsize $ D $}
     \put(-60, 0) {\footnotesize\shadowbox{$1$}}
     \put(-50,60) {\footnotesize\shadowbox{$2$}
         \put(0,0){\line(3,-1){30}}\put(0,0.2){\line(3,-1){30}}}
     }\setlength{\unitlength}{1pt}}
  \end{picture}}
  =:~ (\hat{\bar{D}} \phi_2)(z) \,.
  \ee
As will be discussed in section \ref{sec:complete} below, these operators do not 
specify the defect uniquely. They do, however, contain important information 
about the defect. For example, just like boundary conditions, defects also 
possess excitations that are localised on the defect; these are called 
defect fields. The spectrum of defect fields can be determined by considering 
the trace of $\hat{\bar{D}} \hat{D}$ in $\Hc_1$. Furthermore, defect operators 
determine the correlators of two bulk fields on a sphere which are separated by 
the defect loop. These correlators play an analogous role for topological 
defects as the one-point correlators of bulk fields on the disk do for conformal
boundary conditions. Defect operators 
finally provide the point of view from which 
topological defects were first studied in \cite{Petkova:2000ip}.

Since a topological defect $D$ is transparent to $T$ and $\bar{T}$, 
the corresponding operator commutes with the Virasoro modes, i.e.
  \be
  L_m^{2} \,\hat D = \hat D \,L_m^{1}  \qquad\text{and}\qquad
  \bar{L}_m^{2} \, \hat D = \hat D \, \bar{L}_m^{1} \,,
  \label{eq:D-Vir-com}
  \ee
where $L_m^{i}$ and $\bar{L}_m^{i}$ act on the state space $\Hc_i$. A similar 
statement also holds for $\hat{\bar{D}}$. However, not every map $\hat{D}$ 
satisfying (\ref{eq:D-Vir-com}) arises as the operator of a topological defect.
This is analogous to the case of boundary conditions: (\ref{eq:D-Vir-com}) 
corresponds to the conformal gluing condition, but not every solution of the 
conformal gluing condition defines a consistent boundary state -- for example, 
the boundary states must in addition satisfy the Cardy condition, etc.  

Topological defects can be fused. We denote the fusion of two defects $D_1$ and
$D_2$ by $D_1 \,{*}\, D_2$. In terms of the associated operators $\hat{D}$, 
fusion just corresponds to the composition $\hat{D}_1 \,{\circ}\, \hat{D}_2$. 
Note that in general the fusion of defects is not commutative. Let $D$ be a 
defect between two copies of the same conformal field theory, i.e.\ 
CFT$_1\,{=}\,$CFT$_2$. Then we call $D$ {\em group-like\/} iff fusing
it with the defect of opposite orientation yields the 
trivial defect, $\bar{D} \,{*}\, D \,{=}\, \one$. (The trivial defect between 
two identical conformal field theories corresponds simply to the identity
map.) The group-like defects form a group. It can be shown that this group 
describes internal symmetries for 
CFT correlators on world sheets of arbitrary genus \cite{Frohlich:2006ch}.  

We can also form superpositions of topological
defects, which corresponds to 
adding the associated operators. We call a defect operator $\hat D$ 
{\em fundamental\/} iff it is nonzero and it cannot be written as a sum of 
two other nonzero defect operators.
Finally, a topological defect $D$ is called a {\em duality defect\/}
iff $\bar{D} \,{*}\, D$ decomposes into a superposition of only group-like 
defects (thus group-like defects are a special case of duality
defects). Duality defects relate correlators of the theories CFT$_1$
and CFT$_2$ and describe order-disorder, or Kramers-Wannier 
type, dualities \cite{Frohlich:2004ef,Frohlich:2006ch}.

As we will see, T-duality is generated by a special kind of duality defect. In 
general, duality defects give rise to identities that relate correlators of bulk
fields to correlators of disorder fields. (Disorder fields are those defect 
fields that appear at the end points of defect lines.) In the case of T-duality,
the resulting disorder fields are in fact again local bulk fields. Thus 
T-duality can be understood as a special type of order-disorder duality, an 
observation also made in 
\cite{Shapere:1988zv,Asatani:1996jc} on the basis of lattice discretisations.


\sect{The compactified free boson}\label{sec:comp-FB}

Before discussing topological defects in detail, let us fix our
conventions for the free boson compactified on a circle of radius $R$.
The chiral symmetry of the free boson $\phi(z)$ is a $\Ue$ current
algebra generated by $J(z) \,{=}\, \ii\, \sqrt{\frac{2}{\alpha'}}
\tfrac{\mathrm d}{\mathrm dz} \phi(z)$ with operator product expansion
  \be
  J(z)\,J(0) \,{=}\, z^{-2} +\, \text{reg.}
  \ee 
Its irreducible highest weight representations $\Hc_q$ are uniquely 
characterised by the eigenvalue of the zero mode
$J_0$ on the highest weight state
$|q\rangle$, $J_0 |q\rangle \,{=}\, q\, |q\rangle$. The stress-energy tensor is 
$T(z) \,{=}\, \tfrac12\, \noo J(z)J(z) \noo\,$, so that the conformal
weight of $|q\rangle$ is $h \,{=}\, \tfrac12\, q^2$. 

For the boson of radius $R$, the bulk spectrum can be written as a direct sum
  \be
  \Hc(R) = \bigoplus_{m,w \in \Zb}
  \Hc_{q_{m,w}(R)} \otimes \bar\Hc_{\bar q_{m,w}(R)}
  \labl{eq:FB-bulk-states}
of representations of the left- and right-moving current algebras, where
  \be
  q_{m,w}(R) = \frac{1}{\sqrt{2}}\Big( \frac{\sqrt{\alpha'}}{R}\, m + 
   \frac{R}{\sqrt{\alpha'}}\, w \Big) 
  \,, \qquad
  \bar q_{m,w}(R) = \frac{1}{\sqrt{2}}\Big( 
      \frac{\sqrt{\alpha'}}{R}\, m - \frac{R}{\sqrt{\alpha'}}\, w \Big) \,.
  \ee
The integer $w$ is the winding number, and $m$ is related to the total
momentum $p$ via 
$p \,{=}\, (2\alpha')^{-1/2}(q_{m,w}+\bar q_{m,w})\,{=}\, m/R$.  
The charges $(q,\bar{q})$ that appear in the decomposition 
(\ref{eq:FB-bulk-states}) of $\Hc(R)$ form the {\em charge lattice\/} 
$\Lambda(R)$. It is generated by the two vectors 
  \be
  \frac{\sqrt{\alpha'}}{\sqrt{2}R} \,(1,1) \qquad \hbox{and} \qquad
  \frac{R}{\sqrt{2\alpha'}} \, (1,-1)
  \labl{basis}
which describe a pure momentum and pure winding state, respectively.

\smallskip

The chiral vertex operator corresponding to the highest weight state 
$|q\rangle$ is given by the normal ordered exponential  
$\noo\eE^{\ii \sqrt{2/\alpha'}\, q\, \phi(z)}\noo\,$. 
The bulk field corresponding to the highest weight state in the sector
$(q,\bar{q})\,{\in}\,\Lambda(R)$ is obtained from the product of two such normal
ordered exponentials; it will be denoted by $\phi_{(q,\bar{q})}$. The bulk
fields can be normalised in such a way that the operator products take the form
  \be
  \phi_{(q_1,\bar{q}_1)}^{}(z)\, \phi_{(q_2,\bar{q}_2)}^{}(w)
  = (-1)^{m_1 w_2}_{} \,
  (z{-}w)^{q_1 q_2}_{}\, 
    (z^*{-}w^*)^{\bar{q}_1 \bar{q}_2}_{}\,
  \big( \phi_{(q_1+q_2,\bar{q}_1 + \bar{q}_2)}^{}(w) + O(|z{-}w|) \big)
  \,. 
  \labl{eq:bulk-OPE}
The factor $(-1)^{m_1 w_2}$ is needed for locality; in 
particular, one cannot set all OPE coefficients to 1.\,%
  \footnote{~Consider, for example, two fields $\phi_1$ and $\phi_2$
  with $(m_1,w_1)\,{=}\,(1,0)$ and $(m_2,w_2)\,{=}\,(0,1)$. The operator
  products $\phi_1(z)\,\phi_2(w)$ and $\phi_2(z)\,\phi_1(w)$ must be
  related by analytic continuation. Since in both cases the leading
  singularity is $(z{-}w)^{\frac12} (z^*{-}w^*)^{-\frac12}$, the sign
  arising in the analytic continuation must be compensated by the OPE
  coefficient. Note also that in the convention \erf{eq:bulk-OPE} some
  two-point functions are negative. This can be avoided at the cost of
  introducing imaginary OPE coefficients. Namely, in terms of the
  basis $\phi'_{(q,\bar{q})} \,{=}\, \mathrm i^{mw} \phi_{(q,\bar{q})}
  \,{=}\, \mathrm e^{\mathrm i \pi (q^2-\bar{q}{}^2)/4} \phi_{(q,\bar{q})}$
  the OPE coefficients are $\mathrm i^{m_1 w_2-m_2 w_1}$. 
  }

The conformal field theory of a free boson compactified at
radius $R$ as described above will be denoted by $\FB{R}$. 
It is not difficult to see that the bulk spectrum (\ref{eq:FB-bulk-states})
is invariant under the substitution $R \,{\mapsto}\, \alpha'/R$; this
is the usual T-duality relation for the compactified free boson. 

If applied to the perturbative expansion of the string free energy $F(R,g_s)$, 
one must take into account that T-duality also acts on the dilaton field (see 
e.g.\ \cite{Polchinski:1996na}).  As a result, at the same time as changing the 
radius one must also modify 
the string coupling constant, $F(R,g_s) \,{=}\, F(R',g_s')$ with
  \be
  R' = \frac{\alpha'}{R} \,, ~~\quad
  g_s' = \frac{\sqrt{\alpha'}}R\, g_s \,.
  \labl{eq:T-dual-parameters}
We will recover this change in $g_s$ when analysing the description of T-duality
in terms of topological defects in section \ref{sec:T-duality-rat} below. 

\medskip

In the following we shall set the value of the parameter $\alpha'$ to
  \be
  \alpha' = \Frac12 \,.
  \labl{eq:alpha-12}
With this choice, the radius T-dual to $R$
is $R'\,{=}\,1/(2R)$, and the self-dual radius is 
  \be
  \Rsd = 1/\sqrt2 \,.
  \labl{Rsd}
To restore the $\alpha'$-dependence in the expressions below, one
simply has to substitute all appearances of $R$ by $R/\sqrt{2\alpha'}$.


\sect{Topological defects for the free boson}\label{sec:topdef-sum}

Before going through details of the calculations, let us explain and
summarise the results found in section \ref{sec:rat-def} and
\ref{sec:topdef-gen}. We will explicitly give operators $\hat D$ of
the various topological defects and compute their compositions.

\subsection{Topological defects preserving the $\Ueh$-symmetry}
\label{sec:u1-pres-def}

The simplest topological defects are those that actually preserve more
symmetry than just the two Virasoro symmetries \erf{eq:D-Vir-com}. In
particular, we can demand that the topological defect also intertwines
the $\Ueh$-symmetries up to automorphisms,
  \be
  J_m^{2} \,\hat D = \epsilon\,\hat D \,J_m^{1} \qquad
     \text{and} \qquad
  \bar J_m^{2} \, \hat D = \bar\epsilon\, \hat D \, \bar J_m^{1} 
  \qquad \quad
  \text{for}~~\epsilon,\bar\epsilon \,{\in}\, \{ \pm 1\} \,.
  \labl{eq:D-U1-com}
This condition implies \erf{eq:D-Vir-com}. From the action of the 
zero modes $J_0^{1,2}$ and $\bar J_0^{1,2}$ we see that $\hat D$ has 
to map the highest weight state $(q,\bar{q})\,{\in}\, \Lambda(R_1)$ to the 
highest weight state $(\epsilon q,\bar{\epsilon} \bar{q})\,{\in}\,\Lambda(R_2)$.  
Thus $\hat D$ can only be nonzero in sectors for which
$(q,\bar{q})$ lies in the intersection $\Lambda$ of the two lattices,
  \be
  \Lambda = \Lambda^{\epsilon,\bar\epsilon}(R_2) \cap \Lambda(R_1)
  \,,\qquad \text{where} \qquad
  \Lambda^{\epsilon,\bar\epsilon}(R)
  = \big\{ (\epsilon q,\bar \epsilon \bar{q}) \, \big| \,
  (q,\bar{q}) \,{\in}\, \Lambda(R) \big\} .
  \labl{eq:Lambda-intersect}
To describe this lattice more explicitly, we observe that 
  \be
  \Lambda^{\epsilon,\bar\epsilon}(R) = \Lambda(\RR) \,, \qquad
  \hbox{where} \qquad 
  \RR := \left\{ \begin{array}{ll}
  R & \hbox{if~ $\epsilon\,{=}\,\bar\epsilon$\,,} \\
  \Frac{1}{2R} & \hbox{if~ $\epsilon\,{=}\,{-}\bar\epsilon$\,,}
  \end{array} \right.
  \labl{eq:lattice-Lam-epseps}
as follows directly from (\ref{basis}). Thus the
intersection $\Lambda$ consists of all points such that
  \be
  \frac{m}{2 R_1} \begin{pmatrix} \,1 \\ \,1 \end{pmatrix} 
  + w R_1 \begin{pmatrix} \,1 \\ -1 \end{pmatrix} 
  =
  \frac{m'}{2 \RR_2} \begin{pmatrix} \,1 \\ \,1 \end{pmatrix} 
  + w' \RR_2 \begin{pmatrix} \,1 \\ -1 \end{pmatrix}  
  \qquad \text{for some} ~~ m,w,m',w' \,{\in}\, \Zb \,.
  \ee
In particular, this means that 
$m \,{=}\, (R_1/\RR_2) m'$ and $w \,{=}\, (\RR_2/R_1) w'$. We will treat
separately the  cases that $\RR_2/R_1$ is rational or irrational.


\subsubsection{$\RR_2/R_1$ rational}\label{sec:RR-rat}

Let us write $\RR_2/R_1 \,{=}\, M/N$, where $M$ and $N$ are coprime
positive integers. It then follows from the discussion above that the lattice
$\Lambda$ is spanned by the vectors $N/(2R_1) \cdot (1,1)$ and 
$MR_1 \cdot (1,-1)$. Because of \erf{eq:D-U1-com} the defect operator is fixed 
on each sector $\Hc_{q}\otimes \bar{\Hc}_{\bar{q}}$ of $\Hc(R_1)$ 
once we know its action on the primary bulk fields $\phi_{(q,\bar q)}$
chosen in section \ref{sec:comp-FB}. One finds that the possible
defect operators are parametrised by two complex numbers
$x$ and $y$,  and that they act on the primary bulk fields as
  \be
  \hat D(x,y)^{\epsilon,\bar\epsilon}_{R_2,R_1} \phi_{(q,\bar q)}
  = \sqrt{MN} (\epsilon \bar\epsilon)^{\frac12 q^2 - \frac12 \bar q^2}
  \mathrm e^{2 \pi \mathrm i (x q - y \bar q)} 
  \phi_{(\epsilon q,\bar \epsilon\bar q)}
  \ee
if $(q,\bar q) \in \Lambda$ and as zero otherwise. Note that the exponent of 
the sign $\epsilon \bar\epsilon$ is an integer, since it is the difference 
$h\,{-}\,\bar h$ of the left and right conformal weights of the $\Ueh$-primary 
field with charge $(q,\bar q)$. The complete defect operator can be written as 
  \be
  \hat D(x,y)^{\epsilon,\bar\epsilon}_{R_2,R_1}
  = \sqrt{MN} \sum_{(q,\bar q) \in \Lambda}
      (\epsilon \bar\epsilon)^{\frac12 q^2 - \frac12 \bar q^2}
  \mathrm e^{2 \pi \mathrm i (x q - y \bar q)}
  P^{\epsilon q,\bar \epsilon \bar q}_{q,\bar q}\,,
  \labl{eq:u1-def-op-explicit-simp}
where $P^{\epsilon q,\bar \epsilon \bar q}_{q,\bar q}{:}\ \Hc(R_1) 
\,{\rightarrow}\,\Hc(R_2)$ is the twisted intertwiner uniquely determined by
  \be \bearll&
  P^{\epsilon q,\bar \epsilon \bar q}_{q,\bar q}  
  \phi_{(q',\bar q')} = \delta_{q,q'} \delta_{\bar q,\bar q'}
  \,\phi_{(\epsilon q,\bar \epsilon\bar q)}
  \\{}\\[-.5em]& \text{and} \qquad
  J_m^{}\, P^{\epsilon q,\bar \epsilon \bar q}_{q,\bar q}  
  = \epsilon \, P^{\epsilon q,\bar \epsilon \bar q}_{q,\bar q}  J_m^{} \,,\quad
  \bar J_m^{}\, P^{\epsilon q,\bar \epsilon \bar q}_{q,\bar q}  = 
  \bar\epsilon \, P^{\epsilon q,\bar \epsilon \bar q}_{q,\bar q}  \bar J_m^{} \,.
  \eear \labl{eq:P-defining-eqn}

As will be explained in section \ref{sec:u1-pres-detail}, the specific scalar
coefficients that multiply the maps $P^{\epsilon q,\bar\epsilon\bar q}_{q,
\bar q}$ for each sector in $\Lambda$ can be determined by requiring consistency 
with the bulk OPE \erf{eq:bulk-OPE}. One can also verify that the defects 
\erf{eq:u1-def-op-explicit-simp} give consistent torus amplitudes, i.e.\ integer 
multiplicities in the channel in which the defects run parallel to the euclidean
time direction. Furthermore, when $R_1$ and $R_2$ square to rational numbers, 
some of the defects \erf{eq:u1-def-op-explicit-simp} can be  analysed from 
the point of view of the extended chiral symmetry. This is done in section 
\ref{sec:rat-def}; it demonstrates that at least these defects are consistent 
with all other sewing conditions as well. This leads us to believe that in fact 
all the operators \erf{eq:u1-def-op-explicit-simp} come from consistent defects.

In \erf{eq:u1-def-op-explicit-simp} $x$ and $y$ are a priori arbitrary
{\em complex\/} constants. However, unless $x,y \,{\in}\, \Rb$  the spectrum
of defect changing fields may contain complex conformal weights.  
(This is similar to the situation in \cite{Gaberdiel:2001xm}.)
Furthermore, not all values of $x$ and $y$ lead to distinct defect operators;
in fact it follows directly from \erf{eq:u1-def-op-explicit-simp} that 
  \be
  \hat D(x,y)^{\epsilon,\bar\epsilon}_{R_2,R_1}
  = \hat D(x',y')^{\epsilon',\bar\epsilon'}_{R_2,R_1}
  \quad~\text{iff} \quad ~
  \epsilon=\epsilon' ,~~ \bar\epsilon=\bar\epsilon' ,~~
  (x{-}x',y{-}y') \,{\in}\, \Lambda^* \,,
  \ee
where $\Lambda^*$ is the lattice dual to $\Lambda$, consisting of
all points $(x,y) \,{\in}\, \Rb^2$ such that 
$xq-y\bar q \,{\in}\, \Zb$ for all $(q,\bar q) \,{\in}\, \Lambda$.

\smallskip

It is also straightforward to read off the composition rules
by rewriting the composition of two defect operators as a sum over 
operators of the form \erf{eq:u1-def-op-explicit-simp}. One finds, for
example,   
  \be\begin{array}{rll}
  \hat D(x,y)^{\epsilon,\epsilon}_{R,R} \etb\!\!\circ~ \hat
      D(u,v)^{\nu,\nu}_{R,R} 
   \etb\!\!=~ \hat D(\nu x{+}u, \nu y{+}v)^{\epsilon\nu,\epsilon\nu}_{R,R} \,, 
  \enl
  \hat D(0,0)^{\epsilon,\bar\epsilon}_{R_2,R_1} 
     \etb\!\!\circ~ \hat D(x,y)^{+,+}_{R_1,R_1}
     \etb\!\!=~ \hat D(x,y)^{\epsilon,\bar\epsilon}_{R_2,R_1}
  = \hat D(\epsilon x,\bar\epsilon y)^{+,+}_{R_2,R_2} 
   \circ \hat D(0,0)^{\epsilon,\bar\epsilon}_{R_2,R_1} \,,
  \enl
  \hat D(0,0)^{\epsilon,\bar\epsilon}_{R_1,R_2} \etb\!\!\circ~ 
  \hat D(0,0)^{\epsilon,\bar\epsilon}_{R_2,R_1} 
  \etb\!\!=~ \sum_{m=0}^{M-1} \sum_{w=0}^{N-1} 
  \hat D\big(\tfrac{m}{2MR_1}{+}\tfrac{w R_1}{N}\,,
  \,\tfrac{m}{2M R_1}{-}\tfrac{w R_1}{N}\big)^{\!+,+}_{\!R_1,R_1} \,.
  \eear\labl{eq:u1-def-fusion}
The equalities in the second and third lines imply that a defect with operator
$\hat D(x,y)^{\epsilon,\bar\epsilon}_{R,R}$
is group-like if and only if $M\,{=}\,N\,{=}\,1$. This is the case either if 
$\epsilon\,{=}\,\bar\epsilon$ and $R$ is arbitrary (in this case our results 
have also been confirmed using a geometric realisation of the defect lines 
\cite{Fuchs:2007fw}), or if $\epsilon\,{=}\,{-}\bar\epsilon$ and $R$ is the 
self-dual radius. The group $\mathcal{G}^{\Ue}_R$ of $\Ueh$-preserving
group-like defect operators for $\FB{R}$ is thus
  \be
  \mathcal{G}^{\Ue}_R
  = \begin{cases}
  (\Cb^2/\Lambda(R)) \rtimes (\Zb_2 \times \Zb_2) &\text{if}~~ 
     R = \tfrac{1}{\sqrt{2}}\,, \\
  (\Cb^2/\Lambda(R)) \rtimes \Zb_2 &\text{else} \,, 
  \end{cases}
  \labl{eq:u1-pres-group}
where we use that $\Lambda(R)^* \,{=}\, \Lambda(R)$, and abbreviate 
$\Zb/(a\Zb)$ by $\Zb_a$. The multiplication rule in the two cases is given by
  \be
  \begin{array}{rcll}
  (x,y,\epsilon,\bar\epsilon) \cdot (u,v,\nu,\bar\nu)
  & \!\!=\!\! & (\nu x{+}u,\bar\nu y{+}v,
  \epsilon \nu,\bar\epsilon \bar\nu) \quad & \text{if}~~ R=\tfrac{1}{\sqrt2}\,,
  \\{}\\[-.8em]
  (x,y,\epsilon) \cdot (u,v,\nu)
  & \!\!=\!\! & (\nu x{+}u,\nu y{+}v,\epsilon \nu) & \text{else}  \,, 
  \end{array}
\ee  
so that the $\Zb_2$'s are realised multiplicatively as $\{\pm 1\}$.
Note that the group \erf{eq:u1-pres-group} is non-abelian in all cases.

Furthermore we see from \erf{eq:u1-def-fusion} that {\em all\/} 
$\Ueh$-preserving defects are duality defects. The duality defect
implementing T-duality should take $J\,\bar J$ to $-J\,\bar J$ and be
an isomorphism. From \erf{eq:u1-def-op-explicit-simp} one sees that
this can only happen if $\bar\epsilon\,{=}\,{-}\epsilon$ and 
$\Lambda^{\epsilon,-\epsilon}(R_2) \,{=}\, \Lambda(R_1)$, i.e.\ if 
$R_2 \,{=}\, 2/R_1$. Also, up to the action of group-like defects
(cf.\ the  second line in \erf{eq:u1-def-fusion}), there is a exactly
one T-duality defect operator.


\subsubsection{$\RR_2/R_1$ irrational}\label{sec:RR-nonrat}

If $\RR_2/R_1 \,{\not\in}\, \Qb$ then $\Lambda \,{=}\, \{ (0,0) \}$,
so that up to a multiplicative constant the defect operators can only consist 
of the projection $P^{0,0}_{0,0}$ to the vacuum sector. Such an operator
corresponds to a defect $D$ with a continuous spectrum of defect fields. 
Furthermore, $\hat{\bar D} \,{\circ}\, \hat D$ will decompose  into an
integral of $\Ueh$-preserving fundamental defect operators of $\FB{R_1}$.   
Defects of this type probably exist, but it is difficult to check 
their consistency in detail.


\subsection{General topological defects}

So far we have considered topological defects that actually preserve the full 
$\Ueh$-symmetry. If we only require that the defect intertwines the Virasoro 
algebra, i.e.\ only impose \erf{eq:D-Vir-com}, but not \erf{eq:D-U1-com}, then 
there are also other defects.
In order to understand how they arise, we need
to decompose the various $\Ueh$-representations $\Hc_q$
into representations of the
Virasoro algebra. The result depends in a crucial manner on the value of $q$: 
if $q$ is not an integer multiple of $\frac{1}{\sqrt{2}}$, then $\Hc_q$ is 
irreducible with respect to the Virasoro action. On the other hand, if $q$ is 
an integral multiple of $\frac{1}{\sqrt{2}}$ we have the decomposition 
  \be
  \Hc_{q=\frac{s}{\sqrt{2}}} = 
  \bigoplus_{k=0}^{\infty} \Hc_{h=\frac14(|s|+2k)^2}^{\rm Vir} \,, 
  \labl{eq:u1-into-vir}
where $\Hc_{h}^{\rm Vir}$ denotes the irreducible Virasoro representation of 
highest weight $h$ (at central charge $1$), see e.g.\ \cite{Kac:1987gg}.
Whether or not the bulk state space $\Hc(R)$ contains reducible representations
other than $(0,0)$ depends on the arithmetic
properties of $R$. There are three cases to be distinguished:

\medskip

\noindent
{\bf Case 1:} The equation $x/(2R) + yR \,{=}\, 1/\sqrt{2}$ has no
solution for $x,y\,{\in}\, \mathbb{Q}$. 
(For example, this is the case for $R\,{=}\,1$.) Then there are no integer
solutions to $m/(2R) \,{\pm}\, wR \,{=}\, s/\sqrt{2}$ for any nonzero
$s$. Thus the only Virasoro degenerate representations come from the
vacuum sector $\{(0,0)\} \,{=}\, \Lambda(R)$. 

\medskip

\noindent
{\bf Case 2:} The equation $x/(2R) + yR \,{=}\, 1/\sqrt{2}$ has a solution with
$x,y\,{\in}\,\mathbb{Q}$, and $R$ and $1/R$ are linearly independent over 
$\mathbb{Q}$. (For example, this is the case for $R\,{=}\,\tfrac12 (1+\sqrt{2})$
which is solved by $x\,{=}\,\tfrac{1}{4}$, $y\,{=}\,\tfrac{1}{2}$.) It then 
follows that the solution $x,y$ is unique. Let $L$ be the least common multiple 
of the denominators of $x$ and $y$. Then all integer solutions to 
$m/(2R) + w R \,{=}\, s/\sqrt{2}$ are 
of the form $(m,w,s) \,{\in}\, (xL,yL,L) \Zb$. Consider the two one-dimensional 
sub-lattices $\Lambda_l$ and $\Lambda_r$ of $\Lambda(R)$ given by
  \be
  \Lambda_l := L \binom{1/\sqrt{2}}{x/(2R) - yR} \, \Zb  
      \qquad\text{and}\qquad
  \Lambda_r := L \binom{x/(2R) - yR}{1/\sqrt{2}}\, \Zb \,.
  \ee
The above calculation shows that all sectors $\Hc_q \otimes \bar\Hc_{\bar q}$ of
$\Hc(R)$ for which $\Hc_q$ is degenerate are given by $(q,\bar q) \,{\in}\, 
\Lambda_l$, and all sectors for which $\bar\Hc_{\bar q}$ is degenerate are given
by $(q,\bar q) \,{\in}\, \Lambda_r$. 
Note that $\Lambda_l \,{\cap}\, \Lambda_r \,{=}\, \{ (0,0) \}$.

\medskip\noindent
{\bf Case 3:} The equation $x/(2R) + yR \,{=}\, 1/\sqrt{2}$ has a solution with 
$x,y\,{\in}\, \mathbb{Q}$, and $R$ and $1/R$ are linearly dependent over 
$\mathbb{Q}$, i.e.\ there are nonzero $u,v \,{\in}\, \Qb$ such that $uR+v/R 
\,{=}\, 0$. Then $R^2\,{=}\,{-}v/u$, and hence $R^2$ is rational. We can then 
write $R^2 \,{=}\, P/(2Q)$ for some coprime $P,Q \,{\in}\, \Zb_{>0}$. 
Substituting this into $x/(2R) + yR\,{ =}\, 1/\sqrt{2}$ we conclude that  
$x + yP/Q \,{=}\, R/\sqrt{2}$, from which it follows that $R$ is a rational
multiple of the self-dual radius $R_{\rm s.d.}\,{=}\,1/\sqrt{2}$,
  \be
  R = \frac{E}{F}\, R_{\rm s.d.} 
  \ee
for some coprime non-negative integers $E$ and $F$. (In particular, we
have $P\,{=}\,E^2$ and $Q\,{=}\,F^2$.) It is now easy to check that
$(m,w)$ is an integer solution to $m/(2R) + w R \,{\in}\,2^{-\frac12}\Zb$ if 
and only if $m F^2 + w E^2 \,{\in}\, EF \Zb$. Evaluating the latter condition 
modulo $E$ and modulo $F$, and using that $E$ and $F$ are coprime, shows that 
$(m,w) \,{\in}\, (E \Zb) \,{\times}\, (F \Zb)$ provides all solutions to
$m/(2R) + w R \,{\in}\, 2^{-\frac12} \Zb$. The same set also provides all
solutions to $m/(2R) - w R \,{\in}\, 2^{-\frac12} \Zb$. It follows that 
in a sector $\Hc_q \oti \bar\Hc_{\bar q}$ of $\Hc(R)$
either both $\Hc_q$ and $\bar\Hc_{\bar q}$ are degenerate, 
or neither of them is, and that the Virasoro-degenerate
representations occur precisely for $(q,\bar q) \,{\in}\, \Lambda$ with  
  \be
  \Lambda = \Lambda(R) \cap \Lambda(R_{\rm s.d.})
  = \Big\{
  \frac{k F}{\sqrt{2}} \begin{pmatrix} \,1 \\ \,1 \end{pmatrix} 
  + \frac{l E}{\sqrt{2}} \begin{pmatrix} \,1 \\ -1 \end{pmatrix} 
  \Big| \, k,l \,{\in}\, \Zb \Big\} \,. 
  \labl{eq:case3-lattice}

\medskip

In this paper we shall only analyse case 3. From the results in case
3 one would suspect that any additional defects arising for cases 1
and 2 have a continuous spectrum of defect fields, and that their fusion can 
lead to a continuum of defects, rather than just a discrete superposition,
similar to the situation in section \ref{sec:RR-nonrat}. 
We note in passing that in case~3 the theory is always rational, while case~2 
can never arise for rational theories. (The free boson theory is rational iff 
$R^2$ is a rational multiple of $R_{\rm s.d.}^2\,{=}\,\tfrac{1}{2}$.) On the 
other hand, case~1 may or may not be rational.


\subsection{Virasoro-preserving defects at the self-dual radius}
\label{sec:Vir-self}

It is instructive to consider first the simplest example of case 3 which 
arises when $R \,{=}\, R_{\rm s.d.}$.  The free boson theory is then
equivalent to the $\su(2)$ WZW model at level 1.\,%
\footnote{~Note that the $\suh(2)_1$ modes satisfy
  $[J^3_m,J^\pm_n] \,{=}\,{\pm}J^\pm_{m+n}$ and 
  $[J^3_m,J^3_n] \,{=}\, \tfrac12 m\, \delta_{m,-n}$. In particular, the 
  $\Ue$-current $J(z)$ is related to $J^3(z)$ via 
  $J(z)\,{=}\,\sqrt2\,J^3(z)$.} 
The integrable $\suh(2)_1$-representations can be decomposed into
$\Ueh$-representations (see e.g.\ \cite[sect.\,15.6.2]{DiFrancesco:1997nk}),  
all of which are in turn reducible with respect to the Virasoro algebra, and 
thus decompose as in \erf{eq:u1-into-vir}. The resulting decomposition is 
most easily understood in terms of the zero modes of the affine Lie algebra 
which commute with the Virasoro generators. This allows us to decompose the 
space of states simultaneously with respect to the two Virasoro algebras 
(generated by $L_m$ and $\bar{L}_m$) and the two commuting $\su(2)$ algebras
(generated by $J^a_0$ and $\bar{J}^a_0$), leading to (see for example
\cite[sect.\,2]{Gaberdiel:2001xm}) 
  \be
  \Hc(R_{\rm s.d.}) = \hspace{-2mm}
  \bigoplus_{\scriptstyle s,\bar s=0 \atop \scriptstyle s+\bar s~\text{even}}
  ^\infty \hspace{-2mm} \Hc_{[s,\bar s]} \,,
  \qquad \text{where} \quad
  \Hc_{[s,\bar s]}= V_{s/2} \otimes \bar{V}_{\bar{s}/2} \otimes 
  \Hc^\text{Vir}_{s^2/4} \otimes \bar\Hc{}^\text{Vir}_{\bar s^2/4} \,.
  \labl{eq:Rsd-bulk-decompose}
Here $\Hc^\text{Vir}_{h}$ denotes the irreducible Virasoro
representation of highest weight $h$ (at central charge $1$), while $V_j$ is 
the irreducible $\su(2)$-representation of spin $j$. The spaces denoted by 
a barred symbol give the representation of the anti-holomorphic modes. 

Because of the condition \erf{eq:D-Vir-com}, a topological defect  
has to act as a multiple of the identity on each sector 
$\Hc^\text{Vir}_{s^2/4} \oti \bar{\Hc}^\text{Vir}_{\bar s^2/4}$. 
However, since these spaces now appear with multiplicity 
$\dim(V_{s/2} \oti \bar{V}_{\bar{s}/2})$, one can have a non-trivial
action on these multiplicity spaces. In fact one finds, in close
analogy to the result for conformal boundary conditions
\cite{Gaberdiel:2001xm}, that the action of the defect operator on 
the multiplicity spaces $V_{s/2}\otimes \bar{V}_{\bar{s}/2}$ is
described by a pair of group elements $g,h\,{\in}\, {\rm SL}(2,\Cb)$,
where $g$ acts on each $V_{s/2}$, and $h$ on each
$\bar{V}_{\bar{s}/2}$. Since the action of
${\rm SL}(2,\Cb)$ on these spaces is described in terms of $J^a_0$ and
$\bar{J}^a_0$, respectively, we can write the defect operator compactly as  
  \be
  \hat D(g,h) = \exp( \alpha_a J^a_0 + \beta_b \bar J{}^b_0 )
  \qquad \text{where}\quad
  g = \exp(\alpha_a J^a)\,,~~ h = \exp(\beta_a J^a) \,.
  \labl{eq:Rsd-zeromode-exp}
More explicitly, on the summand $\Hc_{[s,\bar{s}]}$ in 
(\ref{eq:Rsd-bulk-decompose}) the action is 
  \be
  \hat D(g,h) \big|_{\Hc_{[s,\bar s]}}^{} = 
  \rho_{s/2}(g) \otimes \rho_{\bar s/2}(h) \otimes 
  \id_{\Hc^\text{Vir}_{s^2/4}} \otimes 
  \id_{\bar\Hc{}^\text{Vir}_{\bar{s}^2/4}} \,, 
  \labl{eq:Rsd-Vir-pres-def}
where $\rho_{s/2}$ is the representation of ${\rm SL}(2,\Cb)$ on
$V_{s/2}$, and similarly for $\rho_{\bar s/2}$. It is then easy to see
that $\hat D(g,h) \,{=}\, \hat D(g',h')$ if and only if $(g,h) \,{=}\,(g',h')$
or $(g,h)\,{=}\,(-g',-h')$, and that the composition of two such defect 
operators is given by
  \be
  \hat D(g,h) \circ \hat D(g',h') = \hat D(g g', h h') \,.
  \labl{eq:su2-1-def-fuse}
Thus, all fundamental Virasoro-preserving defect operators of 
$\FB{R_{\rm s.d.}}$ are group-like and the corresponding group is
  \be
  \mathcal{G}^{\text{Vir}}_{R_{\rm s.d.}} = 
  \big( {\rm SL}(2,\Cb) \,{\times}\, {\rm SL}(2,\Cb) \big) / \{ \pm 1 \} \,. 
  \labl{eq:su2-1-vir-def-group}
Except for the quotient by the center $\Zb_2$ of the diagonally embedded 
${\rm SL}(2,\Cb)$, which is a direct consequence of the restriction to 
$s\,{+}\,\bar s \,{\in}\, 2\Zb$ in the sum \erf{eq:Rsd-bulk-decompose}, we 
thus find just a doubling of the result found for conformal boundary states 
\cite{Gaberdiel:2001xm}, which are parametrised by ${\rm SL}(2,\Cb)$. In 
contradistinction to the case of the boundary conditions, the multiplicative 
structure of ${\rm SL}(2,\Cb)$ now has a meaning in terms of the
fusion of the defects; similarly, the action of a 
defect operator on a boundary state is given by the 
adjoint action $D(g_1,g_2) |\!| h\rangle\!\rangle \,{=}\, |\!| g_1\,h\, 
g_2^{-1} \rangle\!\rangle$.
Using the description in terms of D-branes of the product theory one  
can prove, using the same methods as in \cite{Gaberdiel:2001xm}, that
these are  the only fundamental defect operators in this case. We
shall give an alternative proof directly in terms of defects in
section \ref{sec:Vir-self-detail}.


\subsection{Virasoro-preserving defects at rational multiples of 
$R_{\rm s.d.}$} 
\label{sec:Vir-rat-mult}

In the previous section we have constructed all fundamental defect operators
of the self-dual theory. Together with the $\Ueh$-preserving defect operators 
that we found in section \ref{sec:RR-rat}, we can now immediately obtain a
class of defect operators between two theories whose radii are
rational multiples of the self-dual radius $R_{\rm s.d.}$, i.e.
  \be
  R_1 = \frac{E_1}{F_1}\, R_{\rm s.d.} \,, \qquad
  R_2 = \frac{E_2}{F_2}\, R_{\rm s.d.} \,.
  \ee
In fact, we can simply compose two $\Ueh$-preserving radius-changing
defects with a general Virasoro-preserving defect at the self-dual radius,  
which for the corresponding operators yields
  \be
  \hat D(0,0)^{+,+}_{R_2,R_{\rm s.d.}} {\circ}\; 
  \hat D(g,h) \circ \hat D(0,0)^{+,+}_{R_{\rm s.d.},R_1}
  =: \hat D(g,h)_{R_2,R_1}^{} \,.
  \labl{eq:R2-R1-vir-pres}
We could also use the $\Ueh$-preserving defects with $(\epsilon,\bar\epsilon) 
\,{\neq}\, (+,+)$ or with $(x,y)\,{\neq}\, (0,0)$, but as shall become clear 
below, they do not generate any additional defect operators. In fact, since the
Virasoro-degenerate representations in $\Hc(R_1)$ are precisely those
representations that $\Hc(R_1)$ has in common with $\Hc(R_{\rm s.d.})$, it is 
reasonable to assume that \erf{eq:R2-R1-vir-pres} does produce all new defect 
operators that appear when imposing \erf{eq:D-Vir-com}, but not 
\erf{eq:D-U1-com}. In section \ref{sec:Vir-self-detail} we present an
argument that this is indeed the case.

It turns out that the operators \erf{eq:R2-R1-vir-pres} are not all distinct. To
describe the identification rule, it is
helpful to introduce, following \cite{Gaberdiel:2001zq}, the matrices 
  \be
  \Gamma_{\!L} := \begin{pmatrix}
  \mathrm e^{\pi \mathrm i /L} & 0 \\
  0 & \mathrm e^{-\pi \mathrm i / L} \end{pmatrix} ~ \in {\rm SU}(2) \,.
  \ee
One then finds (see section \ref{sec:Vir-self-detail} for details) that the
operators \erf{eq:R2-R1-vir-pres} are parametrised by elements of the 
double coset 
  \be
  \raisebox{-.4em}{$\Zb_{E_2}{\times}\Zb_{F_2\!}$} 
  \!{\raisebox{-.2em}{$\big\backslash$}}
  \big( {\rm SL}(2,\Cb) \,{\times}\, {\rm SL}(2,\Cb) / \{ \pm 1 \}\big) 
  \raisebox{-.2em}{$\big/$}
    \raisebox{-.4em}{$\Zb_{E_1}{\times}\Zb_{F_1}$} \,,  
  \labl{eq:vir-preserve-set}
where the quotient by $\{ \pm 1 \}$ is as in \erf{eq:su2-1-vir-def-group}, the
right action of an element $(k,l) \,{\in}\, \Zb_{E_1} \,{\times}\, 
\Zb_{F_1}$ is as $(g,h) \,{\mapsto}\,
(g\,\Gamma_{\!E_1}^k\,\Gamma_{\!F_1}^l,\, 
h\,\Gamma_{\!E_1}^{-k} \,\Gamma_{\!F_1}^{l})$ and the left action of
an element 
$(k,l) \,{\in}\, \Zb_{E_2} \,{\times}\, \Zb_{F_2}$ is given by
$(g,h) \,{\mapsto}\, (\Gamma_{E_2}^{k} \, \Gamma_{\!F_2}^{l} \,g , \,
\Gamma_{E_2}^{-k} \,\Gamma_{\!F_2}^{l} \,h)$.

We also observe that the $\Ueh$-preserving defect operators at $R_{\rm s.d.}$ 
are just special cases of \erf{eq:su2-1-vir-def-group}, for example 
  \be
  \hat D(x,y)^{+,+}_{R_{{\rm s.d.}},R_{{\rm s.d.}}} = \hat D(g,h)
  \quad \text{for} ~~~
  g = \begin{pmatrix}
  \!\mathrm e^{\pi \mathrm i \sqrt{2}\, x}\!\!\! & 0 \\
  0 & \!\!\!\mathrm e^{-\pi \mathrm i \sqrt{2} \,x}\!
  \end{pmatrix} ,~~
  h = \begin{pmatrix}
  \!\mathrm e^{-\pi \mathrm i \sqrt{2}\, y}\!\!\! & 0 \\
  0 & \!\!\!\mathrm e^{\pi \mathrm i \sqrt{2} \,y}\!
  \end{pmatrix} ,
  \labl{eq:diag-Dgh-u1-pres}
see section \ref{sec:Vir-self-detail} below for more details. In
particular,  
$D(\Gamma_{\!L},\Gamma_{\!L'}) \,{=}\, D\big(\tfrac1{\sqrt{2}L}\,,
-\tfrac1{\sqrt{2}L'} \big)^{\!+,+}_{\!R_{\rm s.d.},R_{\rm s.d.}}$. 

The composition law for these defect operators can then be easily obtained by 
combining \erf{eq:u1-def-fusion},
\erf{eq:diag-Dgh-u1-pres} and \erf{eq:su2-1-def-fuse}. One finds 
  \be
  \hat D(g,h)_{R_3,R_2} \circ \hat D(g',h')_{R_2,R_1}
  = \sum_{m=0}^{E_2-1} \sum_{w=0}^{F_2-1}
  \hat D(g \Gamma_{\!E_2}^{\,m} \Gamma_{\!F_2}^{\,w} g', h 
  \Gamma_{\!E_2}^{\,-m} \Gamma_{\!F_2}^{\,w} h')_{R_3,R_1} \,.
  \ee
Using this result we can verify for which $(g,h)$ the defect operator 
$\hat D(g,h)_{R_2,R_1}$ is fundamental. To this end we need to check
whether  the trivial defect appears exactly once in the product 
$\hat D(g^{-1},h^{-1})_{R_1,R_2} \,{\circ}\, \hat D(g,h)_{R_2,R_1}$,
as is  required for fundamental defect operators. To do so, we count
for how many pairs 
$(m,w)$ the elements $g^{-1} \Gamma_{E_2}^{\,m} \Gamma_{F_2}^{\,w} g$ and
$h^{-1} \Gamma_{E_2}^{\,-m} \Gamma_{F_2}^{\,w} h$ are equal to $(e,e)$ 
modulo the identification \erf{eq:vir-preserve-set}.
This in particular requires 
$g^{-1} \Gamma_{E_2}^{\,m}\Gamma_{F_2}^{\,w} g$ (say) to be diagonal,
which in turn is possible only if $(m,w)\,{=}\,(0,0)$ or 
if $g$ is itself either diagonal or anti-diagonal. The same holds for
$h$.  Thus if $g$ and $h$ are neither diagonal nor anti-diagonal, then
$\hat D(g,h) _{R_2,R_1}$ is fundamental. On the other hand, if $g$ and
$h$ are diagonal or  anti-diagonal, then whether 
$\hat D(g,h)_{R_2,R_1}$ is fundamental or not  
depends on the divisibility properties of $R_1$ and $R_2$. For
example, if $g$ 
and $h$ are both diagonal and if $E_1$, $E_2$ are coprime and
$F_1,F_2$ are  
coprime, then $\hat D(g,h)_{R_2,R_1}$ is fundamental. 
On the other hand, if $g$ and $h$ are as in \erf{eq:diag-Dgh-u1-pres}
and $R_1\,{=}\,R_2$, \erf{eq:u1-def-fusion} implies that  
  \be
  \hat D(g,h)_{R_1,R_1}
  = \sum_{m=0}^{F_1-1} \sum_{w=0}^{E_1-1} 
  \hat D\big(\tfrac{m}{\sqrt{2}\,E_1}{+}\tfrac{w}{\sqrt{2}\,F_1}{+}x\,,
  \,\tfrac{m}{\sqrt{2}\,E_1}{-}
     \tfrac{w}{\sqrt{2}\,F_1}{+}y\big)^{\!+,+}_{\!R_1,R_1} \,. 
  \labl{eq:Dgh-decompose}
Thus it follows that for $R_1\,{\neq}\, R_{\rm s.d.}$, 
$\hat D(g,h)_{R_1,R_1}$ is not fundamental if $g$ and $h$ are both
diagonal.


\subsection{Completeness of the set of defect operators}\label{sec:complete}

Before proceeding to the detailed calculations, let us discuss 
whether the results listed in sections 
\ref{sec:u1-pres-def}\,--\,\ref{sec:Vir-rat-mult} provide all topological 
defects for the free boson. There are in fact three related, but distinct, 
questions one can pose. The first and most obvious one is
\begin{list}{-}{\topsep .4em \leftmargin 2.3em}
    \item[{\bf Q1:}] 
What are all topological defects that join the theories $\FB{R_1}$ and 
$\FB{R_2}$?
\end{list}
Just as in the rational case treated in section \ref{sec:rat-def}, the 
collection of all topological defects is probably best described as a suitable 
category, rather than as a set. To address Q1 one must then decide when two 
defects should be regarded as `isomorphic', i.e.\ when they give rise to the 
same correlators involving in particular defect fields, but also bulk fields, 
boundaries, etc. While this will in general be difficult, one can restrict 
oneself to considering defect operators and ask the more concrete question 
\begin{list}{-}{\topsep .4em \leftmargin 2.3em}
    \item[{\bf Q2:}] 
What are all operators $L{:}\ \Hc(R_1) \,{\rightarrow}\, \Hc(R_2)$ that arise 
as defect operator for some topological defect joining $\FB{R_1}$ and $\FB{R_2}$?
\end{list}
The questions Q1 and Q2 are indeed different: In general a defect is not 
determined uniquely by its associated defect operator. For example, while the 
operator does determine the spectrum of defect fields, it is not always possible
to deduce their OPE, or even the representation of the Virasoro algebra on the 
space of defect fields. This aspect can be stressed by asking instead
\begin{list}{-}{\topsep .4em \leftmargin 2.3em}
    \item[{\bf Q2':}] 
Given an operator $L{:}\ \Hc(R_1) \,{\rightarrow}\, \Hc(R_2)$, what are all 
topological defects $D$ 
joining $\FB{R_1}$ and $\FB{R_2}$ such that $L \,{=}\, \hat D$?
\end{list}
As an illustration, we construct in appendix \ref{app:family} a one-parameter 
family of mutually distinct defects which all have the same defect operator. 
This family is obtained by perturbing a superposition of two defects by a 
marginal defect-changing field. In this example the perturbing field is not 
self-adjoint, and the resulting defects are `logarithmic' in the sense that 
the perturbed Hamiltonian is no longer diagonalisable. One can then ask whether
restricting oneself to defects for which the Hamiltonian generating translations
along the cylinder is self-adjoint on all state spaces (for disorder-, defect-, 
or defect-changing fields) makes the assignment ~(defect) $\mapsto$ (defect 
operator)~ injective. We think that this is indeed true for
the free boson, but we do not have a proof.

For the Virasoro-preserving defects, in this paper we will consider 
the following variant of Q2:
\begin{list}{-}{\topsep .4em \leftmargin 2.3em}
    \item[{\bf Q3:}] 
What are all operators $L{:}\ \Hc(R_1) \,{\rightarrow}\, \Hc(R_2)$ that arise 
as defect operator for some topological defect $D$, for which the spectrum of 
defect fields (calculated from the torus with insertion of $\bar D \,{*}\, D$) 
contains a unique (up to normalisation) state of lowest conformal weight 
$h\,{=}\,\bar h\,{=}\,0$, separated by a gap from the rest of the spectrum 
with $h\,{+}\,\bar h\,{>}\,0$~?
\end{list}
Uniqueness of the lowest weight state implies that the defect operator $\hat D$ 
is fundamental. For, suppose that $\hat D \,{=}\, \hat D_1 \,{+}\, \hat D_2$. 
Then the trace contains terms coming from $\bar D_1 \,{*}\, D_1$ and 
$\bar D_2 \,{*}\, D_2$, 
both of which lead to a field with conformal weight $h\,{=}\,\bar{h}\,{=}\,0$ in the
spectrum.

\medskip

In section \ref{sec:Vir-self-detail} (as summarised in section
\ref{sec:Vir-self}) we answer Q3 for $\FB{R_\text{s.d.}}$ by showing that there 
cannot be more defect operators than those listed in \erf{eq:Rsd-Vir-pres-def} 
and that they in fact are all realised as perturbations of the trivial defect. 
Furthermore, we argue that the answer to Q3 for $\FB{R}$, with $R$ a rational 
multiple of $R_\text{s.d.}$, is given by combining the defect operators 
\erf{eq:u1-def-op-explicit-simp} (with $R_1\,{=}\,R_2\,{=}\,R$) and 
\erf{eq:R2-R1-vir-pres}. We lack a proof that these are all possible operators 
(see, however, the short remark at the end of section \ref{sec:Vir-self-detail}).

\medskip

Finally we would like to stress that the same issues we discussed above for 
topological defects also arise in the classification of conformal boundary 
conditions. Indeed, it is in general not true that the boundary state 
determines the boundary condition uniquely. Again an example can be constructed 
by perturbing a superposition of boundary conditions by a (non-selfadjoint) 
boundary changing field.


\sect{Free boson with extended chiral symmetry}\label{sec:rat-def}

In this section we investigate topological defects for the compactified
free boson using the methods developed in
\cite{Fuchs:2002cm,Fuchs:2004dz,Frohlich:2006ch}.
These apply to {\em rational\/} conformal field theories.


\subsection{Chiral symmetry}\label{sec:extend-chiral}

For any $N \,{\in}\, \Zb_{>0}$ the $\Ue$ current algebra can be
extended by the  two vertex operators 
$W^\pm_N(z) \,{=}\, \noo \eE^{\pm 2\ii \sqrt{2N}\,\phi(z)}  
\noo$ of $\Ue$-charge $\pm \sqrt{2N}$ and conformal weight $N$.
The resulting chiral algebra, denoted by $\Ueh_{N}$, is in
fact rational. $\Ueh_{N}$ has $2N$ inequivalent irreducible highest
weight  representations, labelled $U_0,\, U_1, ...\,,\, U_{2N-1}$,
which decompose into  irreducible representations of the $\Ue$
(vertex) subalgebra as 
  \be
  U_k \cong \bigoplus_{m \in \Zb} \Hc_{(k + 2Nm)/\sqrt{2N}} \,.
  \labl{eq:Uk-via-U1}
We denote by $\Uc_{N}$ the category formed by the representations of  
$\Ueh_{N}$. $\Uc_{N}$ is a modular tensor category; thus it is
balanced braided  
rigid monoidal, which means, roughly speaking, that given two objects $U,V$ of 
$\Uc_{N}$ one can take the tensor product $U \oti V$ (monoidal), that there are
two distinct ways to go from $U \oti V$ to $V \oti U$ (braided), that each 
object has a two-sided dual, namely its contragredient representation (rigid) 
and a balancing twist, namely its exponentiated conformal weight (balanced);
and finally the braiding is maximally non-degenerate (modular). 
See e.g.\ \cite[app.\,A.1]{Fjelstad:2005ua} for details and references.


\subsection{Rational compactified free boson}

In the TFT approach to rational CFT \cite{Fuchs:2002cm} one finds that
the algebra of boundary fields for a given single boundary condition determines
the entire CFT, including in particular the bulk spectrum, other
boundary conditions, and also topological defects. 
The boundary conditions and topological defects one finds in this way
are precisely those which preserve the rational chiral algebra.

In terms of the representation category, the algebra of boundary fields of the 
rational free boson theory is a symmetric special Frobenius algebra $A$ in the 
category $\Uc_{N}$. If one requires that there is a unique boundary vacuum 
state, i.e.\ a unique primary boundary field of conformal weight zero, then, as
shown in \cite[sect.\,3.3]{Fuchs:2004dz}, every such algebra is of the form
  \be
  A_r = \bigoplus_{k=0}^{r{-}1} U_{2kN/r}
  \qquad \text{where} ~~ r \,{\in}\, \Zb_{>0} ,~ r~ \text{divides}~ N ,
  \labl{eq:Ar-decomp}
and for each of them the multiplication is unique up to isomorphism. In other 
words, together with \erf{eq:Uk-via-U1}, $A_r$ gives the space of boundary 
fields, and the boundary OPE is unique up to field redefinition.

All other quantities of the full CFT can be computed starting from the algebra 
$A_r$. Consider for example the matrix $Z_{ij}^{(r)}$ which determines the bulk
partition function via $Z \,{=}\, \sum_{i,j=1}^{2N{-}1} Z_{ij}^{(r)}\,
\chi_i(\tau) \,\chi_j({-}\tau^*)$. Here $\chi_k(\tau) \,{=}\, 
\text{tr}_{U_k} \eE^{2\pi\ii\tau(L_0-\frac{1}{24})}$ is the Virasoro character 
of $U_k$. For $A_r$ one finds \cite[sect.\,5.6.1]{Fuchs:2002cm}\,\footnote{%
 ~The notation $A_{2r}$ used in \cite{Fuchs:2002cm} corresponds to $A_{N/r}$
 in the present convention, e.g.\ here $A_1 \,{=}\, U_0$, while in
 \cite{Fuchs:2002cm} one has $A_{2N}\,{=}\,U_0$.}
  \be
  Z_{ij}^{(r)} = \delta_{i+j,0}^{[2N/r]} \, \delta_{i,j}^{[2r]} \,,
  \labl{eq:Z^(r)-matrix}
where $\delta^{[p]}$ is the periodic Kronecker symbol, i.e.\
$\delta^{[p]}_{i,j} \,{=}\, 1$ if $i\,{\equiv}\,j\,\text{mod}\,p$ and
$\delta^{[p]}_{i,j} \,{=}\, 0$ else.
Comparison with \erf{eq:FB-bulk-states} and \erf{eq:Uk-via-U1} shows that
the compactification radius $R$ is related to $r$ and $N$ via
  \be
   R =  r / \sqrt{2N} \,.
  \labl{eq:R-for-Ar}

Conversely, the free boson compactified at a radius of the form 
   $R \,{=}\, \sqrt{P/(2Q)}$ 
with $P,Q$ coprime positive integers contains in its subset of
holomorphic bulk  fields the chiral algebra $\Ueh_{n^2PQ}$, for any
choice of $n\,{\in}\,\Zb_{>0}$. The relevant algebra in $\Uc_{n^2PQ}$
is then $A_r$ with $r\,{=}\,nP$. 

The simplest choice is just to set $n\,{=}\,1$. However, recall that
in the TFT approach one can only describe boundary conditions and
topological defects that  preserve the rational chiral algebra one
selects as a starting point. If we keep $n$ as a parameter, this
chiral algebra is $\Ueh_{n^2PQ}$. Taking $n$ large  
allows us to obtain more boundary conditions and defects. In fact, the 
$\Ueh$ algebra and its extension $\Ueh_{n^2PQ}$ start to differ only from 
$L_0$-eigenvalue $n^2 PQ$ onwards, and the explicit 
calculation shows that in the $n\,{\rightarrow}\,\infty$ limit we obtain all 
boundary conditions and defects that preserve only the $\Ue$ current algebra.


\subsection{Rational topological defects as bimodules}\label{sec:rat-def-bi}

Let us fix a chiral algebra $\Ueh_{N}$. We want to compute the topological 
defects that interpolate between the free boson CFTs described by two algebras 
$A_r$ and $A_s$ in $\Uc_{N}$, i.e.\ those topological defects that are 
transparent to the fields of $\Ueh_{N}$ and link a free boson compactified 
at radius $R_1 \,{=}\, r/\sqrt{2N}$
to a free boson compactified at $R_2 \,{=}\, s/\sqrt{2N}$.
Note that in this description we automatically have $R_1 / R_2 \,{\in}\, \Qb$. 

In the TFT approach the task of finding all topological defects preserving 
$\Ueh_{N}$ is reduced to finding all $A_r$-$A_s$-bimodules
in the category $\Uc_{N}$. We will solve the latter problem in two steps. First
we describe all $A_r$-modules and $A_r$-bimodules, and afterwards
also the $A_r$-$A_s$-bimodules for $r\,{\neq}\, s$.


\subsubsection{Modules and bimodules of the algebra $A_r$}\label{sec:Ar-Ar-bimod}

Since the objects $U_j$ appearing in the decomposition \erf{eq:Ar-decomp} of 
$A_r$ are all simple currents \cite{Schellekens:1989am}, the modules and 
bimodules of $A_r$ can be 
obtained using the methods of \cite{Fuchs:2004dz,Frohlich:2006ch}. 

As for the left $A_r$-modules, one finds that every simple module is isomorphic 
to an induced module $A_r \oti U_k$, and that the induced modules $A_r \oti U_k$
and $A_r \oti U_l$ are isomorphic if and only if $k \,{\equiv}\, l \bmod 2N/r$.
Thus there are $2N/r$ distinct simple (left) $A_r$-modules, which we denote by
$M^{(r)}_\kappa$, $\kappa \,{\in}\, \{0,1,...\,,\tfrac{2N}r{-}1\}$. The 
one-point function of a bulk field on a disk with boundary condition labelled by
$M^{(r)}_\kappa$ is zero for a pure momentum state, so that in string theory 
these boundary conditions correspond to $D1$-branes with equally spaced Wilson 
line parameter.

At this point it is useful to think of the radius as being given and of the 
form 
$R \,{=}\, \sqrt{P/(2Q)}$ (with $P$ and $Q$ coprime), and to describe this CFT 
via the algebra $A_{nP}$ in $\Uc_{n^2PQ}$, i.e.\ we set $r\,{=}\,nP$ and $N\,{=}
\,n^2PQ$.  Then the label $\kappa$ of the simple modules takes values in $\{0,
1,...\,,2 n Q {-}1\}$, and the corresponding 
$D1$-branes have values for their Wilson lines that are equally spaced.

The simple $A_r$-bimodules can be obtained by the methods of
\cite[sect.\,5]{Frohlich:2006ch}; we defer the details to appendix 
\ref{app:ArAr-bi}. The result is that the (isomorphism classes of) simple
$A_r$-bimodules are in one-to-one correspondence to elements of the abelian group
  \be
  G_{n,P,Q} = (\Zb_{nP} \times \Zb_{nQ} \times \Zb) \,\big/ 
  \,\langle \,(1,1,-2)\, \rangle \,,
  \labl{eq:GnPQ-def}
where $\langle (1,1,-2) \rangle$ denotes the subgroup generated
by the element $(1,1,-2)$ of $\Zb_{nP} \,{\times}\, \Zb_{nQ}\,{\times}\, \Zb$.
Note that since every element of $G_{n,P,Q}$ can be written either as $(a,b,0)$ 
or $(a,b,1)$ for suitable $a$ and $b$, $G_{n,P,Q}$ has $2 n^2 PQ \,{=}\, 2N$ 
elements (and thus is in particular finite). We denote these simple 
$A_r$-bimodules by $B\oor_{(a,b,\rho)}$ for $(a,b,\rho)\,{\in}\,G_{n,P,Q}$.

The fusion of two topological defects is obtained by computing the tensor
product over $A_r$ for the corresponding bimodules. 
The calculation is done in appendix \ref{app:ArAr-bi}. The result is that 
the tensor product is just addition in $G_{n,P,Q}$, i.e.\ for any
two elements $(a,b,\rho), (c,d,\sigma) \,{\in}\, G_{n,P,Q}$ one has
  \be
  B\oor_{(a,b,\rho)} \otimes_{\!A_r}^{} B\oor_{(c,d,\sigma)}
  \cong B\oor_{(a+c,b+d,\rho+\sigma)} \,.
  \ee
In particular, all topological defects preserving the extended 
chiral symmetry $\Ueh_{N}$ are group-like. Similarly, the fusion of a
topological defect to a conformal boundary condition is obtained by
the tensor product over $A_r$. One finds
  \be
  B\oor_{(a,b,\rho)} \otimes_{\!A_r}^{} M^{(r)}_{\kappa} 
  \cong M^{(r)}_{\kappa + 2b + \rho} \,.
  \ee
Thus the topological defect labelled by $B_{(a,b,\rho)}$ shifts the Wilson line 
of the $D1$-branes by $2b{+}\rho$ units. As shown by \erf{eq:D_abrho-bulk} 
below, the parameter $a$ amounts to a similar shift in the position 
of the $D0$-branes. Of course, $D0$-branes, for which in the present
convention $J(x) \,{=}\, {-} \bar J(x)$ on the boundary $x\,{\in}\,\Rb$ of 
the upper half plane, do not appear in the present description, as they do not 
preserve the $\Ueh_{N}$. But we can still deduce 
the effect of $a$ by computing the action of topological defects on bulk fields.

Let $\phi_{q,\bar q}$ be a bulk field corresponding to a state in the sector
$\Hc_{q} \,{\otimes}\, \bar\Hc_{\bar q}$ of the space
\erf{eq:FB-bulk-states} of bulk states. The action of the topological defect 
$D\oor_{(a,b,\rho)}$ corresponding to the bimodule $B^{(r)}_{(a,b,\rho)}$ amounts
simply to the multiplication by a phase (see appendix \ref{app:ArAr-bi}):
  \be
  \hat D\oor_{(a,b,\rho)}\, \phi_{q, \bar q}
  = \exp\!\Big\{ {-}2\pi \ii\,\Big(
  \frac{a{+}\rho/2}{nP} \,R\, \big(q \,{+}\, \bar q\big) + 
  \frac{b{+}\rho/2}{nQ} \,\frac1{2R}\, \big(q \,{-}\, \bar q\big) 
  \Big) \Big\} \, \phi_{q, \bar q} \,.
  \labl{eq:D_abrho-bulk}
We see that the parameter $a$ gives a phase shift depending only on the total 
momentum $q \,{+}\, \bar q$ of $\phi_{q, \bar q}$, while $b$ gives a phase 
shift depending on the winding number $q\,{-}\, \bar q$. Also, in the limit 
of large $n$ we find that the phase shifts are parametrised 
by two continuous parameters taking values in $\Rb/\Zb$. This agrees with 
the result stated in section \ref{eq:u1-def-op-explicit-simp}; specifically,
  \be
  \hat D\oor_{(a,b,\rho)} = 
  \hat D\big({-}(a{+}b{+}\rho)/\sqrt{2N},\,(a{-}b)/\sqrt{2N}\big)^{\!+,+}_{\!R,R}
  \,,
  \ee
where $R \,{=}\, r/\sqrt{2N}$. This also shows that at least for these
values of the parameters, the operators $\hat D(x,y)^{+,+}_{R,R}$
are indeed defect operators for a consistent defect.


\subsubsection{$A_r$-$A_s$-bimodules and radius-changing defects}
\label{sec:Ar-As-bi}

Consider two algebras $A_r$ and $A_s$ in $\Uc_{N}$. As shown in appendix 
\ref{app:ArAs-bi}, all simple $A_r$-$A_s$-bi\-mo\-dules can be obtained as 
follows. The algebra $A_{\text{lcm}(r,s)}$ is the smallest algebra that has 
both $A_r$ and $A_s$ as a subalgebra. By embedding $A_r$ and $A_s$ into 
$A_{\text{lcm}(r,s)}$ one defines the structure of an $A_r$-$A_s$-bimodule on 
$A_{\text{lcm}(r,s)}$; we denote this bimodule by $A\ors$. 
Every simple $A_r$-$A_s$-bimodule $B\ors$ can then be written as
  \be
  B\ors \cong B\oor_{(a,b,\rho)} \,{\otimes_{\!A_r}^{}}\, A\ors
  \cong A\ors \,{\otimes_{\!A_s}^{}}\, B\oos_{(c,d,\sigma)} \,,
  \labl{eq:radius-change-one-orbit}
for $B\oor_{(a,b,\rho)}$ and $B\oos_{(c,d,\sigma)}$ appropriate $A_r$- and 
$A_s$-bimodules, respectively. The total number of inequivalent simple 
$A_r$-$A_s$-bimodules is given by $\text{tr}(Z^{(r)} Z^{(s)}) \,{=}\, 2\, 
\text{gcd}(r,s)\, \text{gcd}(\frac Nr,\frac Ns)$, where $Z^{(\cdot)}$ is the  
matrix \erf{eq:Z^(r)-matrix}, see \cite[remark\,5.19]{Fuchs:2002cm}.

Let us refer to a defect as being elementary iff it corresponds to a simple
bimodule. The result above can then be rephrased as the statement that there
are $2\,\text{gcd}(r,s) \text{gcd}(\frac Nr,\frac Ns)$ distinct elementary
topological defects (transparent to fields in the chiral algebra $\Ueh_{N}$) 
that join the free boson compactified at radii $r/\sqrt{2N}$ and $s/\sqrt{2N}$.
All of these can be written as the fusion $D\oor_{(a,b,\rho)} \,{*}\, D\ors$ or 
$D\ors \,{*}\, D\oos_{(c,d,\sigma)}$, where $D\ors$ is the defect corresponding 
to the bimodule $A\ors$. In particular, all such radius-changing topological 
defects lie on a single orbit  with respect to the action of the group-like 
defects of the free boson on either side of the radius-changing defect.

As we did for the defects $D\oor_{(a,b,\rho)}$, let us also give the defect 
operator for the topological defect $D^{(rs)}$. One finds that, up to an overall 
constant, $\hat D^{(rs)}$ projects onto fields with left/right $\Ue$-charges
in the intersection of the charge lattices of the two theories. 
Concretely, if we fix a primary bulk field $\phi_{x,y}^{(r)}$ in each sector\,%
  \footnote{~Note that here the symbol `$\otimes$' does not stand for the tensor 
  product of the category $\Cc$, but rather $U \,{\otimes}\, \ol V$ is an object 
  of the product category (see \cite{Frohlich:2003hm}) 
  $\Cc \,{\boxtimes}\, \ol \Cc$.}
$U_x \,{\otimes}\, \ol U_y$ of $\Hc(R\,{=}\,\frac r{\sqrt{2N}})$
as in appendix \ref{app:ArAs-bi}, we have
  \be
  \hat D\ors \phi_{x,y}^{(s)} = \frac{\text{lcm}(r,s)}{r} \,
  \delta_{x+y,0}^{[2N/\text{gcd}(r,s)]}\,
  \delta_{x,y}^{[2\,\text{lcm}(r,s)]}\, \phi_{x,y}^{(r)} \,.
  \labl{eq:Drs-action}
The relation to the operator given in \erf{eq:u1-def-op-explicit-simp} is\,%
  \footnote{~The additional factor arises from different normalisation
  conventions for the zero-point functions on the sphere. 
  The expression \erf{eq:u1-def-op-explicit-simp} is computed for
  $\langle \one^{(R_2)} \rangle / \langle \one^{(R_1)} \rangle \,{=}\, 1$,
  whereas the TFT approach selects
  $\langle \one^{(R_2)} \rangle / \langle \one^{(R_1)} \rangle \,{=}\, s/r$.
  The analogue of \erf{eq:u1-def-op-explicit-simp} with unfixed normalisations
  is given in \erf{eq:u1-def-op-explicit} below.}
$\hat D\ors  \,{=}\, \sqrt{s/r} \, \hat D(0,0)^{+,+}_{R_2,R_1}$, where 
$R_1 \,{=}\, s/\sqrt{2N}$ and $R_2 \,{= }\, r/\sqrt{2N}$.
  
As opposed to the general non-rational case (compare the discussion in section
\ref{sec:complete}), for defects that preserve the rational chiral algebra
one can show that the defect operator determines the defect uniquely, in the 
sense that it fixes the corresponding bimodule up to isomorphism
\cite[prop.\,2.8]{Frohlich:2006ch}. For example, using \erf{eq:Drs-action} 
together with $D\ors \,{*}\, D\osr \,{=}\, 
D\oor_{\!A\ors\otimes_{\!A_s}^{}\!A\osr}$ it is straightforward to check that
  \be
  A\ors \otimes_{\!A_s}^{} A\osr \cong 
  \bigoplus_{m=0}^{\frac{\tilde \ell}{N/r}-1}
  \bigoplus_{n=0}^{\frac{\ell}{r}-1}
  D\oor_{(\frac{N}{\tilde \ell} m,\frac{N}{\ell} n,0)} 
  \qquad \text{with}~~
  \tilde \ell \,{=}\, \text{lcm}(\tfrac{N}{r},\tfrac{N}{s}) \,,~~
  \ell = \text{lcm}(r,s) \,.
  \ee
Acting with the defect $D\ors$ on a boundary condition of the free boson
at radius $s/\sqrt{2N}$ results in a boundary condition of the free
boson at radius $r/\sqrt{2N}$. It is easy to compute the corresponding
tensor product $A\ors \,{\otimes_{\!A_s}^{}}\, M^{(s)}_\kappa \,{\cong}\,
A\ors \,{\otimes_{A_s}^{}}\, A_s \,{\otimes}\, U_\kappa \,{\cong}\, 
A\ors \,{\otimes}\, U_\kappa$. Comparing the respective decompositions
into simple objects of $\Uc_{N}$ one finds that
  \be
  A\ors \otimes_{\!A_s}^{} M^{(s)}_\kappa 
  \cong \bigoplus_{m=0}^{\frac{N/r}{\tilde g}-1}\! M^{(r)}_{\kappa+2m\tilde g}
  \qquad ~ \text{with} ~~
  \tilde g \,{=}\, \text{gcd}(\tfrac{N}{r},\tfrac{N}{s}) 
  \ee
as $A_r$-modules.

{}From \erf{eq:Drs-action} it is easy to see that $D^{(rs)}$ cannot give an
equivalence of theories unless $r\,{=}\,s$, in which case $\hat D\ors$ is the 
identity. For $r \,{\neq}\, s$, $\hat D\ors$ will map some of the bulk fields to 
zero.


\subsection{T-duality}\label{sec:T-duality-rat}

On the CFT level, T-duality amounts to the statement that the free boson CFTs 
at radius $R$ and $R' \,{=}\, \alpha'/R \,{\equiv}\, 1/(2R)$ are isomorphic; the
isomorphism inverts the sign of $J\,\bar J$. To find a defect which implements 
this isomorphism, it is thus not sufficient to look only among the defects 
transparent to the $\Ue$ currents, as was done in section \ref{sec:rat-def-bi}.
Instead, we work with the chiral algebra $\Ueh/\Zb_2$, which consists of fields
invariant under $J \,{\mapsto}\, {-}J$, and its rational extension 
$\Ueh_{N}/\Zb_2$ by a field  of conformal weight $N$. Denote the category of 
representations of $\Ueh_{N}/\Zb_2$ by $\Dc_{N}$.  

Let us treat the case $R \,{=}\, 1/\sqrt{2N}$ and $R'\,{=}\,\sqrt{N/2}$ as an 
example. The details for how to arrive at the statements below can be found in 
appendix \ref{app:morita}. $\Dc_{N}$ contains two algebras $\tilde A_1$ and 
$\tilde A_N$ which describe the compactifications to $R$ and $R'$, respectively. 
The algebras $\tilde A_1$ and $\tilde A_N$ are in fact Morita equivalent, i.e.\ there 
exist an $\tilde A_1$-$\tilde A_N$-bimodule $X$ and an $\tilde A_N$-$\tilde 
A_1$-bimodule $X'$, both coming from the twisted sector of the orbifold 
(see \erf{defSig}), such that one has isomorphisms
  \be
  X \otimes_{\!\tilde A_N}^{} X' \,\cong\, \tilde A_1 \qquad{\rm and}\qquad
  X' \otimes_{\!\tilde A_1}^{} X \,\cong\, \tilde A_N 
  \ee
of bimodules. The corresponding defects $D_{X}$ and $D_{X'}$ interpolating 
between these two T-dual CFTs obey
  \be
  \hat D_X^{} \hat D_X' \phi = \hat D_{X{\otimes_{\!\tilde A_N}}X'} \phi
  = \hat D_{\!\tilde A_1} \phi = \phi
  \ee
for any bulk field $\phi$ of the $\tilde A_1$-theory, and vice versa, so that 
$\hat D_X$ indeed provides an isomorphism between the 
bulk state spaces of the two CFTs.

To relate the correlators of the two compactifications we can use the result
\cite[sect.\,3.3]{Frohlich:2006ch} that
  \be
  \text{Cor}_{\!\tilde A_1}(\Sigma) = \gamma^{-\chi(\Sigma)} \,
  \text{Cor}_{\!\tilde A_N}(\Sigma')
  \quad~\text{with}~~ \gamma = \dim(\tilde A_N){/}{\dim}(X) = \sqrt{N} .
  \labl{eq:Morita-rel}
Here $\Sigma$ denotes a world sheet decorated with data
for the CFT described by $\tilde A_1$, i.e.\
the free boson theory at radius $R \,{=}\,1/\sqrt{2N}$. This world sheet may 
have a boundary, as well as insertions of bulk and boundary fields. The world 
sheet $\Sigma'$ is equal to $\Sigma$ as a surface, but it is decorated with data for
the free boson compactified at radius $R'$, and the field insertions,
boundary conditions and defect lines
of $\Sigma$ are replaced by the ones obtained via the action of the 
interpolating defect $X'$. By $\text{Cor}_{\!\tilde A_1}$ and 
$\text{Cor}_{\!\tilde A_N}$ we then mean the correlators for the respective world 
sheets. In the prefactor $\gamma^{-\chi(\Sigma)}$, $\chi(\Sigma)$ is the 
Euler character of $\Sigma$, while $\gamma$ is the quotient of
quantum dimensions for $\Dc_{N}$ as stated in the formula.

Since in the perturbative expansion of the string free energy, the CFT 
correlator $\text{Cor}_{\!\tilde A_1}(\Sigma)$ appears with the prefactor 
$g_s^{-\chi(\Sigma)}$, we see that on the right hand side of the equality 
\erf{eq:Morita-rel} the combination $(g_s \gamma)^{-\chi(\Sigma)}$ appears. In 
the present example we have $\gamma \,{=}\,\sqrt{N}\,{=}\,1/(\sqrt2 R)$, 
so that we obtain precisely the expected identity \erf{eq:T-dual-parameters}. 
Note that the derivation of the change in $g_s$ in terms of defects can be 
carried out entirely on the level of the world sheet CFT, without explicitly 
mentioning the dilaton field.


\subsection{Truly marginal deformations}\label{sec:true-marg}

One limitation when working within rational CFT is the absence of continuous 
moduli. That is, there are no continuous families of CFTs with a fixed rational 
chiral algebra, and no continuous families of boundary conditions or topological
defects preserving this chiral algebra \cite[sect.\,3.1]{Fuchs:2004dz}. However,
one can deduce the existence of moduli by looking for truly marginal fields. 

For a boundary field $\psi$ there is a simple sufficient criterion 
to ensure that it leads to a truly marginal perturbation: $\psi$ needs to have
conformal weight one and be {\em self-local\/} \cite{Recknagel:1998ih}, i.e., 
exchanging the position of two adjacent boundary fields
$\psi(x)\,\psi(y)$ by analytic continuation in $x$ and $y$ does not modify
the value of a correlator. Via the folding trick \cite{Wong:1994pa}, which 
relates conformal defects of one CFT to conformal boundary conditions of the 
product theory, this results in a corresponding condition for defect fields. 
One obtains in this way a sufficient criterion for a conformal defect to stay 
conformal under a perturbation. We are more specifically interested in a 
condition for a topological defect to stay topological. This leads to the 
following definition.

Let $D$ be a topological defect and let $\Hc_D^{(1,0)}$ and $\Hc_D^{(0,1)}$ be
the spaces of all defect fields living on the defect $D$ of left/right conformal
weight $(1,0)$ and $(0,1)$, respectively. A subspace $\mathcal L$ 
of $\Hc_D^{(1,0)} \,{\oplus}\, \Hc_D^{(0,1)}$ is called {\em self-local\/} iff 
for all defect fields $\varTheta, \varTheta' \,{\in}\, \mathcal{L}$ we have
  \be
  \raisebox{-20pt}{
  \begin{picture}(130,40)
    \put(0,5){ \scalebox{1}{\includegraphics{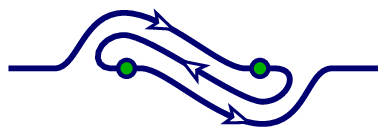}} }
    \put(0,5){
     \setlength{\unitlength}{1pt}\put(-8,-11){
     \put(44,15)  {$ \varTheta $}
     \put(80,33)  {$ \varTheta' $}
     \put(25,39)  {$ D $}
     }\setlength{\unitlength}{1pt}}
  \end{picture}}
  =
  \raisebox{-5pt}{
  \begin{picture}(120,10)
    \put(0,5){ \scalebox{1}{\includegraphics{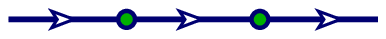}} }
    \put(0,5){
     \setlength{\unitlength}{1pt}\put(-8,-24){
     \put(44,15)  {$ \varTheta $}
     \put(80,15)  {$ \varTheta' $}
     \put(25,35)  {$ D $}
     }\setlength{\unitlength}{1pt}}
  \end{picture}}
  \labl{eq:top-def-marg}
inside every correlator. In other words, exchanging the order of $\varTheta$ and
$\varTheta'$ along the defect $D$ is equivalent to analytic continuation 
(in any correlator) of $\varTheta$ past $\varTheta'$. 
Note that the vertically reflected version of \erf{eq:top-def-marg} holds as 
well, as can be seen by simultaneously moving the insertion points of $\varTheta$ 
and $\varTheta'$ on both sides of the equality such that the defect on the left 
hand side becomes a straight line.

Using the same regularisation procedure as in 
\cite{Recknagel:1998ih} one can check that a perturbation of $D$ by a defect 
field $\varTheta \,{\in}\, \mathcal{L}$ is again a topological defect.

\medskip

Let us now consider the free boson compactified at radius $R\,{=}\,\sqrt{P/(2Q)}$ 
in terms of the algebra $A_{nP}$ in $\Uc_{N}$, $N\,{=}\,n^2 PQ$. We will also
assume $n\,{>}\,1$ (this avoids the special cases $N\,{=}\,1,2$ which require a 
separate treatment). The three representations of $\Ueh_{n^2PQ}$ that contain  
states of weight one are $U_0$, $U_{2a}$ and $U_{2(N{-}a)}$ with $a \,{=}\, n 
\sqrt{PQ}$. Since $P$ and $Q$ are coprime, $a$ can only be an integer if 
$P \,{=}\, E^2$ and $Q\,{=}\,F^2$ for some coprime $E,F \,{\in}\, \Zb_{>0}$.

Suppose the space of defect fields living on a topological defect $D$ contains
a field $\varTheta$ in the sector $U_{2a} \,{\otimes}\, \ol U_0$, and suppose
that $\varTheta$ has weight $(1,0)$. Then the subspace $\Cb\varTheta$
is {\em always\/} self-local (see appendix \ref{app:self-loc}). 
The same holds for fields of overall weight one in $U_0\,{\otimes}\, \ol U_0$,
$U_{2(N{-}a)} \,{\otimes}\, \ol U_0$, $U_0 \,{\otimes}\, \ol U_{2a}$,
and $U_0 \,{\otimes}\, \ol U_{2(N{-}a)}$. In other words, every field in
$\Hc_D^{(1,0)} {\oplus}\, \Hc_D^{(0,1)}$ with well-defined $\Ue$-charge
gives rise to a truly marginal perturbation that leaves $D$ topological.

However, if we perturb a defect $D$ by a field of definite non-zero 
$\Ue$-charge, the resulting theory is no longer unitary. To see this consider a 
cylinder $S^1 \,{\times}\, \Rb$ with the defect $D$ inserted on the line 
$\{ \alpha \} \,{\times}\, \Rb$ for some $\alpha \,{\in}\, S^1$.
The perturbation by a defect field $\varTheta$ amounts to an insertion of
$\exp\big( \lambda\! \int_{-\infty}^\infty\! \varTheta(\alpha,x)\,{\rm d}x )$ 
for some $\lambda \,{\in}\, \Rb$. The Hamiltonian generating translations along 
the cylinder is $H(\lambda) \,{=}\, H_0 \,{+}\, \lambda\, \varTheta(\alpha,0)$, 
where $H_0$ is the unperturbed Hamiltonian. Since we start from a unitary 
theory, we have $H_0^\dagger \,{=}\, H_0$. For the perturbed Hamiltonian 
$H(\lambda)$ to be self-adjoint we need the perturbing operator 
$\varTheta(\alpha,0)$ to be self-adjoint.  However, if $\varTheta$ has 
left/right $\Ue$-charge $(\sqrt{2},0)$, say, then $\varTheta^\dagger$ has charge
$(-\sqrt{2},0)$. Thus we need to perturb by appropriate linear combinations 
of defect fields of charges $(\sqrt{2},0)$ and $(-\sqrt{2},0)$. 

Incidentally, perturbing a topological defect (or a conformal boundary 
condition) by a marginal field of left and right $\Ue$-charges $(\sqrt{2},0)$, 
say, leads to a {\em logarithmic\/} theory, i.e.\ the perturbed Hamiltonian is 
no longer diagonalisable. For example, in \cite{Gaberdiel:2005sz}, the operator 
$I_\lambda(a,b) \,{=}\, \exp(\tfrac{\lambda}{2\pi}\!\int_a^b\!J^+(x)\,{\rm d}x)$
was considered for the $\suh(2)_1$ WZW model. This operator can be
understood as a topological defect running from $a$ to $b$,
obtained by a perturbation of the trivial defect by the field 
$J^+(z)$. Correlators of $I_\lambda(a,b)$ are found to contain logarithms.

The same will happen e.g.\ for conformal boundary conditions of the
free boson at the self-dual radius. As shown in \cite{Gaberdiel:2001xm}, 
these are parametrised by elements of ${\rm SL}(2,\Cb)$. Correlators involving
boundary changing fields that join boundary conditions belonging to different
${\rm SU}(2)$ cosets in ${\rm SL}(2,\Cb)$ may contain logarithms. Another 
example of a non-logarithmic bulk theory which allows for boundary fields with
logarithmic correlators was given in \cite{Creutzig:2006wk}.

\medskip

To obtain unitary defect perturbations we thus need to find a self-local 
subspace that is pointwise fixed with respect to charge conjugation. Rather 
than trying to classify all such cases, we will consider as a particular 
example the decomposition of $\hat D(g,h)$ 
for $g\,{=}\,h\,{=}\,e$ as obtained from \erf{eq:Dgh-decompose}.
Let $D$ be the defect given by the superposition
  \be
  D = \sum_{k=0}^{E-1} \sum_{l=0}^{F-1} D^{(nP)}_{(ka,la,0)} \,.
  \labl{eq:D-superpos}
Note that $ka \equiv 0 \,\text{mod}\, nP$ is equivalent to $knEF \,{\equiv}\, 0 
\,\text{mod}\, n E^2$, i.e.\ $k \,{\equiv}\, 0 \,\text{mod}\, E$, as $E$ and $F$
are coprime.  For the same reason, $la \,{\equiv}\, 0 \,\text{mod}\, nQ$ is 
equivalent to $l \,{\equiv}\, 0 \,\text{mod}\, F$; hence all the elementary 
defects appearing in the sum \erf{eq:D-superpos} are distinct.

The space of defect fields living on $D$ is given by (see appendix 
\ref{app:self-loc})
  \be
  \Hc_D = \bigoplus_{i,j=0}^{2N-1}
  \big( U_i \,{\otimes}\, \ol U_j \big)^{\oplus Z^D_{ij}} \qquad\text{with}
  \qquad Z^D_{ij} 
   = E F \,  \delta_{i{+}j,0}^{[2nF]}\, \delta_{i{-}j,0}^{[2nE]} \,.
  \labl{eq:def-field-on-D}
Since $\Hc_D$ contains the representations $U_0 \,{\otimes}\, U_0$, 
$U_{2a} \,{\otimes}\, U_0$, etc., with multiplicity $EF$, the 
space $\Hc_D^{(1,0)} \,{\oplus}\, \Hc_D^{(0,1)}$ has dimension $6EF$. It 
contains a six-dimensional self-local subspace $\mathcal{L}$ pointwise fixed
under charge conjugation. The subspace $\mathcal{L}$
is not unique, but maximal in the sense that there is no self-local
subspace of $\Hc_D^{(1,0)} {\oplus}\, \Hc_D^{(0,1)}$ of which $\mathcal{L}$
is a proper subspace (see appendix \ref{app:self-loc}).

This is in accordance with the results summarised in section 
\ref{sec:Vir-rat-mult}, where for a rational multiple of the self-dual radius a
six-dimensional moduli space of topological defects is found. The defect 
operators described there are fundamental, except possibly 
when $g$ and $h$ are diagonal or anti-diagonal. In particular, for
$(g,h)\,{=}\,(e,e)$ on obtains the superposition \erf{eq:D-superpos}.


\sect{General topological defects for the free boson}\label{sec:topdef-gen}

In this section we present details of the calculations that lead to the results 
stated in section \ref{sec:topdef-sum} for topological defects joining two free 
boson CFTs $\FB{R_1}$ and $\FB{R_2}$. 
We first give the relation between the defect operators
$\hat D$  and $\hat{\bar D}$ (section \ref{sec:rel-DDbar}). Then we consider 
defects that preserve the $\Ueh$-symmetry up to an automorphism (section  
\ref{sec:u1-pres-detail}). In section \ref{sec:Vir-self-detail} we
investigate  defects that preserve only the Virasoro algebra for the
free  boson at the self-dual radius.


\subsection{The defect operators $\hat D$ and $\hat{\bar D}$}\label{sec:rel-DDbar} 

The two operators $\hat D$ and $\hat{\bar D}$ associated to a defect $D$ are 
related in a simple manner. Let $\{ \varphi^{(1)}_i \}$ be a basis of $\Hc_1$ 
and $\{ \varphi^{(2)}_j \}$ a basis of $\Hc_2$. 
Let the two-point functions on the sphere in CFT$_1$ and CFT$_2$ be given by
  \be
  \big\langle \varphi^{(a)}_i(z)\, \varphi^{(a)}_j(w) \big\rangle
  = G^{(a)}_{ij} (z{-}w)^{-2h_i} (z^*{-}w^*)^{-2\bar h_i}
  \quad~ \text{for}~~a=1,2\,,
  \ee
respectively. Note that $G^{(a)}_{ij}$ is related to the OPE coefficient 
$C_{ij}^{(a)\,\one}$ via $G^{(a)}_{ij} \,{=}\, C_{ij}^{(a)\,\one} \langle 
\one^{(a)} \rangle$, where $\one^{(a)}$ is the identity field of CFT$_a$.

Consider now a two-point correlator on the sphere 
where on one hemisphere we have CFT$_1$ with an insertion of
$\varphi^{(1)}_i(z)$, while the other hemisphere supports CFT$_2$ with
an insertion $\varphi^{(2)}_j(w)$, separated by the topological defect
$D$. We may then deform the defect into a tight circle around
either $\varphi^{(1)}_i$ or $\varphi^{(2)}_j$; this results in the identity
  \be
  \big\langle (\hat D\varphi^{(1)}_i)(z)\, \varphi^{(2)}_j(w) \big\rangle
  = \big\langle \varphi^{(1)}_i(z)\, (\hat{\bar D}\varphi^{(2)}_j)(w) \big\rangle
  \,.
  \labl{eq:D-from-Dbar}
In terms of matrix elements, i.e.\ writing
$\hat D \varphi^{(1)}_i \,{=}\, \sum_k d_{ik}^{} \varphi^{(2)}_k$ and
$\hat{ \bar D} \varphi^{(2)}_j \,{=}\, \sum_l \ol d_{jl}^{} \varphi^{(1)}_l$,
this becomes
  \be
  \sum_k G^{(2)}_{kj}\, d_{ik}^{} = \sum_l G^{(1)}_{il}\, \ol d_{jl}^{} \,. 
  \labl{eq:Dbar-matrix}
Since $G^{(1)}$ and $G^{(2)}$ are invertible, this fixes $\ol d$ uniquely
in terms of $d$.

The relation between $\hat D$ and $\hat{\bar D}$ is further simplified 
if the operator $\hat D{:}\ \Hc_1 \,{\rightarrow}\, \Hc_2$ is invertible 
and preserves the two-point function on the sphere in the sense that
  \be
  \big\langle (\hat D \varphi)(z)\, (\hat D \varphi')(w) \big\rangle
  = \xi \, \big\langle \varphi(z)\, \varphi'(w) \big\rangle
  \quad~ \text{for all}~\,\varphi,\varphi' \in \Hc_1
  \ee
for some $\xi \,{\in}\, \Cb$. In this situation one has 
$\hat{\bar D} = \xi \, \hat D^{-1}$.


\subsection{Defects preserving the {$\Ueh$}-symmetry}\label{sec:u1-pres-detail}

Let us start by considering topological defects $D$ within the free 
boson at a given radius $R$ that obey \erf{eq:D-U1-com} with 
$\epsilon \,{=}\, \bar\epsilon \,{=}\, 1$. In other words, $D$ is transparent
to the currents $J$ and $\bar J$. This implies that the disorder fields
starting the defect $D$ carry a representation of the left and right copies of 
the $\Ueh$-symmetry. Let $\varTheta$ be a $\Ueh$-primary disorder field starting
$D$ of left and right $\Ue$-charges $(q,\bar q)$ and suppose that $(q,\bar q) 
\,{\in}\, \Lambda(R)$. Let $w$ be a point on the defect $D$ such that there is 
no defect field insertion on $D$ between $w$ and the insertion point $z$ of 
$\varTheta$. Then there exists a $\Ueh$-primary disorder field $\mu$
starting $D$ with charges $(0,0)$ such that
  \be
  \theta(z) = \phi_{(q,\bar q)}(z) \, \mu(w)
  \labl{eq:detach-D}
inside every correlator.

In words equation \erf{eq:detach-D} states that the defect $D$ can be 
detached from the disorder field $\theta(z)$, leaving a bulk field of 
the same charge at the point $z$. The new end point $w$ of the defect 
is marked by a disorder field $\mu(w)$. Since $\mu$
has charge zero it obeys $L_{-1} \mu = 0 = \bar L_{-1} \mu$ and as a
consequence correlators do not depend on the insertion point $w$.

The validity of \erf{eq:detach-D} can be established as follows. Assuming for
simplicity that $z\,{=}\,0$, we consider the product of two bulk fields
with $\theta(0)$.
Taking the operator product of $\phi_{(-q,-\bar q)}$ and $\theta$, we have
  \be
  \phi_{(q,\bar q)}(u) \, \phi_{(-q,-\bar q)}(v) \, \theta(0)
  = \phi_{(q,\bar q)}(u) \sum_{l,r=0}^\infty 
  v^{-q^2+l} (v^*)^{-\bar q{}^2+r} M_{l,r} \eta(0) \,,
  \labl{eq:detach-D-aux1}
where $\eta(0)$ is a disorder field of charge $(0,0)$ and
$M_{r,l}$ denotes the appropriate combination of $\Ueh$-modes of total
left/right weight $(l,r)$. The crucial point is now that
  \be
  M_{l,r} \eta(0) = M_{l,r} \one(0)\, \eta(w) \,.
  \labl{eq:detach-D-aux2}
To see this we observe that any combination of $\Ueh$-modes in
$M_{l,r}$ can be obtained as a suitable contour integral of the currents
$J$ and $\bar{J}$. Furthermore, by assumption the
field $\eta$ does not have any poles with these currents, and hence it may be 
moved through the contours at no cost. Finally, since
$L_{-1}\eta \,{=}\,0\,{=}\, \bar L_{-1}\eta$, the correlator does not actually 
depend on the precise location of $\eta$, and hence we may move $\eta$ from 
$0$ to $w$. The modes $M_{l,r}$ now act on the identity field at 0. Since the 
field $\eta$ does not create a branch cut for the currents $J$
and $\bar{J}$, we can apply the same contour arguments for correlation functions
involving $\eta$ as for those involving the identity field ${\bf 1}$.
Thus the operator product of two bulk fields results in the same combinations 
$M_{l,r}$ of modes as in the product in \erf{eq:detach-D-aux1},
  \be
  \phi_{(-q,-\bar q)}(v) \, \phi_{(q,\bar q)}(0)
  = C_{-q,q} \sum_{l,r=0}^\infty 
  v^{-q^2+l} (v^*)^{-\bar q{}^2+r} M_{l,r} \phi_{(0,0)}(0) \,,
  \labl{eq:detach-D-aux3}
where the structure constant $C_{-q,q}$ is the sign factor given in 
\erf{eq:bulk-OPE}. Combining \erf{eq:detach-D-aux1}, \erf{eq:detach-D-aux2}
and \erf{eq:detach-D-aux3} we can write
  \be
  \phi_{(q,\bar q)}(u) \, \phi_{(-q,-\bar q)}(v) \, \theta(0)
  = \phi_{(q,\bar q)}(u) \, \phi_{(-q,-\bar q)}(v) \, \phi_{(0,0)}(0) \,
  \big( (C_{-q,q})^{-1} \eta(w) \big) \,.
  \ee
Finally, applying the limit $\lim_{u\rightarrow v} |u-v|^{q^2}(\cdots)$
to both sides of this equality amounts to the replacement of 
$\phi_{(q,\bar q)}(u) \, \phi_{(-q,-\bar q)}(v)$ by
the identity field. Setting $\mu \,{=}\, (C_{-q,q})^{-1} \eta$
then results in \erf{eq:detach-D}.

\medskip

Let now $D$ be a topological defect joining the theories $\FB{R_1}$  
and $\FB{R_2}$, and obeying \erf{eq:D-U1-com} for arbitrary $\epsilon$ and
$\bar\epsilon$. Suppose further that $D$ has the properties stated in Q3 in 
section \ref{sec:complete}, in particular that there is just a single defect 
field of left/right weight $(0,0)$ living on $D$. Let $\phi_{(q,\bar q)}(z)$ 
be a $\Ueh$-primary bulk field of $\FB{R_1}$ such that $(\epsilon q,
\bar \epsilon \bar q)$ lies in the charge lattice of $\FB{R_2}$, 
i.e.\ $(q,\bar q) \,{\in}\, \Lambda(R_1) \,{\cap}\, 
\Lambda^{\epsilon,\bar\epsilon}(R_2)$. Then
  \be
  \raisebox{-40pt}{
  \begin{picture}(85,60)
    \put(10,0){\scalebox{.35}{\includegraphics{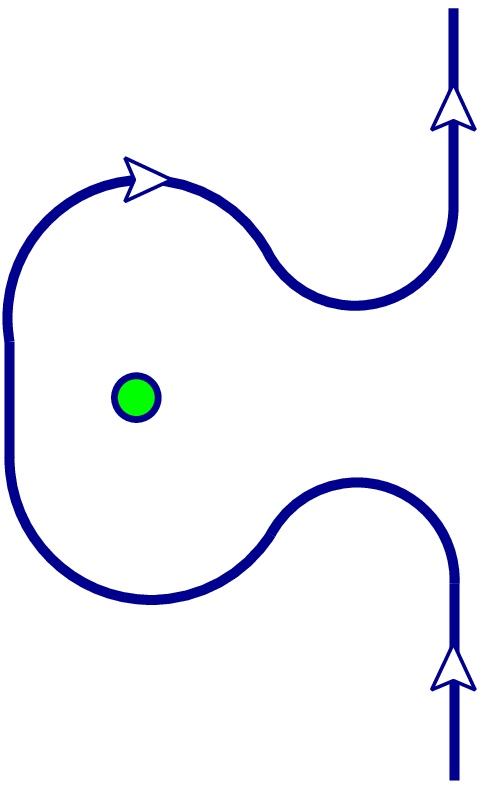}}}
    \put(10,0){
     \setlength{\unitlength}{.35pt}\put(0,0){
     \put(52,110)  {\scriptsize $ \phi_{(q,\bar q)} $}
     \put(102,10)  {\scriptsize $ D $}
     \put(-40,180) {\footnotesize\shadowbox{$R_2$}}
     \put(150,180) {\footnotesize\shadowbox{$R_1$}}
     }\setlength{\unitlength}{1pt}}
  \end{picture}}
  = ~
  \raisebox{-40pt}{
  \begin{picture}(85,85)
    \put(0,0){\scalebox{.35}{\includegraphics{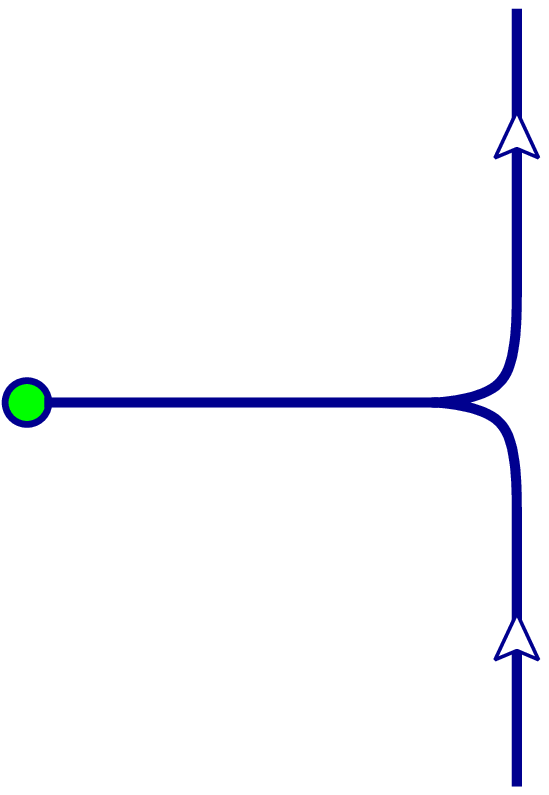}}}
    \put(0,0){
     \setlength{\unitlength}{.35pt}\put(0,0){
     \put(-5,85)  {\scriptsize $ \theta $}
     \put(47,123)  {\scriptsize $ \bar D{*}D $}
     \put(155,155)  {\scriptsize $ D $}
     \put(155, 59)  {\scriptsize $ D $}
     \put( 50,180) {\footnotesize\shadowbox{$R_2$}}
     \put(170,180) {\footnotesize\shadowbox{$R_1$}}
     }\setlength{\unitlength}{1pt}}
  \end{picture}}
  = ~~~
  \raisebox{-40pt}{
  \begin{picture}(90,85)
    \put(0,0){\scalebox{.35}{\includegraphics{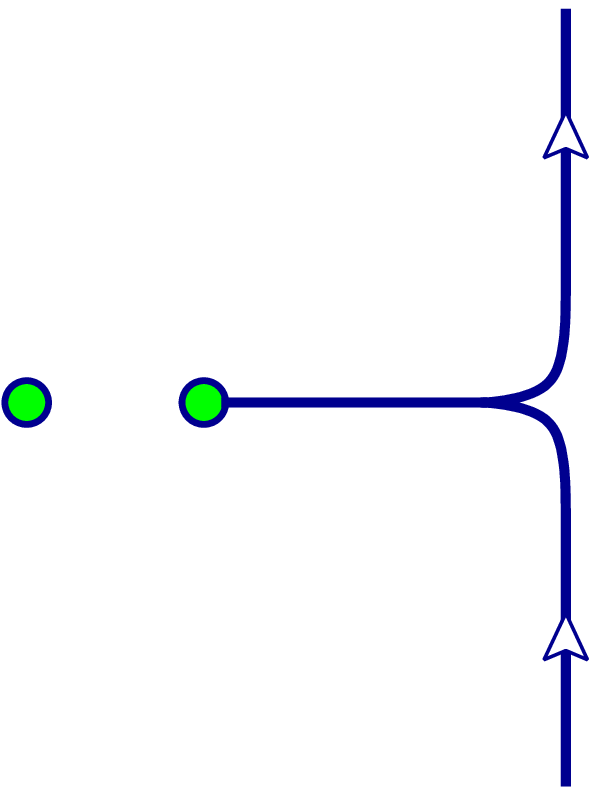}}}
    \put(0,0){
     \setlength{\unitlength}{.35pt}\put(0,0){
     \put(-35,88)  {\scriptsize $ \phi_{(\epsilon q,\bar \epsilon \bar q)} $}
     \put(55,88)  {\scriptsize $ \mu $}
     \put(75,123)  {\scriptsize $ \bar D{*}D $}
     \put(170,155)  {\scriptsize $ D $}
     \put(170, 59)  {\scriptsize $ D $}
     \put( 70,180) {\footnotesize\shadowbox{$R_2$}}
     \put(190,180) {\footnotesize\shadowbox{$R_1$}}
     }\setlength{\unitlength}{1pt}}
  \end{picture}}
  = ~\lambda_D^{\,-1} ~
  \raisebox{-40pt}{
  \begin{picture}(100,85)
    \put(0,0){\scalebox{.35}{\includegraphics{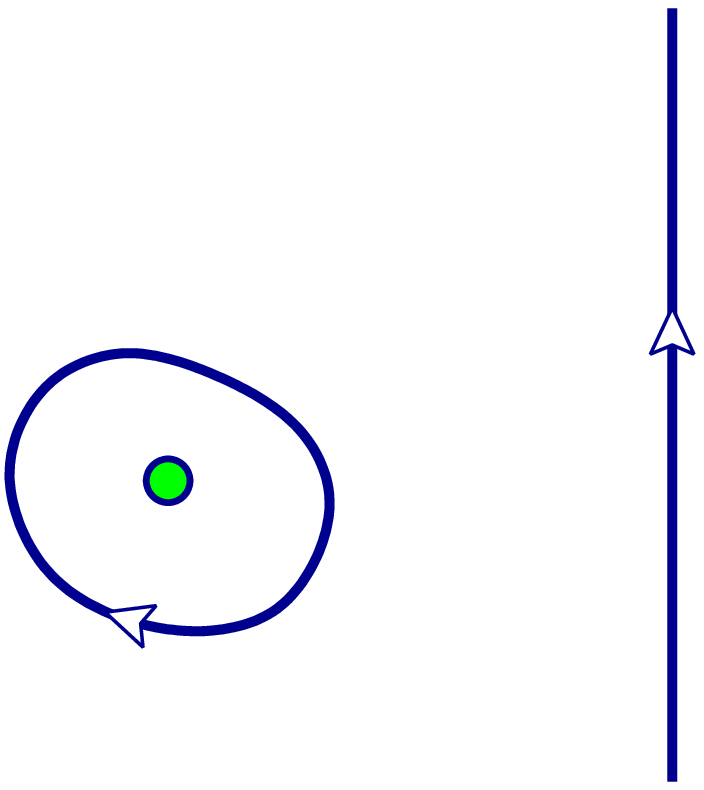}}}
    \put(0,0){
     \setlength{\unitlength}{.35pt}\put(0,0){
     \put(25,68)  {\scriptsize $ \phi_{(q,\bar q)} $}
     \put(88,40)  {\scriptsize $ D $}
     \put(166,103) {\scriptsize $ D $}
     \put( 80,180) {\footnotesize\shadowbox{$R_2$}}
     \put(220,180) {\footnotesize\shadowbox{$R_1$}}
     }\setlength{\unitlength}{1pt}}
  \end{picture}}
  \labl{eq:D-past-phi}
for some non-zero constant $\lambda_D$. To see this, first fuse
the part of $D$ surrounding $\phi_{(q,\bar q)}(z)$, which results in
a defect line $\bar D \,{*}\, D$ ending on a disorder field $\theta(z)$.
Since $\bar D \,{*}\, D$ obeys \erf{eq:D-U1-com} with 
$\epsilon \,{=}\, \bar\epsilon \,{=}\,1$ and since $\theta(z)$ has
charges $(\epsilon q,\bar \epsilon \bar q)\,{\in}\,\Lambda(R_2)$
we can apply the relation \erf{eq:detach-D}. This results in an insertion of
$\phi_{(\epsilon q,\bar \epsilon \bar q)}(z)$ and of a weight zero
disorder field $\mu(w)$. But a disorder field starting $\bar D *D$
is the same as a defect field living on $D$, and by assumption
every weight zero defect field on $D$ is proportional to the
identity field on $D$. Thus we can write $\mu \,{=}\, a_{q,\bar q} \one_D$
for some constant $a_{q,\bar q} \,{\in}\, \Cb$. 
Altogether we hence obtain the equality
  \be
  D \, \phi_{(q,\bar q)}(z) = a_{q,\bar q} \,
  \phi_{(\epsilon q,\bar \epsilon \bar q)}(z) \,D \,,
  \labl{eq:D-past-phi-aux2}
where the position of the symbol $D$
indicates on which side of the defect line the bulk
field is inserted. Closing the contour of $D$
in \erf{eq:D-past-phi-aux2} to a loop then results in the equality
  \be
  \hat D \phi_{(q,\bar q)} = \lambda_D \, a_{q,\bar q} \,
  \phi_{\epsilon q,\bar \epsilon \bar q} \,,
  \labl{eq:D-past-phi-aux1}
where the constant $\lambda_D$ is determined by the action of $\hat D$
on the identity field,
  \be
  \hat D \, \one^{(R_1)} = \lambda_D \, \one^{(R_2)}\,.
  \ee
{}From \erf{eq:D-past-phi-aux1} we see that $\lambda_D$ has to be
non-zero, or else the defect operator would vanish identically, contradicting
the assumptions in Q3. Substituting \erf{eq:D-past-phi-aux1} into
\erf{eq:D-past-phi-aux2} finally gives the last equality in \erf{eq:D-past-phi}.

\medskip

We would now like to use \erf{eq:D-past-phi} to determine the operator $\hat D$ 
as accurately as possible. As already noted in section \ref{sec:u1-pres-def},
$\hat D$ is necessarily of the form
  \be
  \hat D = \sum_{(q,\bar q) \in \Lambda} d(q,\bar q) \,
  P(R_2,R_1)^{\epsilon q,\bar \epsilon \bar q}_{q,\bar q} \,,
  \labl{eq:u1-Dhat-expand}
where the maps $P(R_2,R_1)^{\epsilon q,\bar \epsilon \bar q}_{q,\bar q}$
are those that we denoted by $P^{\epsilon q,\bar \epsilon \bar q}_{q,\bar q}$ 
in \erf{eq:P-defining-eqn}. The operator 
for the action of $\bar D$ can be obtained from \erf{eq:Dbar-matrix} to be
  \be
  \hat{\bar D} =
  \frac{\langle \one^{(R_2)} \rangle}{\langle \one^{(R_1)} \rangle}
   \sum_{(q,\bar q) \in \Lambda} 
  d(-\epsilon q,- \bar \epsilon \bar q) \,
  P(R_1,R_2)^{\epsilon q,\bar \epsilon \bar q}_{q,\bar q} \,.
  \labl{eq:u1-Dbarhat-expand}

Recall the definition of $\RR_2$ in \erf{eq:lattice-Lam-epseps}. We will 
investigate the case that $\RR_2/R_1$ is rational, so that $\Lambda$ 
contains an infinite number of points. As in section \ref{sec:RR-rat} we set 
$\RR_2/R_1 \,{=}\, M/N$ for coprime positive integers $M$ and $N$, and note 
that the lattice $\Lambda$ is spanned by the vectors
$e_1 \,{=}\, N/(2R_1)\,{\cdot}\,(1,1)$ and $e_2 \,{=}\, MR_1\,{\cdot}\,(1,-1)$.
  
We select two vectors $(p,\bar p) \,{=}\, a e_1 \,{+}\, b e_2$ and 
$(q,\bar q) \,{=}\, c e_1 \,{+}\, d e_2$ in $\Lambda$ and consider 
two insertions $\phi_{(p,\bar p)}(z)$ and $\phi_{(q,\bar q)}(0)$ of bulk 
fields of $\FB{R_1}$ surrounded by a loop of defect $D$. Invoking the identity 
\erf{eq:D-past-phi} we conclude that
  \be
  \hat D\big( \phi_{(p,\bar p)}(z)\, \phi_{(q,\bar q)}(0) \big)
  = \lambda_D^{\,-1}\, (\hat D \phi_{(p,\bar p)})(z) \,
    (\hat D \phi_{(q,\bar q)})(0)  \,.
  \labl{eq:u1-D-loop}
On the left hand side of this equality we can take the operator product
of the theory
$\FB{R_1}$, while on the right hand side we use the operator product
of $\FB{R_2}$. Taking care of the sign factors in the 
OPE \erf{eq:bulk-OPE}, in terms of the coefficients $d(\,\cdot\,,\,\cdot\,)$ 
in the decomposition \erf{eq:u1-Dhat-expand} the condition reads
  \be
  d(p{+}q,\bar p {+} \bar q)
  = \lambda_D^{\,-1}\, (\epsilon\bar\epsilon)^{(ad-bc)MN}_{} 
  d(p,\bar p)\, d(q,\bar q) \,.
  \labl{eq:u1-D-d-coeff}
Every solution to \erf{eq:u1-D-d-coeff} is of the form
  \be
  d(q,\bar q) \,{=}\, \lambda_D \,
  (\epsilon\bar\epsilon)^{\frac12(q^2-\bar q{}^2)} \,
  \alpha^q \beta^{\bar q}
  \labl{eq:d_qq-res}
for some $\alpha,\beta \,{\in}\, \Cb^\times$. To determine the constant $\lambda_D$ 
we compute the torus partition function with an insertion of the defect loop 
$\bar D \,{*}\, D$ via a trace and use the modular transformation $\tau 
\,{\mapsto}\, {-}1/\tau$ to deduce the spectrum of defect operators on $D$:
  \bea
  \text{tr}_{\Hc(R_1)}^{}\big( \hat{\bar D} \,\hat D \, 
  \rmq^{L_0-\frac{1}{24}} \,
  (\rmq^*)^{\ol L_0-\frac{1}{24}} \big)
  \enL \qquad\qquad
  =  
  \frac{\langle \one^{(R_2)} \rangle}{\langle \one^{(R_1)} \rangle}
  \lambda_D^{\;2}
  \sum_{(p,\bar p) \in \Lambda} \!\!
  \frac{\rmq^{\frac12 p^2} (\rmq^*)^{\frac12 \bar p^2}}{\eta(\rmq)\,\eta(\rmq^*)}
  = 
  \frac{\langle \one^{(R_2)} \rangle}{\langle \one^{(R_1)} \rangle}
  \frac{\lambda_D^{\;2}}{MN}
  \sum_{(p,\bar p) \in \Lambda^*_{\phantom|}} \!\!
  \frac{{\tilde{\rmq}}^{\frac12 p^2} (\tilde{\rmq}^*)^{\frac12 \bar p^2}}{
  \eta(\tilde{\rmq})\,\eta(\tilde{\rmq}^*)} \,.
  \eear\labl{eq:D-defect-spectrum}
Here $\Lambda^*$ is the lattice dual to $\Lambda$, while $\rmq\,{=}\,\mathrm e
^{2\pi \mathrm i \tau}$ and 
$\tilde{\rmq} \,{=}\, \mathrm e^{2\pi \mathrm i (-1/\tau)}$ give the dependence 
on the modular parameter of the torus. For $\hat D$ to be fundamental, the
multiplicity space of weight zero fields should be one-dimensional, so that we 
need $\lambda_D^{\;2} \,{=}\, MN \langle\one^{(R_1)}\rangle / \langle\one^{(R_2)}
\rangle$. It remains to determine the sign of $\lambda_D$. This is done in 
appendix \ref{app:defect-sign}. One finds a remaining ambiguity which
amounts to  
choosing, once and for all, a square root of
$\langle\one^{(R)}\rangle$ for each 
$R\,{>}\,0$. To summarise the findings so far, the fundamental defect
operators from $\Hc(R_1)$ to $\Hc(R_2)$ and obeying \erf{eq:D-U1-com}
are given by 
  \be 
  \hat D(x,y)^{\epsilon,\bar\epsilon}_{R_2,R_1}
  = \sqrt{\Frac{\langle\one^{(R_1)}\rangle\, MN}{\langle\one^{(R_2)}\rangle}}
  \sum_{(q,\bar q) \in \Lambda}
      (\epsilon\bar\epsilon)^{\frac12(q^2-\bar q{}^2)} \,
  \mathrm e^{2 \pi \mathrm i (x q - y \bar q)}
     P(R_2,R_1)^{\epsilon q,\bar \epsilon \bar q}_{q,\bar q} \,.
  \labl{eq:u1-def-op-explicit}
The form given in \erf{eq:u1-def-op-explicit-simp} is 
then obtained by choosing $\sqrt{\langle \one^{(R)} \rangle} \,{\equiv}\, 1$.


\subsection{Virasoro preserving defects} \label{sec:Vir-self-detail}

Let us start with the free boson compactified at the self-dual radius. To arrive
at the formula stated in \erf{eq:Rsd-Vir-pres-def} for the fundamental defect 
operators preserving only the Virasoro symmetry we could proceed analogously 
to \cite{Gaberdiel:2001xm}, which uses factorisation of bulk two-point 
correlators on the upper half plane as in \cite{Cardy:1991tv}. In terms of the 
decomposition \erf{eq:Rsd-bulk-decompose} of $\Hc(R_\text{s.d.})$, this amounts
to picking Virasoro primaries $\phi, \tilde\phi \,{\in}\, \Hc_{[s,\bar s]}$ and
$\psi, \tilde\psi \,{\in}\, \Hc_{[t,\bar t]}$, and considering the complex 
plane with a defect $D$ running along the real axis and bulk field insertions 
of $\phi(z)$, $\tilde\phi(z^*)$, $\psi(w)$ and $\tilde\psi(w^*)$. Comparing the 
limits $\Im(z),\Im(w) \,{\rightarrow}\, 0$ and $|z{-}w| \,{\rightarrow}\, 0$ 
one finds constraints on the defect operator $\hat D$, which can be solved in 
terms of representation matrices of ${\rm SL}(2,\Cb)$. However, when describing 
the defect just as a boundary condition in the folded system, not all ways of 
analysing topological defects can be applied. This is for instance the case for
the method used below, since it relies on the deformation and fusion of defect lines.

\medskip

We start by investigating the properties of topological defects 
$X$ that can end on a disorder field $\mu$ which is Virasoro-primary of weights 
$(h,\bar h) \,{=}\, (\tfrac14,\tfrac14)$. To this end we consider the 
monodromies of the $\su(2)$ currents $J^a$ and $\bar J^a$
around $\mu$. By factorisation it is 
enough to investigate the monodromy of the two three-point correlators
  \be
  f(\tilde\mu;z) = 
  \big\langle J^a(z)\, \mu(0)\, \tilde\mu(-L) \big\rangle
  \qquad \text{and} \qquad
  g(\tilde\mu;z) = 
  \big\langle \bar J^a(z)\, \mu(0)\, \tilde\mu(-L) \big\rangle
  \ee
on the complex plane with a defect $X$ stretched from $0$ to $-L$,
and $\tilde\mu$ an arbitrary disorder field that can terminate a $X$-defect.
We will show that $f(\tilde\mu;z)$ and $g(\tilde\mu;z)$ are single-valued on 
$\Cb\,{\setminus}\{0,-L\}$ for all choices of $\tilde\mu$.
It is enough to consider $\tilde\mu$ that are Virasoro primary, since going to 
descendents does not affect $f(\tilde\mu;z)$ and $g(\tilde\mu;z)$ being 
single-valued or not. Next note that $J^a\,{\in}\,\Hc_{[2,0]}$ and so the fusion 
rules dictated by the Virasoro null vectors imply that $f(\tilde\mu;z)$ can be 
nonzero only if $\tilde\mu$ has conformal weights $(h,\bar h) \,{=}\, 
(s^2/4,\ol s^2/4)$ with $(s,\ol s) \,{=}\, (1,1)$ or $(s,\ol s) \,{=}\, (3,1)$. 
Thus $f(\tilde\mu;z)$ is proportional to $z^{-1} \, (z+L)^{-1} \, L^{\frac12}$
for $(s,\ol s) \,{=}\, (1,1)$, proportional to $z \, (z+L)^{-3} \, L^{-\frac32}$
for $(s,\ol s) \,{=}\, (3,1)$, and zero otherwise. In particular
$f(\tilde\mu;z)$ is single-valued for all choices of $\tilde\mu$. A 
similar argument shows that $g(\tilde\mu;z)$ is single-valued. 
It follows that the $\su(2)$ currents are single valued close to $\mu(0)$.

Suppose now that $X$ is of the form $\bar D \,{*}\, D$ for some defect $D$ 
obeying the properties stated in Q3. The positivity assumption made there 
implies that the disorder field $\mu$ can be written as a sum of disorder fields
which are $(J_0,\bar J_0)$-eigenstates. 
To see this let $U$ be the space of disorder fields starting the
defect $X$ that are Virasoro-primary of weight $(\tfrac14,\tfrac14)$. By positivity,
these fields are annihilated by the positive $\hat{u}(1)$-modes, and the
action of $L_0$ and $\bar{L}_0$ is thus given by $L_0=\tfrac{1}{2} J_0 J_0$
and $\bar{L}_0=\tfrac{1}{2} \bar{J}_0 \bar{J}_0$, respectively. 
Since $L_0$, $\bar L_0$ and $J_0$, $\bar J_0$ commute, the modes $J_0$ and 
$\bar J_0$ map the $L_0$,$\bar{L}_0$-eigenspace $U$ to itself. We can
always bring $J_0$ and $\bar{J}_0$ into Jordan normal form, and since the
action of $L_0$ and $\bar{L}_0$ is diagonal with non-zero eigenvalue, 
it follows that $J_0$ and $\bar{J}_0$ must also be diagonalisable.
We conclude that $\mu$ is a 
sum of disorder fields with $\Ue$-charges $(\pm 1/\sqrt{2},\pm 1/\sqrt{2})$, 
all of which are contained in the charge lattice $\Lambda(R_\text{s.d.})$.
Defects ending on the disorder field $\mu$ therefore behave as those associated
to $(q,\bar q)\,{\in}\,\Lambda(R)$ in the previous subsection and we can thus
use the same arguments which lead to \erf{eq:D-past-phi} to deduce that for a 
primary bulk field $\phi \,{\in}\, \Hc_{[1,1]}$ we have
  \be
  \raisebox{-40pt}{
  \begin{picture}(105,60)
    \put(10,0){\scalebox{.35}{\includegraphics{pic-cf1.eps}}}
    \put(10,0){
     \setlength{\unitlength}{.35pt}\put(0,0){
     \put(50,113)  {\scriptsize $ \phi $}
     \put(103,10)  {\scriptsize $ D $}
     \put(-66,180) {\footnotesize\shadowbox{$R_\text{s.d.}$}}
     \put(150,180) {\footnotesize\shadowbox{$R_\text{s.d.}$}}
     }\setlength{\unitlength}{1pt}}
  \end{picture}}
  = ~\lambda_D^{\,-1} ~~
  \raisebox{-40pt}{
  \begin{picture}(110,85)
    \put(0,0){\scalebox{.35}{\includegraphics{pic-cf4.eps}}}
    \put(0,0){
     \setlength{\unitlength}{.35pt}\put(0,0){
     \put(25,71)  {\scriptsize $ \phi $}
     \put(87,40)  {\scriptsize $ D $}
     \put(168,104) {\scriptsize $ D $}
     \put( 80,180) {\footnotesize\shadowbox{$R_\text{s.d.}$}}
     \put(220,180) {\footnotesize\shadowbox{$R_\text{s.d.}$}}
     }\setlength{\unitlength}{1pt}}
  \end{picture}}
  \labl{eq:D-past-Vir-phi}
where $D$ is an arbitrary defect with the properties stated in Q3
(not necessarily one that can end on $\mu$ as above).

One can check by recursion that every Virasoro-primary bulk field appears in 
the repeated fusion of primary 
fields in the sector $\Hc_{[1,1]}$, cf.\ \cite{Gaberdiel:2001xm}.
We can use the same recursion to show that \erf{eq:D-past-Vir-phi} applies to 
all bulk fields. Indeed, suppose \erf{eq:D-past-Vir-phi} holds for $\psi 
\,{\in}\, \Hc_{[s,\bar s]}$. Let us write \erf{eq:D-past-Vir-phi} symbolically
as $D\, \phi(z) \,{=}\, \lambda_D^{\,-1}\, (\hat D\phi)(z)\, D$. Then
  \be
  D \,\phi(z)\, \psi(w) 
  = \lambda_D^{\,-2}\, (\hat D \phi)(z)\, (\hat D \psi)(w) \,D
  \labl{eq:D-past-phi-symb}
for $\phi \,{\in}\, \Hc_{[1,1]}$. Taking the OPE on both sides and comparing 
terms implies that \erf{eq:D-past-Vir-phi} also holds for the primary fields in
$\Hc_{[s \pm 1,\bar s \pm 1]}$.

The defect operator $\hat D$ obeys \erf{eq:D-Vir-com} and thus maps the sector 
$\Hc_{[s,\bar s]}$ to itself. It follows from \erf{eq:D-past-phi-symb} with the
same arguments as in section \ref{sec:u1-pres-detail} that
$\lambda_D^{-1} \hat D$ has to be a homomorphism of the bulk OPE.
Since the bulk fields are generated by the elements of $\Hc_{[1,1]}$, 
the operator $\hat D$ is uniquely determined by its action on $\Hc_{[1,1]}$. 

The restriction of $\hat D$ to $\Hc_{[1,1]}$ can be written as
  \be
  \hat D \big|_{\Hc_{[1,1]}}
  = 
  \lambda_D \, R \otimes 
  \id_{\Hc{}^\text{Vir}_{1/4} \otimes \bar\Hc{}^\text{Vir}_{1/4}}
  ~\quad \text{with} \quad
  R \in \text{End}(V_{1/2} \oti \ol V_{1/2}) \,.
  \labl{DtoR}
Consistency with the bulk OPE poses constraints on the linear map $R$. These are
analysed in appendix \ref{app:RinSLxSL}, with the result that we can always find
$g,h\,{\in}\, \mathrm{SL}(2,\Cb)$ such that
  \be   
  R = g \otimes h \,.
  \labl{eq:R=gxh}

The restriction of $\hat{D}$ to $\Hc_{[1,1]}$ defines $\hat{D}$ uniquely, and
thus we could in principle construct the full defect operator inductively from 
\erf{DtoR}. It is, however, simpler to obtain this operator in a different
manner. To this end we consider the family of topological defects that is 
obtained by perturbing the trivial defect by the bulk field
  \be
  \phi = a J^+ + b J^3 + c J^- + 
  \tilde a \bar J^+ + \tilde b \bar J^3 + \tilde c \bar J^- .
  \ee
This field is clearly self-local and thus generates a truly marginal 
deformation (see section \ref{sec:true-marg}). The corresponding defect
operator is simply the exponential of the corresponding zero modes
as given in \erf{eq:Rsd-zeromode-exp},
  \be
  \hat D = \exp\big( 
  a J^+_0 + b J^3_0 + c J^-_0 + 
  \tilde a \bar J^+_0 + \tilde b \bar J^3_0 + \tilde c \bar J^-_0\big) \,.
  \ee
By construction, $J^a_0$ acts on the $V_{s/2} \oti V_{\bar s/2}$ part 
of $\Hc_{[s,\bar s]}$ via the representation $R_{s/2}(J^a) \otimes \id$ of 
the Lie algebra $sl(2,\Cb)$, and similarly $\bar J^a_0$ acts as 
$\id \oti R_{\bar s/2}(J^a)$.
By exponentiating we obtain the representations $\rho_{s/2}$ and 
$\rho_{\bar s/2}$ of the Lie group, so that altogether we arrive at
the operator \erf{eq:Rsd-Vir-pres-def}.
It is then immediate to deduce the rule for orientation reversal as
$\bar D(g,h) \,{=}\, D(g^{-1},h^{-1})$. In particular, all defect
operators  $\hat D(g,h)$ are fundamental, since the torus amplitude  
with an insertion of $\bar D \,{*}\, D$ is equal to the torus
amplitude without  
defects, which contains the vacuum with multiplicity one.

The operators $\hat D(g,h)$ in fact constitute all fundamental Virasoro 
preserving defect operators for $\FB{R_\text{s.d.}}$. This follows since the
restriction of $\hat D(g,h)$ to $\Hc_{[1,1]}$ is $g \oti h \oti \id$. Combining  
this with the previous result \erf{eq:R=gxh} we see that for any given 
fundamental $\hat D$ we can find $g,h$ such that $\hat D \,{=}\, \lambda_D 
\hat D(g,h)$. The amplitude of a torus with insertion of $\bar D \,{*}\, D$ is  
then equal to $\lambda_D^{\,2} Z(R_\text{s.d.})$. Using the assumptions in Q3,
we conclude $\lambda_D\,{=}\,{\pm}1$. Finally, $\lambda_D\,{=}\,{-}1$ would lead
to all coefficients in the torus amplitude with an insertion of a singe 
$D$-defect being negative. Hence $\lambda_D\,{=}\,1$. 

\medskip

Since the set \erf{eq:su2-1-vir-def-group} gives all fundamental defect 
operators we can recover the $\Ueh$-pre\-ser\-ving operators for $R_1\,{=}\,R_2
\,{=}\,R_\text{s.d.}$. Note that $(\sqrt{2}R_2)^{\epsilon\bar\epsilon}/\sqrt{2}
\,{=}\, R_\text{s.d.}$, so that we are in the case treated in section 
\ref{sec:RR-rat}, with $\RR_2/ R_1\,{=}\,1$. It is enough to compare the action
on the subspace $\Hc_{[1,1]}$. The Virasoro-pri\-maries in this space are just 
the $\Ueh$-primary fields $\phi_{(\pm 1/\sqrt{2},\pm 1/\sqrt{2})}$. In the 
representation $V_{\frac12}$
of ${\rm SL}(2,\Cb)$ we choose the standard basis 
$e_1 \,{=}\, |j{=}\tfrac12, m{=}\tfrac12\rangle$ and $e_2 \,{=}\, |j{=}\tfrac12,
m{=}{-}\tfrac12\rangle$. Fixing in addition an appropriate vector 
$\varphi \,{\in}\, \Hc^\text{Vir}_{s^2/4} \oti \bar\Hc{}^\text{Vir}_{\bar s^2/4}$,
we can write\,\footnote{
  The minus sign in the last term comes from \erf{eq:bulk-OPE}. To see this note
  that $e_2 \,{=}\, J^-_0 e_1$ and $\bar e_2 \,{=}\, \bar J^-_0 \bar e_1$, where
  we chose $J^-(z) \,{=}\, \phi_{(-\sqrt{2},0)}(z)$ and
  $\bar J^-(z) \,{=}\, \phi_{(0,-\sqrt{2})}(z)$.}
  \be \begin{array}{ll} \displaystyle
  e_1 \oti \bar e_1 \oti \varphi = \phi_{(1/\sqrt{2},1/\sqrt{2})}  \,, \etb
  e_1 \oti \bar e_2 \oti \varphi = \phi_{(1/\sqrt{2},-1/\sqrt{2})} \,, \enl
  e_2 \oti \bar e_1 \oti \varphi = \phi_{(-1/\sqrt{2},1/\sqrt{2})} \,,\quad \etb
  e_2 \oti \bar e_2 \oti \varphi = -\phi_{(-1/\sqrt{2},-1/\sqrt{2})}\,.
  \eear\ee
Comparing \erf{eq:u1-def-op-explicit-simp} and \erf{eq:Rsd-Vir-pres-def} then 
establishes \erf{eq:diag-Dgh-u1-pres}, as well as, e.g.,
  \be
  D(g,h) = D(x,y)^{+,-}_{R_{{\rm s.d.}},R_{{\rm s.d.}}} \,, \quad~
  g = \begin{pmatrix}
  \!\mathrm i\, \mathrm e^{\pi \mathrm i \sqrt{2}\, x}\!\!\! & 0 \\
  0 & \!\!\!\!-\mathrm i\,\mathrm e^{-\pi \mathrm i \sqrt{2} \,x}\!
  \end{pmatrix} , ~~~
  h = \begin{pmatrix}
  0 & \!\!\!-\mathrm i\,\mathrm e^{\pi \mathrm i \sqrt{2}\, y} \\
  \!-\mathrm i\,\mathrm e^{-\pi \mathrm i \sqrt{2} \,y}\!\!\! & 0
  \end{pmatrix} .
  \ee

\medskip

Up to now we have concentrated on $\FB{R_\text{s.d.}}$. Let us now also give 
some details on how to arrive at the parametrisation \erf{eq:vir-preserve-set} 
for the defect operators \erf{eq:R2-R1-vir-pres} defined at rational multiples 
of $R_\text{s.d.}$. On a vector $|j,m\rangle$ in the spin-$j$
representation $V_j$ of ${\rm sl}(2,\Cb)$, $\Gamma_{\!L}$ acts as 
  \be
  \rho(\Gamma_{\!L}) \, |j,m\rangle = \mathrm e^{2 \pi \mathrm i m/L}\,
  |j,m\rangle
  \ee
($j \,{\in}\, \Zb/2$, $m+j \,{\in}\, \Zb$, $|m| \,{\leq}\, j$). Let us denote a 
basis of Virasoro-highest weight states in the sector $\Hc_{[s,\bar s]}$ (cf.\ 
the decomposition \erf{eq:Rsd-bulk-decompose}) as $|\tfrac{s}{2},m;
\tfrac{\bar s}{2},\bar m\rangle$. Selecting the maximal torus of ${\rm SU}(2)$ 
that consists of diagonal matrices as the one corresponding to the $\Ue$-charge,
the state $|\tfrac{s}{2},m;\tfrac{\bar s}{2},\bar m\rangle$ has $\Ue$-charges 
$\sqrt{2}\,(m,\bar m)$. This state is in the image of $\hat{D}(0,0)^{+,+}_
{R_{\rm s.d.},R_1}$ iff $\sqrt{2}(m,\bar m) \,{\in}\, \Lambda(R_1)$. Comparing 
with the expression \erf{eq:case3-lattice} for $\Lambda$ we see that this is 
the case if and only if $(m,\bar m) \,{=}\, \tfrac12(kF_1{+}lE_1,kF_1{-}lE_1)$
for some $k,l \,{\in}\, \Zb$. Altogether, we therefore have
  \be
  \sqrt{2}\,(m,\bar m) \in \Lambda(R_1)
  \quad \Leftrightarrow \quad
  m\,{+}\,\bar m \,{\in}\, F_1 \Zb ~~\text{and}~~
  m\,{-}\,\bar m \,{\in}\, E_1 \Zb \,.
  \ee
This implies that for $|\tfrac{s}{2},m;\tfrac{\bar s}{2},\bar m\rangle$ in the 
image of $\hat{D}(0,0)^{+,+}_{R_{\rm s.d.},R_1}$ (these are the only states 
on which $\hat{D}(g,h)$ in (\ref{eq:R2-R1-vir-pres}) actually acts) we have
  \be
  \hat D(\Gamma_{\!E_1}^{},\Gamma_{\!E_1}^{-1}) 
  |\tfrac{s}{2},m;\tfrac{\bar s}{2},\bar m\rangle
  = \mathrm e^{2 \pi \mathrm i (m-\bar m)/{E_1}}\,
  |\tfrac{s}{2},m;\tfrac{\bar s}{2},\bar m\rangle
  = |\tfrac{s}{2},m;\tfrac{\bar s}{2},\bar m\rangle \,.
  \ee
The same holds for $\hat D(\Gamma_{\!F_1},\Gamma_{\!F_1})$, so that we conclude that
  \be\bearll
  \hat D(\Gamma_{\!E_1}^k,\Gamma_{\!E_1}^{-k}) \circ 
      \hat D(0,0)^{+,+}_{R_{\rm s.d.},R_1}
  \!\!\!\etb= \hat D(0,0)^{+,+}_{R_{\rm s.d.},R_1} \,, \enl
  \hat D(\Gamma_{\!F_1}^k,\Gamma_{\!F_1}^{k})  \circ 
     \hat D(0,0)^{+,+}_{R_{\rm s.d.},R_1}
  \!\!\!\etb= \hat D(0,0)^{+,+}_{R_{\rm s.d.},R_1}
  \eear\ee
for all $k \,{\in}\, \Zb$. In the same way one can show that
  \be\bearll
  \hat D(0,0)^{+,+}_{R_2,R_{\rm s.d.}}  \circ 
     \hat D(\Gamma_{\!E_2}^k,\Gamma_{\!E_2}^{-k}) 
  \!\!\etb= \hat D(0,0)^{+,+}_{R_2,R_{\rm s.d.}} , \enl
  \hat D(0,0)^{+,+}_{R_2,R_{\rm s.d.}}  \circ  
     \hat D(\Gamma_{\!F_2}^k,\Gamma_{\!F_2}^{k}) 
  \!\!\etb= \hat D(0,0)^{+,+}_{R_2,R_{\rm s.d.}}
  \eear\ee
for all $k \,{\in}\, \Zb$. Combining these observations with the composition law 
\erf{eq:su2-1-def-fuse} for the defect operators at the self-dual radius, 
we see that $\hat D(g,h)_{R_2,R_1} \,{=}\, \hat D(g',h')_{R_2,R_1}$ if 
  \be
  (g',h') = 
  (\Gamma_{E_2}^{k_2} \Gamma_{\!F_2}^{l_2}g \Gamma_{\!E_1}^{k_1}
   \Gamma_{\!F_1}^{l_1},
  \Gamma_{E_2}^{-k_2}\Gamma_{\!F_2}^{l_2}h \Gamma_{\!E_1}^{-k_1}
   \Gamma_{\!F_1}^{l_1})
  \labl{eq:vir-pres-ident}
for some $k_1,k_2,l_1,l_2 \,{\in}\, \Zb$. One can convince oneself that these 
identifications, together with $(g',h') \,{=}\, (-g,-h)$ are the only cases for 
which $\hat D(g,h)_{R_2,R_1} \,{=}\, \hat D(g',h')_{R_2,R_1}$,
so that we arrive at the formula \erf{eq:vir-preserve-set}.

\medskip

One can also argue, similarly as in \cite{Gaberdiel:2001zq}, that the defect 
operators \erf{eq:R2-R1-vir-pres} together with the $\Ueh$-preserving defect 
operators \erf{eq:u1-def-op-explicit} describe already all fundamental
defect operators (in the sense of question Q3 of section \ref{sec:complete}). In
particular, one can obtain the individual projectors onto the
Virasoro-irreducible sectors which occur in both $\Hc(R_1)$ and $\Hc(R_2)$
as appropriate integrals over these defect operators. Any additional
defect operator can therefore be written as a sum or integral over
these operators, and hence is not fundamental.

\vskip 2em

\noindent {\bf Acknowledgements} \\[1pt]
IR thanks G\'erard Watts for helpful discussions.
JF is partially supported by VR under project no.\ 621-2006-3343. MRG is
supported in parts by the Swiss National Science Foundation and the
Marie Curie network ``Constituents, Fundamental Forces and Symmetries
of the Universe'' (MRTN-CT-2004-005104). IR is partially supported by the
EPSRC First Grant EP/E005047/1 and the PPARC rolling grant PP/C507145/1, and
CS by the Collaborative Research Centre 676 ``Particles, Strings and the
Early Universe - the Structure of Matter and Space-Time''. 


\appendix

\sect{A family of defects with common defect operator}\label{app:family}
   
Here we present an example of a marginal perturbation of a defect $D$ by a 
non-selfadjoint defect field that leads to a family of distinct defects 
$D(\lambda)$ which nonetheless all have the same defect operator 
$\hat D(\lambda) \,{=}\, \hat D$. 
In particular, the defects $D(\lambda)$ can then not be distinguished
in correlators involving only defect lines and bulk fields, but no
disorder or defect fields. 

For the free boson theory $\FB{R}$, let the defect $D$ be a superposition of 
the trivial defect and a $\Ueh$-preserving defect as investigated in section 
\ref{sec:RR-rat},
  \be
  D = D_0 + D_1
  \qquad \text{with} ~~~D_0 = D(0,0)^{+,+}_{R,R}~,~~ 
  D_1 = D(\sqrt{2},0)^{+,+}_{R,R} \,.
  \ee
The spectrum of defect changing fields that change $D_0$ to $D_1$
(when passing  
along the defect in the direction of its orientation) is the same as the 
spectrum of disorder fields that start the defect $D_1$; it can be computed as 
in \erf{eq:D-defect-spectrum}. We find
  \be
  \text{tr}_{\Hc(R)}^{}\big( \hat D_1 \, \rmq^{L_0-\frac{1}{24}} \,
  (\rmq^*)^{\ol L_0-\frac{1}{24}} \big)
  = \sum_{(p,\bar p) \in \Lambda(R)} 
  \frac{{\tilde{\rmq}}^{\frac12 (p+\sqrt{2})^2} (\tilde{\rmq}^*)^{\frac12 \bar p^2}}
  {\eta(\tilde{\rmq})\,\eta(\tilde{\rmq}^*)} \,.
  \ee
Let us for simplicity assume that $R \,{>}\, 2 R_\text{s.d.}$. Then the field of 
lowest conformal weight is the $\Ueh$-primary defect changing field with 
$\Ue$-charges $(q,\bar q) \,{=}\, (\sqrt{2},0)$ and conformal weights 
$(h,\bar h) \,{=}\, (1,0)$; we denote this field by $\varTheta$. This is one of 
the fields that according to the results of section \ref{sec:true-marg} is truly
marginal.\,%
  \footnote{
  While in section \ref{sec:true-marg} only $R^2 \,{\in}\, \Qb$ was considered,
  $\varTheta$ is truly marginal for all values of $R$. For example, it has
  regular (in fact, vanishing) operator product with itself.}
Denote by $\one_0$ the identity field on the defect $D_0$ (which is the same as 
the identity field in the bulk) and by $\one_1$ the identity field on $D_1$. 
Consider a patch of local coordinates on the world sheet where the defect $D$ 
runs along the real axis. Then by construction, for $x\,{>}\,y\,{>}\,z$ we have
  \be
  \one_1(x) \, \varTheta(y) = \varTheta(y) = \varTheta(y) \, \one_0(z)
  \,, \qquad
  \one_0(x) \, \varTheta(y) = 0 = \varTheta(y) \, \one_1(z) \,.
  \ee
In particular, combining these relations with the associativity of the OPE we 
see that $\varTheta$ has vanishing operator product with itself,
  \be
  \varTheta(x)\,\varTheta(z) = \varTheta(x) \,\one_1(y) \,\varTheta(z) = 0 \,.
  \labl{eq:OPE-zero}
Let now $D(\lambda)$ be the defect obtained by perturbing $D$ by 
$\lambda \varTheta$ for some $\lambda \,{\in}\, \Cb$. Correlators involving 
$D(\lambda)$ are obtained from correlators involving $D$ by inserting the 
operator $\exp(\lambda \int_D \varTheta(x) \,{\rm d}x)$. It is then easy to see 
that $\hat D(\lambda) \,{=}\, \hat D$: Consider a loop of defect $D$; expanding
the exponential gives 
  \be
  \one_D + \lambda  \int_{\!D}\! \varTheta(x) \,\mathrm dx
  + \tfrac12 \lambda^2 \int_{\!D} \int_{\!D}\! 
  \varTheta(x)\, \varTheta(y) \,\mathrm dx \,\mathrm dy \, + \ldots
  \ee  
In this expansion all terms with more than one insertion of $\varTheta$ vanish
owing to \erf{eq:OPE-zero}. The term with one $\varTheta$-insertion vanishes 
because we can replace $\varTheta(x)$ by $\varTheta(x)\, \one_0$ and then drag 
the field $\one_0$ around the loop so as to arrive at the product 
$\one_0\, \varTheta(x)$, which is zero.

One can now ask whether the defects $D(\lambda)$ are at all different from 
$D\,{=}\,D(0)$. There are at least two ways to see that this is indeed the case.
The first is to note that $D(\lambda)$, while still transparent to $T(z)$, is no
longer transparent to $J(z)$ if $\lambda \,{\neq}\, 0$. The second way is to 
compute the Hamiltonian generating translations along a cylinder with 
$D(\lambda)$ running along the euclidean time direction; this Hamiltonian turns 
out to be non-diagonalisable for $\lambda \,{\neq}\, 0$. 

Let us elaborate on the second issue. Consider the cylinder obtained by the
identification $w \,{\sim}\, w{+}2\pi \mathrm i$ on the complex plane, with the 
defect $D(\lambda)$ put on the (equivalence class of the) real axis. This 
geometry can be mapped to the full complex plane with coordinate $z$ by 
$z \,{=}\, \mathrm e^{w}$. In the $z$-coordinates, the defect $D(\lambda)$ runs 
along the positive real axis; the Hamiltonian then takes the form
  \be
  H(\lambda) = L_0 + \bar L_0 - \tfrac{1}{12} + \lambda \, \varTheta(1) \,.
  \ee
The Hamiltonian $H(\lambda)$ acts on the space $\Hc_D$ of disorder fields which
create the defect $D(\lambda)$. To expose the non-trivial Jordan cell structure 
of $H(\lambda)$, consider the three vectors
  \be\begin{array}{ll}
  |a_0\rangle \etb\!\!\!= |0\rangle - \lambda \int_0^1 \!\!\!\mathrm dx\,
  \varTheta(x)\, |0\rangle \,,\enl
  |a_1\rangle \etb\!\!\!= \varTheta(0)\, |0\rangle \,,\enL
  |b_1\rangle \etb\!\!\!= J_{-1} |0\rangle - \lambda \, J_{-1} \!\int_0^1\!\!\!
  \mathrm dx \,\varTheta(x) \,|0\rangle + \lambda\, \sqrt{2} \!\int_0^1 \!\!\!
  \mathrm dx \, x^{-1} \big(\varTheta(x)-\varTheta(0) \big) |0\rangle \,.
  \eear\labl{eq:jordan-basis}
First note that the integrals are finite. For $\lambda\,{=}\,0$ the three 
vectors form a basis of the $(L_0,\bar L_0)$ eigenspaces of $\Hc_D$ with 
eigenvalues $(0,0)$ and $(1,0)$. Using $[J_m,\varTheta(x)] \,{=}\, \sqrt{2} \, 
x^m \, \varTheta(x)$ and $[L_0+\bar L_0,\varTheta(x)] \,{=}\, 
\tfrac{\partial}{\partial x}\big( x \varTheta(x) \big)$, one verifies that
  \be
  H(\lambda) |a_0\rangle = -\tfrac{1}{12} |a_0\rangle \,,~~
  H(\lambda) |a_1\rangle = \big(1{-}\tfrac{1}{12}\big) |a_1\rangle \,,~~
  H(\lambda) |b_1\rangle = \big(1{-}\tfrac{1}{12}\big) |b_1\rangle
    - \lambda \sqrt{2}|a_1\rangle \,.
  \ee
Thus the operator $H(\lambda)$ acts on the vectors 
$\{ |a_0\rangle, |a_1\rangle, |b_1\rangle \}$ as the $3{\times}3$-matrix
  \be
  \begin{pmatrix} 
  0 & 0 & 0 \\
  0 & 1 & \!\!-\lambda\sqrt{2}  \\
  0 & 0 & 1  
  \end{pmatrix} \,-\, \frac{1}{12} \, \mathbf1_{3{\times}3} \,,
  \ee
which for $\lambda \,{\neq}\, 0$ contains a non-trivial Jordan-block.


\sect{Representation theory of algebras in $\Uc_{N}$}

In this appendix we present details of the calculations referred to
in section \ref{sec:rat-def}. We assume some familiarity with the
methods of \cite{Fuchs:2002cm,Fuchs:2004dz,Frohlich:2006ch}.

\subsection{The modular tensor category $\Uc_{N}$}\label{app:U_N}

To define the braided tensor category $\Uc_{N}$ we describe the tensor product 
and braiding in a basis. Our conventions are given in 
\cite[sect.\,2.2]{Fuchs:2002cm}. The index set for the simple objects $U_k$ of 
$\Uc_{N}$ is $\Ic \,{=}\, \{0,1,\dots,2N{-}1 \}$. The fusion rules are $U_k\oti
U_\ell \,{\cong}\, U_{[k{+}\ell]}$, where $k,\ell \,{\in}\, \Ic$ and we set
  \be 
  [k{+}\ell] := k{+}\ell \bmod 2N \,\in\Ic \,. 
  \labl{def.modbracket}
The dual of $k \,{\in}\, \Ic$ is $\ol k \,{=}\, 2N{-}k$, and 
all simple objects have unit quantum dimension, $\dim(U_k)\,
   $\linebreak[0]$
{=}\, 1$. The relevant fusing matrices can be found in 
\cite{Brunner:2000wx} or \cite[sect.\,2.5.1]{Fuchs:2002cm}. With a suitable 
choice of bases in the morphism spaces $\Hom(U_i \,{\otimes}\, U_j, U_k)$, 
which for $\Uc_{N}$ are all one-dimensional, the twists $\theta$,
the fusing matrices $\Fs$ and the braiding matrices $\Rs$ are given by
  \be
  \theta_k = \mathrm e^{- \pi \mathrm i k^2 / (2N) } ,~~~~
  \Rs^{(k\ell)[k{+}\ell]} = \mathrm e^{-\pi \mathrm i k \ell / (2N) } ,~~~~
  \Fs^{(rst)[r{+}s{+}t]}_{[s{+}t]\,[r{+}s]} = (-1)^{r\, \sigma(s{+}t)} ,
  \labl{eq:tRF}
where $k,\ell,r,s,t \,{\in}\, \Ic$, 
and $\sigma(k{+}\ell)$ is $0$ if $k{+}\ell \,{<}\, 2N$ and $1$ otherwise.
We denote the bases of $\Hom(U_i \,{\otimes}\, U_j, U_k)$ for which 
these equalities hold by $\lambda_{(i,j)k}$.

{}From these data the $s$-matrix (related to the modular $S$-matrix by
$s_{i,j} \,{=}\, S_{i,j} / S_{0,0}$) is found to be
  \be
  s_{k,\ell}^{} = \mathrm e^{-\pi \mathrm i \, k \ell / N} .
  \ee


\subsection{Frobenius algebras in $\Uc_{N}$}\label{app:Frob}

Every haploid special symmetric Frobenius algebra
in $\Uc_{N}$ is isomorphic to one of the algebras $A_r$ defined as follows
\cite[sect.\,3.3]{Fuchs:2004dz}. As an object in $\Uc_{N}$ we have
  \be
  A_r = \bigoplus_{a=0}^{r-1} U_{2a\,N/r}
  \qquad \text{where $r \,{\in}\, \Zb_{>0}$ divides $N$} \,.
  \ee
For $a \,{\in}\, \Zb_r$ we denote by 
$e_a \,{\in}\, \Hom(U_{2aN/r},A_r)$ and $r_{\!a}\,{\in}\, \Hom(A_r,U_{2aN/r})$ 
embedding and restriction morphisms for the subobject $U_{2aN/r}$ of $A_r$
(hence $r_{\!a}\,{\circ}\,e_a \,{=}\, \id_{U_{2aN/r}}$). 
One can choose the multiplication and unit morphism of $A_r$ to be
  \be
  m = \sum_{a,b=0}^{r-1} e_{a+b} \circ 
  \lambda_{(2aN/r,2bN/r)\,[2(a{+}b)N/r]} \circ (r_{a} \,{\otimes}\, r_{b})
  \qquad \text{and} \qquad
  \eta = e_0 \,.
  \labl{eq:Ar-mult}
With the help of the fusing matrices in \erf{eq:tRF} it is straightforward to 
verify associativity of $m$ and the unit property. The isomorphism class of 
$A_r$ is specified by the Kreuzer-Schellekens bihomomorphism (KSB) $\Xi^{(r)}$, 
which is given by the scalars
$\Xi^{(r)}(b,a)$ that are defined by the equality \cite[sect.\,3.4]{Fuchs:2004dz}
  \be
  m \circ c_{A_r,A_r}^{} \circ (e_a \otimes e_b) 
  = \Xi^{(r)}(b,a) \, m \circ (e_a \otimes e_b) \,,
  \ee
where $c_{A_r,A_r}^{}$ is the self-braiding of $A_r$.
For the multiplication stated in \erf{eq:Ar-mult} this gives
  \be
  \Xi^{(r)}(b,a) = \mathrm e^{- 2\pi \mathrm i\, N ab/r^2} \,.
  \labl{eq:Xi}


\subsection{$A_r$-bimodules in $\Uc_{N}$}\label{app:ArAr-bi}

According to \cite[prop.\,5.16]{Frohlich:2006ch}, simple $A_r$-bimodules 
in $\Uc_{N}$ are labelled by elements of the abelian group
$G^{(r,N)} \,{=}\, H^* \,{\times_H}\, \text{Pic}(\Uc_{N})$, where 
$\text{Pic}(\Uc_{N}) \,{=}\, \Zb_{2N}$ and $H \,{=}\, \Zb_r$ is embedded in 
$\text{Pic}(\Uc_{N})$ via $\iota(a)\,{=}\,2aN/r$.
Accordingly, $h\,{\in}\, H$ acts on $k\,{\in}\,\text{Pic}(\Uc_{N})$
by $h.k \,{=}\, k\,{+}\,\iota(h)$; the action of $h$ on
$\psi \,{\in}\, H^*$ is given by $(h.\psi)(a) \,{=}\, \psi(a)\, \Xi^{(r)}(a,h)$. 

If we label the elements of $H^*$ by $x \,{\in}\, \Zb_r$ via
  \be
  \psi_x(a) = \mathrm e^{-2 \pi \mathrm i x a / r} 
  \labl{psix}
then with the value \erf{eq:Xi} of the KSB we find 
  \be
  (h.\psi_x)(a) = \mathrm e^{-2 \pi \mathrm i x a / r}
  \mathrm e^{- 2\pi \mathrm i\, N ha/r^2} = \psi_{x+hN/r}(a)
  \ee
for $h \,{\in}\, \Zb_r$. The action of $H$ on $\text{Pic}(\Uc_{N})$ is simply 
$h.k \,{=}\, k+2hN/r$. Thus the group $G^{(r,N)}$ is given by 
$G^{(r,N)} \,{=}\, \Zb_r \,{\times_{\Zb_r}}\, \Zb_{2N}$ where the product over 
$\Zb_r$ means that $(a,k+2N/r) \,{=}\, (a+N/r,k)$.
Denote by $t_\psi$ the automorphism of the algebra $A_r$ associated to 
$\psi \,{\in}\, H^*$, see \cite[sect.\,5.1]{Frohlich:2006ch}.
A simple bimodule corresponding to an element $(\psi,k) \,{\in}\, G^{(r,N)}$ 
is provided by $\alpha^+_{\!A_r}(U_k)_{t_\psi}$, i.e.\ an alpha-induced 
bimodule for which the right action of $A_r$ is twisted by the automorphism 
$t_\psi$, see \cite[sect.\,5.2]{Frohlich:2006ch} for definitions. By
\cite[prop.\,5.16]{Frohlich:2006ch}, the fusion of two such bimodules is
given by addition in $G^{(r,N)}$.

It is convenient to label the $A_r$-bimodules by elements of the group 
$G_{n,P,Q}$ defined in \erf{eq:GnPQ-def}, where $N \,{=}\, n^2 PQ$ and 
$r \,{=}\, nP$. The map 
  \be
  \varphi:\quad (\alpha,\beta,\gamma) \mapsto (\alpha{-}\beta,2\beta{+}\gamma)
  \labl{a-b,2b+c}
from $\Zb\,{\times}\,\Zb\,{\times}\,\Zb$ to $\Zb\,{\times}\,\Zb$ induces a 
surjective group homomorphism from $\Zb\,{\times}\,\Zb\,{\times}\,\Zb$ to 
$G^{(r,N)}$ whose kernel is given by the subgroup generated by the elements
$(r,0,0)$, $(0,N/r,0)$ and $(1,1,-2)$. By comparison with the definition
of $G_{n,P,Q}$ this means that $\varphi$ induces a group isomorphism 
  \be
  \hat\varphi:\quad G_{n,P,Q} \overset{\cong}{\longrightarrow} G^{(r,N)} .
  \labl{eq:GnPQ-iso}

Next we compute the action of the corresponding defects on bulk fields. 
Bulk fields are labelled by bimodule morphisms
$\phi \,{\in}\, \Hom_{A_r,A_r}(U_x \,{\otimes^+} A_r \,{\otimes^-} U_y, A_r)$ 
(see \cite[sect.\,2]{Frohlich:2006ch} for definitions). The action of the
defect $D_B$ corresponding to the bimodule $B \,{=}\, \alpha^+_{A_r}(U_k)_{t_{\psi_a}}$
is defined in \cite[eq.\,(2.30)]{Frohlich:2006ch}, and an expression in terms
of quantities in $\Uc_{N}$ can be found in \cite[eq.\,(4.14)]{Frohlich:2006ch}.
A short calculation, which uses in particular various properties of the algebra
$A_r$ and the fact that $\phi$ is a morphism of bimodules in order to get rid of
the $A_r$-loop, shows that
  \be
  \hat D_B(\phi) = \frac{s_{x,k}}{s_{x,0}}\, t_{\psi_a} \circ \phi \circ 
  (\id_{U_x} \,{\otimes}\, t_{\psi_a}^{-1} \,{\otimes}\, \id_{U_y})
  = \mathrm e^{- \pi \mathrm i( (k+a)x + a y )/N } \phi \,.
  \ee

If, as in section \ref{sec:Ar-Ar-bimod},
we use an element $(a,b,\rho) \,{\in}\, G_{n,P,Q}$ to label the defect, then 
according to the isomorphism \erf{eq:GnPQ-iso} we have
  \be
  \hat D^{(r)}_{(a,b,\rho)}(\phi) = 
  \mathrm e^{- \pi \mathrm i( a(x+y) + b(x-y) + \rho x)/N } \phi \,.
  \labl{eq:D_abrho_xy}
Conversely, substituting $q \,{=}\, x/\sqrt{2N}$, $\bar q \,{=}\, y/\sqrt{2N}$, 
$R \,{=}\, \sqrt{P/(2Q)}$ and $N \,{=}\, n^2 PQ$ into \erf{eq:D_abrho-bulk}, 
one recovers the formula \erf{eq:D_abrho_xy}.


\subsection{$A_r$-$A_s$-bimodules in $\Uc_{N}$}\label{app:ArAs-bi}

For $r$ and $s$ divisors of $N$ the algebra $A_\ell$, with 
$\ell \,{=}\, \text{lcm}(r,s)$, has $A_r$ and $A_s$ as subalgebras. 
As in section \ref{sec:Ar-As-bi}, let $A\ors$ be the $A_r$-$A_s$-bimodule 
obtained by considering $A_\ell$ as an $A_\ell$-$A_\ell$-bi\-module
over itself and restricting the left action to $A_r$ and the right
action to $A_s$. We abbreviate $X \,{=}\, A\ors$.

Let $\psi^{(r)}_x \,{\in}\, (\Zb_r)^*_{}$ be defined as in \erf{psix}.
Given two characters $\psi^{(r)}_x \,{\in}\, (\Zb_r)^*_{}$ and 
$\psi^{(s)}_y \,{\in}\, (\Zb_s)^*$ we set $X_{x,y} \,{=}\, {}_{t_x}X{}_{t_y}$, 
with $t_x \,{=}\, t_{\psi^{(r)}_x}$ and $t_y \,{=}\, t_{\psi^{(s)}_y}$. 
Thus $X_{x,y}$ is the bimodule obtained from $X$ by twisting the left and right 
action by the algebra automorphisms defined by the characters 
$\psi^{(r)}_x$ and $\psi^{(s)}_y$, see \cite[sect.\,5.1]{Frohlich:2006ch}.

We would like to compute the space $H_{x,y} \,{:=}\, \Hom_{A_r,A_s}(X,X_{x,y})$
of bimodule intertwiners. Since $X \,{\cong}\, A_\ell$ as 
objects of $\Uc_{N}$, an element $f \,{\in}\, H_{x,y}$ can be expanded as
  \be
  f = \sum_{a=0}^{\ell-1} f_a \, e_a \circ r_a \,,
  \ee
where $f_a \,{\in}\, \Cb$ and $e_a$ and $r_a$ are the embedding and restriction 
morphisms for $U_{2aN/\ell}$ as a subobject of $A_\ell$, as introduced in 
appendix \ref{app:Frob}. Denote by $e_{A_r}$ the embedding of $A_r$ into 
$A_\ell$, by $e_{A_s}$ that of $A_s$ into $A_\ell$, and by $m$ the multiplication
of $A_\ell$. Then the condition for $f$ to be a bimodule intertwiner is
  \be
  f \circ m \circ (m \otimes \id_{A_\ell})
  \circ ( e_{A_r} \otimes \id_{A_\ell} \otimes e_{A_s} )
  =
  m \circ (m \otimes \id_{A_\ell})
  \circ ( (e_{A_r} \circ t_x) \otimes f \otimes (e_{A_s}\circ t_y) ) \,.
  \labl{eq:ArAs-aux1}
Evaluating this equality for appropriate choices of simple 
subobjects, one finds that it holds if and only if
  \be
  f_{a \ell/r + b \ell/s + d} = 
  \psi^{(r)}_x(a) \, \psi^{(s)}_y(b) \, f_d \qquad \text{for all} ~~ 
  a \,{\in}\, \Zb_r, \,~ b \,{\in}\, \Zb_s \,,~ d \,{\in}\, \Zb_\ell \,.
  \ee
The space of solutions of this equality is one-dimensional if $x\,{-}\,y 
\,{\in}\, g\Zb$, with $g\,{=}\,\text{gcd}(r,s)$, and zero-di\-men\-si\-onal 
otherwise, i.e.\, $\dim_\Cb(H_{x,y})\,{=}\,\delta^{[g]}_{x,y}$. In particular, 
$X$ is simple. Similarly one finds that $X_{x,y} \,{\cong}\, X_{u,v}$ if and 
only if $x\,{-}\,u \,{\equiv}\, y\,{-}\,v \, \text{mod} \,g$.

\medskip

The object $A_r \oti A_s$ is an $A_r$-$A_s$-bimodule in the obvious way. 
It is easy to check that the morphism
  \be
  ( t_x^{-1} \oti t_y^{-1} ) \circ ( r_{\!A_r} \oti r_{\!A_s} ) \circ \Delta \,,
  \ee
where $r_{\!A_r}$ and $r_{\!A_s}$ are the restriction morphisms corresponding
to $e_{A_r}$ and $e_{A_s}$ and $\Delta$ is the coproduct of $A_\ell$,
constitutes a nonzero intertwiner of bimodules from $X_{x,y}$ to $A_r \oti A_s$.
Moreover, for $x\,{=}\,0,1,...\,,g{-}1$ the bimodules $X_{x,0}$ are mutually 
non-isomorphic, and hence we can decompose $A_r\oti A_s$ as a bimodule as
  \be
  A_r \otimes A_s \cong\, \bigoplus_{x=0}^{g-1} X_{x,0} \, \oplus Y
  \labl{eq:ArAs-decompos}
for some $A_r$-$A_s$-bimodule $Y$. Computing the quantum dimensions on both 
sides one finds $r s \,{=}\, g \ell + \dim(Y)$, which implies 
$\dim(Y) \,{=}\, 0$ and hence $Y\,{=}\,0$.

Since every simple $A_r$-$A_s$-bimodule $B$ is a sub-bimodule of $A_r \oti U_k 
\oti A_s$ for some $k$, the decomposition \erf{eq:ArAs-decompos} implies that 
it is also a sub-bimodule of $U_k \,{\otimes^+} X_{x,0}$ and of 
$X_{x,0} \,{\otimes^+}\, U_k$ for some $k$ and $x$ (the same holds with 
$\otimes^-$). It is straightforward to see that this implies the 
isomorphisms \erf{eq:radius-change-one-orbit}, e.g.\ $\,X_{x,0} \,{\otimes^+}\,
U_k \,{=}\, X_{x,0} \,{\otimes_{\!A_s}} \alpha^+_{\!A_s}(U_k)
\,{=}\, X_{0,0} \,{\otimes_{\!A_s}} \alpha^+_{\!A_s}(U_k)_{t_x'}
\,{=}\, A^{(rs)} \,{\otimes_{\!A_s}} B^{(s)}_{a,b,\rho}$.

\medskip

Let us finally sketch the computation of the action of the defect corresponding 
to $X$ on bulk fields. Fix a basis $\phi_{x,y}^{(r)} \,{\in}\, 
\Hom_{A_r,A_r}(U_x \,{\otimes^+} A_r \,{\otimes^-} U_y, A_r)$ by setting
  \be
  \phi_{x,y}^{(r)} \circ ( \id_{U_x} \otimes \eta \otimes \id_{U_y} )
  = e_{[x+y]r/(2N)}^{(r)} \circ \lambda_{(x,y)\,[x+y]}
  \ee
where $e_a^{(r)} \,{\in}\, \Hom(U_{2aN/r},A_r)$. (This differs from the basis 
chosen in section \ref{sec:comp-FB} by phases.) Since the morphism spaces 
$\Hom_{A_r,A_r}(U_x \,{\otimes^+} A_r \,{\otimes^-} U_y, A_r)$ are either zero 
or one-dimensional, we have $\hat D_{X}\,\phi_{x,y}^{(s)} \,{=}\, \xi\, 
\phi_{x,y}^{(r)}$ for some $\xi \,{\in}\, \Cb$. To determine $\xi$ we evaluate 
relation (4.14) of \cite{Frohlich:2006ch} (with $X_\nu \,{=}\, X$, 
$X_\mu \,{=}\, A \,{=}\, A_r$, and $B \,{=}\, A_s$). After a while we obtain
  \be
  \xi = \delta^{[2N/r]}_{x{+}y,0}\, \delta^{[2N/s]}_{x{+}y,0}\,
  \delta^{[2s]}_{x{-}y,0}\, \Frac{1}{r} \sum_{m=0}^{\ell-1}
  \exp\!\big({-}2 \pi \mathrm i \,\Frac{m}\ell\, \Frac{x-y}2 \big) \,.
  \ee
This gives rise to the relation \erf{eq:Drs-action}
(recall that $\ell \,{=}\, \text{lcm}(r,s)$).


\subsection{The $A_r$-bimodule $D$ and self-locality}\label{app:self-loc}

Consider the free boson compactified at radius $R = 2^{-1/2} E/F$ 
(i.e.\ $P=E^2$ and $Q=F^2$) in terms of 
the algebra $A_{nE^2}$ in $\Uc_{(nEF)^2}$. Set $a = nEF$ and $r=nE^2$.

Let us first check that every defect field $\varTheta$ of conformal weight 
$(h,\bar h)\,{=}\,(1,0)$ which is also a $J_0$-eigenvector is self-local. Such a
$\varTheta$ is an element of the sector $U_{k} \,{\otimes}\, \ol U_0$, for 
$k\,{=}\,0$, $k\,{=}\,2a$, or $k\,{=}\,2N{-}2a$. The important point to 
note is that according to \erf{eq:tRF} the braiding is trivial,
  \be
  c_{U_k,U_k}^{} = \id_{U_k \otimes U_k}
  \qquad \text{for} ~~ k\,{=}\,0,\,2a,\,2N{-}2a \,.
  \ee
Using the TFT formalism it is then straightforward to verify the identity 
\erf{eq:top-def-marg} (see e.g.\ \cite[sect.\,4.1]{Frohlich:2006ch} 
for the TFT representation of some defect correlators). The case that
$\varTheta$ has weight $(0,1)$ can be treated analogously.

\medskip

Next let us compute the spectrum of defect fields on the defect $D$ defined in 
\erf{eq:D-superpos}. Denote the bimodule labelling the defect $D$ by $B$, and 
recall that according to \erf{a-b,2b+c} the bimodule labelling the 
elementary defect $D^{(r)}_{(c,d,\rho)}$ is the twisted alpha-induced bimodule 
$\alpha_{A_r}^+(U_{2d+\rho})_{t_{c-d}}$, where again 
$t_x \,{\equiv}\, t_{\psi_{x}^{(r)}}$. Thus we have the decomposition 
  \be
  B = \bigoplus_{k=0}^{E-1} \bigoplus_{l=0}^{F-1} 
  \alpha_{A_r}^+(U_{2al})_{t_{a(k-l)}} \,.
  \labl{eq:bimod-B-decompos}
Defect fields on $D$ in the sector $U_i \,{\otimes}\, \ol U_j$ are labelled by 
elements of the morphism space $H_{ij} \,{=}\, \Hom_{A_r,A_r}
(U_i \,{\otimes^+_{}}\, B \,{\otimes^-_{}}\, U_j, B)$, so that the 
multiplicity $Z_{ij}^D$ in formula \erf{eq:def-field-on-D} is given by
$Z_{ij}^D \,{=}\, \dim_\Cb(H_{ij})$. Using \erf{eq:bimod-B-decompos}, $H_{ij}$
can be written as a direct sum of spaces of the form
  \be
  \Hom_{A_r,A_r}(U_i \,{\otimes^+} \alpha^+_{A_r}(U_k)_{t_x} \,{\otimes^-} U_j, 
  \alpha^+_{A_r}(U_\ell)_{t_y}) \,\cong\, \Hom_{A_r,A_r}(X,A_r)
  \ee
with
  \be
  X = (U_i \,{\otimes^+} \alpha^+_{A_r}(U_k)_{t_x} \,{\otimes^-} U_j)
  \,{\otimes_{A_r}} (\alpha^+_{A_r}(U_\ell)_{t_y})^\vee
  \cong \alpha^+_{A_r}(U_{[i+j+k-\ell]})_{\chi_{U_j}^{-1}t_x t_y} \,,
  \ee
where the second isomorphism, with $\chi_{U_j}^{}(2bN/r) \,{=}\, S_{j,2bN/r}/
S_{j,0} \,{=}\, \mathrm e^{-\pi\mathrm i \,jb/r}$, holds owing to 
\cite[prop.\,5.8 and 5.9]{Frohlich:2006ch}. Putting these results together, 
we find
  \be 
  Z^D_{ij} 
  = \sum_{k,k'=0}^{E-1} \sum_{l,l'=0}^{F-1} d_{2la,2l'a}^{a(k-l),a(k'-l')}
  \ee
with $d_{k,\ell}^{x,y} \,{=}\, \dim_\Cb(\Hom_{A_r,A_r}(X,A_r))$. Using in 
addition \cite[prop.\,5.10]{Frohlich:2006ch} and \erf{eq:Xi}
we can write $d_{k,\ell}^{x,y}$ as
  \be
  d_{k,\ell}^{x,y} = \delta^{[2N/r]}_{i+j+k-\ell,0}\,
  \delta_{\Xi_{A_r}(\cdot,(i+j+k-\ell)r/(2N))t_x t_y\,,\,\chi_{U_j}^{}}
  = \delta^{[2N/r]}_{i+j+k-\ell,0} \, \delta^{[2r]}_{i-j+k-\ell+2(x-y),0} \,.
  \ee
After a short calculation one then indeed obtains formula 
\erf{eq:def-field-on-D} (recall that $N\,{=}\,a^2$).
  
\medskip

What remains to complete the demonstrations of the claims in section 
\ref{sec:true-marg} is to give a maximal self-local subspace of defect fields 
on the defect $D$. As a first step we construct a basis for the spaces
  \be
  L_x := \Hom_{A_r,A_r}(U_{2ax} \,{\otimes^+} B,B) \qquad\text{and}\qquad
  R_x := \Hom_{A_r,A_r}(B \,{\otimes^-} U_{2ax},B) 
  \ee
with $x \,{\in}\, \{0,1,a{-}1\}$ (note that $a^2\,{=}\,N$). Denote by $e_{k,l}$ 
and $r_{k,l}$ the bimodule intertwiners furnishing the embedding and restriction 
morphisms for the simple sub-bimodule $\alpha_{A_r}^+(U_{2al})_{t_{a(k-l)}}$
of $B$. Let $e_a$ and $r_a$ be as in
\erf{eq:Ar-mult} and denote by $\bar\lambda^{(i,j)k}$ the basis vector in
$\Hom(U_k,U_i \otimes U_j)$ dual to $\lambda_{(i,j)k}$ in the sense that
$\lambda_{(i,j)k} \,{\circ}\, \bar\lambda^{(i,j)k} \,{=}\, \id_{U_k}$. Consider 
the morphisms
  \begin{eqnarray}&
  \alpha(x)_{k,l}^{k',l'} \!\!\!\etb := e_{k',l'} \circ 
  \big( m \oti \id_{U_{2al'}} \big) \circ 
       \big( \id_{A_r} \oti e_{2a(l-l'+x)r/N} \oti \id_{U_{2al'}}\big) 
  \nonumber\enL& \etb \quad 
       \circ\, \big( \id_{A_r} \otimes (
       \bar\lambda^{([2a(l-l'+x)],2al')[2a(l+x)]}
       \circ \lambda_{(2ax, 2al)[2a(l+x)]}) \big) 
  \nonumber\enL& \etb \quad
       \circ\, \big( c_{U_{2ax}^{},A_r} \oti \id_{U_{2al}} \big) \circ
       \big( \id_{U_{2ax}} \oti  r_{k,l}\big)
  \, \in \Hom( U_{2ax} \oti B, B) \,,
  \nonumber\enl&
  \beta(x)_{k,l}^{k',l'} \!\!\!\etb := e_{k',l'} \circ 
  \big( m \oti \id_{U_{2al'}} \big) \circ 
       \big( \id_{A_r} \oti e_{2a(l-l'+x)r/N} \oti \id_{U_{2al'}}\big) 
  \nonumber\enL& \etb \quad
       \circ\, \big( \id_{A_r} \otimes (  
       \bar \lambda^{([2a(l-l'+x)],2al')[2a(l+x)]}
       \circ \lambda_{(2al, 2ax)[2a(l+x)]} ) \big)
  \nonumber\enL& \etb \quad
       \circ\, \big( r_{k,l} \oti \id_{U_{2ax}} \big)
  \quad \in \Hom(B \oti U_{2ax} , B) \,.
  \end{eqnarray}
These are nonzero if and only if $U_{2a(l-l'+x)}$ is a subobject of $A_r$, i.e.\
iff $a(l{-}l'{+}x)r/N \,{\in}\, \Zb$. Furthermore it is easy to check that 
these morphisms intertwine the left action of $A_r$. For the right action a 
small calculation is needed, the conclusion being that $\alpha(x)_{k,l}^{k',l'}$
intertwines the right action iff $k'\,{\equiv}\,k{+}x\,\text{mod}\,E$,
and that $\beta(x)_{k,l}^{k',l'}$ intertwines the right action iff 
$k'\,{\equiv}\, k{-}x \,\text{mod}\,E$. For $u \,{\in}\, \Zb$ let $[u]^K$ 
be the element of $\{0,1,...\,,K{-}1\}$ that is equal to $u$ modulo $K$. (The 
relation to the bracket notation $[\,\cdot\,]$ as introduced in 
\erf{def.modbracket} is thus $[u]\,{\equiv}\,[u]^{2N}$.) Then altogether we have
  \be\bearll
  \alpha(x)_{k,l} \equiv
     \alpha(x)_{k,l}^{[k+x]^E,[l+x]^F}\, \in L_x \qquad\text{and}
  \enL
  \beta(x)_{k,l} \equiv \beta(x)_{k,l}^{[k-x]^E,[l+x]^F} ~ \in R_x 
  \eear\labl{eq:Lx-Rx-basis}
for $k \,{\in}\, \{0,1,...\,,E{-}1\}$ and $l \,{\in}\, \{0,1,...\,,F{-}1\}$.
The morphisms $\alpha(x)_{k,l}$ and $\beta(x)_{k,l}$ are all nonzero. It is 
also easy to verify that they are linearly independent (compose a zero linear 
combination with $e_{k,l}$ from the right to isolate the individual terms). In 
fact, the $\alpha(x)_{k,l}$ and $\beta(x)_{k,l}$ provide bases of $L_x$ and 
$R_x$, respectively, as can be checked directly or by 
using that by \erf{eq:def-field-on-D} $\dim(L_x)\,{=}\,\dim(R_x)\,{=}\,EF$.

Using the TFT representation of correlators, one can verify that for
defect fields labelled by 
  \be
  \varTheta \in \Hom_{A_r,A_r}(U_{2ax} \,{\otimes^+} B \,{\otimes^-}\, U_{2ay}, B)
  \quad \text{and} \quad  
  \varTheta' \in\Hom_{A_r,A_r}(U_{2ax'}\,{\otimes^+} B \,{\otimes^-}\, U_{2ay'},B) 
  \ee
with $x,y,x',y' \,{\in}\, \{0,1,a{-}1\}$ such that at least one of $x$, $y$ is 
zero and at least one of $x'$, $y'$ is zero, the self-locality condition
\erf{eq:top-def-marg} is equivalent to
  \be
  \varTheta \circ \big( \id_{U_{2ax}} \oti \varTheta' \oti \id_{U_{2ay}} \big) \circ
  \big( c_{U_{2ax},U_{2ax'}}^{~~-1} \oti \id_B \oti c_{U_{2ay},U_{2ay'}} \big)
  =
  \varTheta' \circ \big( \id_{U_{2ax'}} \oti \varTheta \oti \id_{U_{2ay'}} \big)
  \,.
  \labl{eq:self-loc-Theta}
Let us analyse this condition using the bases \erf{eq:Lx-Rx-basis}. For a 
collection $t\,{=}\,\{ t_{k,l} \,|\, k\,{=}\,0,1,...\,,
    $\linebreak[0]$
E{-}1\,,~l=0,1,...\,,F{-}1 \}$ of $EF$ numbers we set
  \be
  \varTheta[x,t]_L := \sum_{k=0}^{E-1}\sum_{l=0}^{F-1} t_{k,l}\, \alpha(x)_{k,l}
  \qquad \text{and} \qquad
  \varTheta[x,t]_R := \sum_{k=0}^{E-1}\sum_{l=0}^{F-1} t_{k,l}\, \beta(x)_{k,l} \,.
  \ee
Set now $\varTheta\,{=}\,\varTheta[x,t]_L$ and $\varTheta'\,{=}\,\varTheta[x',t']_L$ 
in \erf{eq:self-loc-Theta}. To evaluate the resulting condition on $t$ and $t'$
compose both sides with $e_{k,l}$ to remove the summation. Then rewrite the morphisms
on either side using the $\Fs$ and $\Rs$ matrices as given in \erf{eq:tRF}.
This step is simplified by the fact that 
  \be
  \Fs^{(2e,2f,2g)[2e+2f+2g]}_{[2f+2g]\,[2e+2f]} = 1 \qquad \text{and} \qquad
  \Rs^{(2ae,2af)[2a(e+f)]} = 1 
  \ee
for all $e,f,g$. One finds that \erf{eq:self-loc-Theta} holds if and only if 
  \be
  t_{[k+x']^E,[l+x']^F} \cdot t'_{k,l} = t_{k,l}^{} \cdot t'_{[k+x]^E,[l+x]^F}
  \labl{eq:sl-cond1}
for all $k,l$. Similar conditions result when setting $\varTheta \,{=}\, 
\varTheta[x,t]_L$, $\varTheta' \,{=}\, \varTheta[x',t']_R$ and 
$\varTheta \,{=}\, \varTheta[x,t]_R$, $\varTheta' \,{=}\, \varTheta[x',t']_R$ in
\erf{eq:self-loc-Theta}. All these conditions are fulfilled if $t_{k,l}$ and 
$t'_{k,l}$ are independent of $k$ and $l$. We therefore introduce the vector space
  \be
  \mathcal{L} := \text{span}_\Cb\big\{
  \varTheta[0,t]_L, \varTheta[1,t]_L, \varTheta[a{-}1,t]_L, 
  \varTheta[0,t]_R, \varTheta[1,t]_R, \varTheta[a{-}1,t]_R 
  ~\text{with}~ t_{k,l}\,{\equiv}\,1 \big\}\,. 
  \ee
$\mathcal{L}$ is a six-dimensional self-local subspace of
of $L_0 \,{\oplus}\, L_1 \,{\oplus}\, L_{a-1} \,{\oplus}\, R_0 \,{\oplus}\, 
R_1 \,{\oplus}\, R_{a-1}$.

Let now $\varTheta[x,t]_L \,{\in}\, L_x$ be arbitrary and suppose that the 
space $\text{span}_\Cb\{ \mathcal{L} , \varTheta[x,t]_L \}$ is self-local. Then 
in particular \erf{eq:sl-cond1} has to hold for $t'_{k,l} \,{\equiv}\, 1$ and 
$x'\,{=}\,1$, i.e., for all $k,l$ we have $t_{k,l} \,{=}\, t_{[k+1]^E,[l+1]^F}$.
This implies $t_{k,l} \,{=}\, t_{[k-l]^E,0}$ and $t_{k,0}\,{=}\,t_{[k+mF]^E,0}$
for all $m\,{\in}\,\Zb$. Since $E$ and $F$ are coprime, the latter condition 
leads to $t_{k,0} \,{=}\, t_{0,0}$ and therefore $t_{k,l} \,{=}\, t_{0,0}$ for 
all $k,l$. Thus already $\varTheta[x,t]_L \,{\in}\, \mathcal{L}$. A similar 
argument applies to $\varTheta[x,t]_R$. Thus $\mathcal{L}$ is maximal.


\sect{Morita equivalence of $\tilde A_1$ and $\tilde A_N$ in $\Dc_{N}$}
\label{app:morita}

Let us first describe the fusion rules in the modular tensor category $\Dc_{N}$,
which can be extracted from \cite{Dijkgraaf:1989hb}. 
The category $\Dc_{N}$ has $N{+}7$ 
isomorphism classes of simple objects. We denote a choice of representatives by
  \be
  \{ \one, J,\, V_1, V_2, ...\,, V_{N-1},\, W_0, W_1 \} \cup 
  \{ \sigma_0, \sigma_1, \tau_0, \tau_1 \} \,.
  \ee
We also use the notation $J_\alpha$, $\sigma_\alpha$, $\tau_\alpha$ for
$\alpha \,{\in}\, \Zb_2$ and $V_r$ for $r\,{=}\,0,1,...\,,N$, where we set
  \be
  J_0 = \one \,,~~~~ J_1 = J \,,~~~~ V_0 = \one \oplus J \,,~~~~
  V_N = W_0 \oplus W_1 \,.
  \ee
The twist eigenvalues (given by $\mathrm e^{-2 \pi \mathrm i \Delta}$ with 
$\Delta$ the conformal weight) and the quantum dimensions are 
  \be
  \begin{tabular}{c|ccccc}
         & $J_\alpha$ & $W_\alpha$ & $V_r$ & $\sigma_\alpha$ & $\tau_\alpha$ \\
  \hline   
  \\[-.8em]      
  $\theta$ & $1$ & $\mathrm i^{-N}$ & $\mathrm e^{-\pi \mathrm i r^2/(2N)}$
           & $\mathrm e^{-\pi \mathrm i /8}$ & $\mathrm e^{-\pi \mathrm i 9/8}$
  \\[0.2em]
  $\dim$ & $1$ & $1$ & $2$ & $\sqrt{N}$ & $\sqrt{N}$ 
  \end{tabular}
  \ee 
for $\alpha \,{\in}\, \Zb_2$ and $r\,{=}\,0,1,...\,,N$.
For $u \,{\in}\, \{0,1,...\,,2N\}$ we set 
  \be
  \{u\} := \left\{ \bearll u & {\rm if}~\, u\,{\le}\,N \,, \\[2pt]
  2N\,{-}\,u & {\rm if}~\, u \,{>}\,N \,. \eear \right.
  \ee
The fusion rules in the untwisted sector are 
  \be\begin{array}{lll}
  V_r \otimes V_s \cong V_{\{r+s\}} \oplus V_{|r-s|} ~,~~ &
  J_\alpha \otimes V_r \cong V_r  ~,~~&
  W_\alpha \otimes V_r \cong V_{N-r} ~,
  \\[.5em]
  J_\alpha \otimes J_\beta \cong J_{\alpha+\beta}  \,,~~&
  J_\alpha \otimes W_\beta \cong W_{\alpha+\beta}  \,,~~&
  W_\alpha \otimes W_\beta \cong J_{\alpha+\beta+N} 
  \eear\ee
for $\alpha,\beta \,{\in}\, \Zb_2$ and $r,s \,{\in}\, \{0,1,...\,,N\}$.
The fusion rules involving fields from the twisted sector are given by
  \bea
  J_\alpha \otimes \sigma_\beta \cong 
      \delta^{[2]}_{\alpha,0}\, \sigma_\beta \,\oplus\,
      \delta^{[2]}_{\alpha,1}\, \tau_\beta \,,\qquad
  W_\alpha \otimes \sigma_\beta \cong 
       \delta^{[2]}_{\alpha,\beta+N}\, \sigma_{\beta+N} \,\oplus\,
       \delta^{[2]}_{\alpha,\beta+N+1}\, \tau_{\beta+N} \,,
  \\[.7em]
  V_r \otimes \sigma_\alpha \cong \sigma_{\alpha+r} \oplus \tau_{\alpha+r} \,,
  \\[.6em]
  \sigma_\alpha \otimes \sigma_\beta
    \cong \delta^{[2]}_{\alpha,\beta}\, W_\alpha
      \,\oplus\, \delta^{[2]}_{\alpha+\beta+N,0}\,
      \big(\one \oplus \bigoplus_{k=1, k\,\text{even}}^{N-1}\! V_k \big)
      \,\oplus\, \delta^{[2]}_{\alpha+\beta+N,1}\,
      \big( \!\bigoplus_{k=1, k\,\text{odd}}^{N-1}\! V_k \big) \,.
  \eear\ee
By verifying which fusions contain the tensor unit $\one$, one finds the
duality to be
  \be
  (V_r)^\vee \cong V_r \,,~~
  (J_\alpha)^\vee \cong J_\alpha \,,~~
  (W_\alpha)^\vee \cong W_{\alpha+N} \,,~~
  (\sigma_\alpha)^\vee \cong \sigma_{\alpha+N} \,,~~
  (\tau_\alpha)^\vee \cong \tau_{\alpha+N} \,.
  \ee

Let us now consider the algebras $\tilde A_1$ and $\tilde A_N$ in $\Dc_{N}$.
The algebra $\tilde A_1 \,{=}\, \one \,{\oplus}\, J$ corresponds to the 
decomposition of the chiral algebra $\Ueh_{N}$ into representations of its 
subalgebra $\Ueh_{N}/\Zb_2$. Conversely, the category of local left modules of 
$\tilde A_1$ in $\Dc_{N}$ can be made into a modular tensor category $(\Dc_{N})
^{\ell\rm oc}_{\tilde A_1}$ (see e.g.\ \cite[sect.\,3.4]{Frohlich:2003hm} for 
details and references), which is in fact equivalent to $\Uc_{N}$. 
The free boson compactified at radius $R\,{=}\,1/\sqrt{2N}$ can be described 
by either using the algebra $A_1$ with the chiral algebra $\Ueh_N$ or using 
$\tilde A_1$ with $\Ueh_{N}/\Zb_2$.

Consider the simple induced $\tilde A_1$-module 
  \be
  \Sigma := \tilde A_1 \,{\otimes}\, \sigma_0
  \labl{defSig}
(one can also use $\sigma_1$ in place of $\sigma_0$). The object $\Sigma^\vee 
{\otimes_{\!\tilde A_1}} \Sigma$ carries again a natural structure of
simple symmetric Frobenius algebra \cite[prop.\,2.13]{Fuchs:2003id}; 
we denote this algebra by $\tilde A_N$. The algebra $\tilde A_N$ constructed in
this way is Morita equivalent to $\tilde A_1$ \cite[thm.\,2.14]{Fuchs:2003id}. 
The $\tilde A_1$-$\tilde A_N$-bimodule $X$ and 
$\tilde A_N$-$\tilde A_1$-bimodule $X'$ which furnish the 
Morita-context are $X \,{=}\, \Sigma$ and $X'\,{=}\,\Sigma^\vee$. In particular,
  \be
  \dim(X) = \dim(X') = \dim(\tilde A_1) \, \dim(\sigma_0) = 2 \sqrt{N} \,.
  \ee
Using the braiding to take $\tilde A_1$ past $\sigma_0$, one can turn $\tilde A_N$
into a left $\tilde A_1$-module (this can be done in two ways, it does not matter 
which of them one chooses). Using that $\tilde A_1$ is
commutative and that $\sigma_0^\vee \,{\otimes}\, \sigma_0$ is transparent to
$\tilde A_1$, it is not hard to check that this way $\tilde A_N$
becomes in fact a local $\tilde A_1$-module. 
Let us denote this local module by $B_N$. 
It turns out that the multiplication can be lifted as well, so that
$B_N$ becomes an algebra in the category $(\Dc_{N})^{\ell\rm oc}_{\tilde A_1}$.
The object in $\Dc_{N}$ underlying $B_N$ is 
  \be
  \tilde A_N \cong \tilde A_1 \otimes \sigma_0^\vee \otimes \sigma_0
  \cong \sum_{m = 0}^{N-1} V_{\{2m\}} \,.
  \ee
This also shows that $\dim(\tilde A_N) \,{=}\, 2N$. Via the equivalence
$(\Dc_{N})^{\ell\rm oc}_{\tilde A_1} \,{\simeq}\;\Uc_{N}$ the corresponding
object in $\Uc_{N}$ is $\bigoplus_{m=0}^{N-1} U_{2m}$. Up to isomorphism,
this object carries a unique structure of a special symmetric Frobenius algebra,
namely $A_N$. The image of $B_N$ under the equivalence is therefore isomorphic 
to $A_N$ as a symmetric Frobenius algebra. Thus all correlators which can be 
described using the algebra $A_N$ with the chiral algebra $\Ueh_N$ can 
alternatively be described using $\tilde A_N$ and $\Ueh_{N}/\Zb_2$.


\sect{Sign of $\Ueh$-preserving defect operators}\label{app:defect-sign}

Here we investigate the sign of the parameter $\lambda_D$ which appears
in formula \erf{eq:d_qq-res}. From evaluating the trace 
\erf{eq:D-defect-spectrum} we know that we can write the parameter $\lambda_D$
of the defect operator $\hat D(x,y)^{\epsilon,\bar\epsilon}_{R_2,R_1}$ as
  \be
  \lambda_D = \sigma(x,y)^{\epsilon,\bar\epsilon}_{R_2,R_1} 
  \sqrt{MN} \, \sqrt{\langle \one^{(R_1)} \rangle} / 
  \sqrt{\langle \one^{(R_2)} \rangle}
  \quad~\text{with} ~~
  \sigma(x,y)^{\epsilon,\bar\epsilon}_{R_2,R_1} \in \{ \pm 1 \} \,.
  \labl{eq:app-lambdaD-sigma}
For this expression to be unambiguous, let us agree that $\sqrt{MN}\,{>}\,0$ 
and let us choose, once and for all, a square root 
$\sqrt{\langle \one^{(R)}\rangle}$ for each value of $R$.

To determine the signs $\sigma(x,y)^{\epsilon,\bar\epsilon}_{R_2,R_1}$ first 
note that if $R_1\,{=}\,R_2\,{=:}\,R$, then one can consider instead of 
\erf{eq:D-defect-spectrum} an analogous trace with the insertion of only a 
single defect operator $\hat D(x,y)^{\epsilon,\bar\epsilon}_{R,R}$. This leads 
to an overall factor of $\lambda_D$; positivity of the coefficients in 
the dual channel then enforces
  \be
  \sigma(x,y)^{\epsilon,\bar\epsilon}_{R,R} = 1 \,.
  \ee
Note that acting with $\hat D(x,y)^{\epsilon,\bar\epsilon}_{R,R}$ on the identity 
field $\one^{(R)}$ of $\FB{R}$ results in $\sqrt{MN} \, \one^{(R)}$. In 
particular, the coefficient is positive.

Next consider two defects $D_1 \,{=}\, D(x,y)^{\epsilon,\bar\epsilon}_{R_2,R_1}$
and $D_2 \,{=}\, D(u,v)^{\nu,\bar\nu}_{R_1,R_2}$. The fused defect
$D \,{=}\, D_2 \,{*}\, D_1$ acts on the identity field of $\FB{R_1}$ as
  \be
  \hat D \, \one^{(R_1)} = 
  \sigma(u,v)^{\nu,\bar\nu}_{R_1,R_2}
  \sigma(x,y)^{\epsilon,\bar\epsilon}_{R_2,R_1} MN \, \one^{(R_1)} .
  \ee
Since $D$ is a $\Ueh$-preserving defect of $\FB{R_1}$, by the above result the 
coefficient appearing here is positive. This is possible for all choices of 
parameters only if $\sigma(x,y)^{\epsilon,\bar\epsilon}_{R_2,R_1}$ is 
independent of $x$, $y$ and of $\epsilon$, $\bar\epsilon$. Thus
  \be
  \sigma(x,y)^{\epsilon,\bar\epsilon}_{R_2,R_1} = \sigma_{\!R_2,R_1}^{}
  \qquad \text{and} \qquad \sigma_{\!R,R}^{} = 1 \,.
  \ee

Consider the (rather big and non-connected) graph $\Gamma$ obtained by taking 
a vertex for every positive real number $R$ and a directed edge $e(R_1,R_2)$ 
between two vertices $R_1$ and $R_2$ if either $R_1/R_2$ or $R_1 R_2$ is 
rational (thus for each edge $e(R_1,R_2)$ there is also an edge 
$e(R_2,R_1)$). To a directed edge $e(R_1,R_2)$ assign the value 
$\sigma_{R_1,R_2}$. Let $\gamma$ be a closed path in $\Gamma$. Let 
$R_1, R_2,...\,, R_{n+1}\,{=}\,R_1$ be the vertices traversed by the path. By 
the same argument as above, acting with the fused defect
  \be
  D(x_{n},y_{n})^{\epsilon_{n}\bar\epsilon_n}_{R_{1}R_n} * \cdots *
  D(x_{2},y_{2})^{\epsilon_{2}\bar\epsilon_2}_{R_{3}R_2} *
  D(x_{1},y_{1})^{\epsilon_{1}\bar\epsilon_1}_{R_{2}R_1}
  \ee
on $\one^{(R_1)}$ shows that $\sigma_{R_{1} R_n} \,{\cdots}\, \sigma_{R_3 R_2}
\, \sigma_{R_2 R_1} \,{=}\,1$. One can therefore write $\sigma_{R_2 R_1}\,{=}\,
\sigma_{R_1} \,{\cdot}\, \sigma_{R_2}$ for some function $\sigma {:}\ \Rb_{>0} 
\,{\rightarrow}\, \{\pm 1\}$. (In choosing $\sigma$ one has the freedom of an
over-all sign on each connected component of $\Gamma$.)

Comparing the result found so far, $\sigma(x,y)^{\epsilon,\bar\epsilon}_{R_2,R_1} 
\,{=}\, \sigma_{R_2} \sigma_{R_1}$ with \erf{eq:app-lambdaD-sigma} we see that 
the freedom that is left in determining the signs of the numbers $\lambda_D$ is 
precisely the freedom to choose the square roots $\sqrt{\langle \one^{(R)}\rangle}$.


\section{Constraints on Virasoro preserving defects}\label{app:RinSLxSL}

In this appendix we show that one can always choose the parametrisation 
\erf{eq:R=gxh} of the linear map $R$ in \erf{DtoR}. Select a highest weight 
state $\varphi_{s,\bar s}$ in each of the spaces $\Hc{}^\text{Vir}_{s^2/4} 
\oti \bar\Hc{}^\text{Vir}_{\bar s{}^2/4}$. The Virasoro-highest weight states 
in $\Hc_{[s,\bar s]}$ can then be written as $u \oti \varphi_{s,\bar s}$ with 
$u \,{\in}\, V_{s/2} \oti \ol V_{\bar s/2}$. Denote the corresponding primary 
field by $[u \oti \varphi_{s,\bar s}](z)$. We identify $V_{0} \oti \ol V_{0}$ 
with $\Cb$ and take $\varphi_{0,0}$ to be the identity field; we also abbreviate
$W \,{=}\, V_{1/2}\oti \ol V_{1/2}$.

The leading terms in the OPE of two primary fields in $\Hc_{[1,1]}$ are of the 
form
  \be\bearll
  \big[u\oti\varphi_{1,1}\big](z) \, \big[v\oti\varphi_{1,1}\big](w) \!\!\etb
  =\, A_{0,0}(u,v) \, r^{-1} \, \varphi_{0,0}(w)  
  \,+\, \mathrm e^{\mathrm i\vartheta} \, [A_{2,0}(u,v) \oti \varphi_{2,0}](w)  
  \enL \etb \hspace*{6.5em}
  \,+\, \mathrm e^{-\mathrm i\vartheta} \, [A_{0,2}(u,v) \oti \varphi_{0,2}](w)  
  \,+\, O(r) \,,
  \eear  
  \labl{eq:OPE-phiphi-to-one}
where $r\,{=}\,|z{-}w|$, $\vartheta \,{=}\, \text{arg}(z{-}w)$ and
$A_{s,\bar s} {:}\ W \,{\times}\, W \,{\rightarrow}\, V_{s/2} \oti 
\ol V_{\bar s/2}$ are bilinear maps. Compatibility with the action of the 
$\suh(2)$ zero modes requires $A_{s,\bar s}$ to be an intertwiner,
  \be
  A_{s,\bar s}((g\oti h) u, (g\oti h) v) = 
  \big(\rho_{s/2}(g) \oti \rho_{s/2}(h)\big) A_{s,\bar s}(u,v)
  \ee
for all $g,h \,{\in}\, {\rm SL}(2,\Cb)$ and $u,v \,{\in}\, W$.

The defect operator $\hat D$ restricted to $\Hc_{s,\bar s}$ is of the form
  \be
  \hat D \big|_{\Hc_{[s,\bar s]}}
  = \lambda_D R_{s,\bar s} \oti 
  \id_{\Hc^\text{Vir}_{s^2/4} {\otimes} \bar\Hc{}^\text{Vir}_{\bar s{}^2/4}}
  ~\quad \text{with} \quad
  R_{s,\bar s} \in \text{End}(V_{s/2} \oti \ol V_{\bar s/2}) \,.
  \ee
To match the notation in \erf{DtoR} we abbreviate $R_{1,1} \,{=:}\, R$.
Applying the compatibility condition \erf{eq:D-past-phi-symb} between the 
action of the defect and the OPE of bulk fields to the OPE 
\erf{eq:OPE-phiphi-to-one} yields the conditions 
  \be
  A_{s,\bar s}(R u , R v ) = R_{s,\bar s} \,{\circ}\, A_{s,\bar s}(u,v) 
  \qquad \text{for~all}~~u,v\in W .
  \labl{eq:app-ARR-RA}
on the linear maps $R_{s,\bar s}$.

Let us start by analysing this condition for $A_{0,0}$. First of all, 
non-degeneracy of the bulk two-point correlator implies that $A_{0,0}$ furnishes
a non-degenerate pairing on $W \,{\times}\, W$. Furthermore, $A_{0,0}$ is 
invariant with respect to the action of ${\rm SL}(2,\Cb) \,{\times}\, 
{\rm SL}(2,\Cb)$, and all such bilinear forms on 
$W\,{\times}\, W$ are symmetric. Denote by $\mathrm O(W)$ be the group of all 
endomorphisms of $W$ that leave $A_{0,0}$ invariant. 

{}From \erf{eq:D-past-Vir-phi} it follows that $\hat D \varphi_{0,0} \,{=}\, 
\lambda_D\, \varphi_{0,0}$. This in turn implies $R_{0,0} \,{=}\, 1$. As a
consequence, $A_{0,0}(Ru,Rv)\,{=}\,A_{0,0}(u,v)$, i.e.\ $R\,{\in}\,\mathrm O(W)$.
Since $W$ is a complex vector space, and since $A_{0,0}$ is non-degenerate and 
symmetric, there is a basis $\{v_1,v_2,v_3,v_4\}$ such that $A_{0,0}(v_i,v_j) 
\,{=}\,\delta_{i,j}$. The isomorphism $f {:}\ \Cb^4 \,{\rightarrow}\, W$ given 
by $f(e_i)\,{=}\,v_i$ then gives rise to a group isomorphism 
$f_*{:}\ \mathrm O(4,\Cb)\,{\rightarrow}\,\mathrm O(W)$. Since $A_{0,0}$ is 
invariant, we also get a group homomorphism $b {:}\ {\rm SL}(2,\Cb) \,{\times}\,
{\rm SL}(2,\Cb)\,{\rightarrow}\,\mathrm O(W)$, which takes $g \,{\times}\, h$ 
to the linear map that acts as $u\,{\mapsto}\,(g\otimes h)u$. The kernel of $b$ 
is $\{ (e,e), (-e,-e) \}$. The image of $b$ is equal to the image of 
$\mathrm{SO}(4,\Cb)$ under $f_*$. In fact, the composition $f^{-1}_* \,{\circ}\,
b$ gives rise to the group isomorphism
  \be
  \big( {\rm SL}(2,\Cb) \,{\times}\, {\rm SL}(2,\Cb) \big) / \{ (e,e),(-e,-e) \}
  \,\cong\, \mathrm{SO}(4,\Cb) \,.
  \ee
We already know that $R \,{\in}\, \mathrm O(W)$. In order to show \erf{eq:R=gxh} 
it is therefore enough to prove that $R$ is in the image of $\mathrm{SO}(4,\Cb)$ 
under $f_*$, i.e.\ that $\det(R) \,{=}\, 1$. 
We will show that $\det(R)\,{=}\,{-}1$ 
contradicts \erf{eq:app-ARR-RA} for $s\,{=}\,2$, $\bar s\,{=}\,0$.

Let $e_{\pm} \,{=}\, |j{=}\tfrac12,m{=}{\pm}{\tfrac12}\rangle$ be the standard basis
of $V_{1/2}$ and let $e_{\pm\pm} \,{=}\, e_\pm \oti e_\pm$ be the corresponding 
basis of $W$. By $\Ue$-charge conservation, the operator product of $[e_{++} \oti 
\varphi_{1,1}](z)$ and $[e_{+-} \oti \varphi_{1,1}](z)$ must lie in 
$\Hc_{[2,0]}$. Therefore, $A_{2,0}(e_{++},e_{+-}) \,{\neq}\, 0$, or else the 
operator product would vanish identically (which it does not). On the other 
hand, again by charge conservation, $A_{2,0}(e_{++},e_{-+}) \,{=}\, 0$. Define 
the linear map $S$ via
  \be
  S\, e_{++} = e_{++} \,,~~~ 
  S\, e_{+-} = e_{-+} \,,~~~ 
  S\, e_{-+} = e_{+-} \,,~~~ 
  S\, e_{--} = e_{--} \,.
  \ee
One can check that $S \,{\in}\, \mathrm O(W)$ and $\det(S)\,{=}\,{-}1$. Then 
$A_{2,0}(S e_{++}, S e_{-+}) \,{=}\, A_{2,0}(e_{++}, e_{+-}) \,{\neq}\, 0$, 
while $A_{2,0}(e_{++}, e_{-+}) \,{=}\,  0$. It is therefore impossible to 
satisfy \erf{eq:app-ARR-RA} for the choice $R\,{=}\,S$. Let now $R$ be an 
arbitrary element of $\mathrm O(W)$ with $\det(R)\,{=}\,{-}1$. Since 
$\mathrm O(W)$ has two connected components, the image of $\mathrm{SO}(4,\Cb)$ 
under $f_*$ contains an element $x$ such that $R\,x \,{=}\, S$. Then for 
$u\,{=}\,x e_{++}$ and $v \,{=}\, x e_{-+}$ we have $A_{2,0}(Ru,Rv) \,{=}\, 
A_{2,0}(S e_{++}, S e_{-+}) \,{\neq}\, 0$, while $A_{2,0}(u , v) 
  $\linebreak[0]$
\,{=}\, A_{2,0}(x e_{++}, x e_{-+}) \,{=}\, A_{2,0}(e_{++}, e_{-+}) \,{=}\, 0$, 
so that it is again impossible to satisfy \erf{eq:app-ARR-RA}. Thus the linear 
map $R$ appearing in \erf{eq:R=gxh} has necessarily $\det(R)\,{=}\,1$.



\small

\begin{thebibliography}{10}
\setlength{\itemsep}{.2em}

\bibitem{Petkova:2000ip} V.B.~Petkova and J.B.~Zuber,
{\it Generalised twisted partition functions},
Phys.\ Lett.\ B {\bf 504} (2001) 157 {\tt [hep-th/0011021]}.

\bibitem{Petkova:2001ag} V.B.~Petkova and J.B.~Zuber,
  {\it The many faces of Ocneanu cells},
  Nucl.\ Phys.\ B {\bf 603} (2001) 449  {\tt [hep-th/0101151]}.

\bibitem{Chui:2001kw} C.H.O.~Chui, C.~Mercat, W.P.~Orrick and P.A.~Pearce,
{\it Integrable lattice realizations of conformal twisted boundary
conditions}, Phys.\ Lett.\ B {\bf 517} (2001) 429 {\tt [hep-th/0106182]}.

\bibitem{Coquereaux:2001di} R.~Coquereaux and G.~Schieber,
{\it Twisted partition functions for ADE boundary conformal field
    theories and Ocneanu algebras of quantum symmetries},
  J.\ Geom.\ Phys.\ {\bf 42} (2002) 216 {\tt [hep-th/0107001]}.

\bibitem{Bachas:2001vj} C.~Bachas, J.~de Boer, R.~Dijkgraaf and H.~Ooguri,
  {\it Permeable conformal walls and holography},
  JHEP {\bf 0206} (2002) 027 {\tt [hep-th/0111210]}.

\bibitem{Quella:2002ct} T.~Quella and V.~Schomerus,
{\it Symmetry breaking boundary states and defect lines},
JHEP {\bf 0206} (2002) 028 {\tt [hep-th/0203161]}.

\bibitem{Fuchs:2002cm} J.~Fuchs, I.~Runkel and C.~Schweigert,
{\it TFT construction of RCFT correlators\ I: Partition functions},
Nucl.\ Phys.\ B {\bf 646} (2002) 353 {\tt  [hep-th/0204148]}.

\bibitem{Graham:2003nc} K.~Graham and G.M.T.~Watts,
{\it Defect lines and boundary flows},
JHEP {\bf 0404} (2004) 019 {\tt [hep-th/0306167]}.

\bibitem{Frohlich:2004ef} J.~Fr\"ohlich, J.~Fuchs, I.~Runkel and 
C.~Schweigert, {\it Kramers-Wannier duality from conformal defects},
  Phys.\ Rev.\ Lett.\  {\bf 93} (2004) 070601 {\tt [cond-mat/0404051]}.

\bibitem{Bachas:2004sy} C.~Bachas and M.~Gaberdiel,
  {\it Loop operators and the Kondo problem},
  JHEP {\bf 0411} (2004) 065 {\tt [hep-th/0411067]}.

\bibitem{Frohlich:2006ch} J.~Fr\"ohlich, J.~Fuchs, I.~Runkel and 
C.~Schweigert, {\it Duality and defects in rational conformal field theory},
Nucl.\ Phys.\ B {\bf 763} (2007) 354 {\tt [hep-th/0607247]}.

\bibitem{Quella:2006de} T.~Quella, I.~Runkel and G.M.T.~Watts,
{\it Reflection and transmission for conformal defects},
JHEP {\bf 0704} (2007) 095 {\tt [hep-th/0611296]}.

\bibitem{Wong:1994pa} E.~Wong and I.~Affleck,
  {\it Tunneling in quantum wires: A boundary conformal field theory
    ap\-proach}, Nucl.\ Phys.\ B {\bf 417} (1994) 403.

\bibitem{Friedan} D.~Friedan, 
  {\it The space of conformal boundary conditions for the $c=1$
  Gaussian model}, private notes (1999), 
  {\it The space of conformal boundary conditions for the $c=1$
  Gaussian model (more)}, private notes (2003), 
  both available from\, {\tt http:/$\!$/www.physics.rutgers.edu/\~{}friedan}

\bibitem{Gaberdiel:2001xm} M.R.~Gaberdiel, A.~Recknagel and G.M.T.~Watts,
{\it The conformal boundary states for SU(2) at level 1},
Nucl.\ Phys.\  B {\bf 626} (2002) 344 {\tt [hep-th/0108102]}.
  
\bibitem{Gaberdiel:2001zq} M.R.~Gaberdiel and A.~Recknagel,
{\it Conformal boundary states for free bosons and fermions},
JHEP {\bf 0111} (2001) 016 {\tt [hep-th/0108238]}.

\bibitem{Janik:2001hb} R.A.~Janik,
  {\it Exceptional boundary states at $c=1$},
  Nucl.\ Phys.\ B {\bf 618} (2001) 675 \\ {\tt [hep-th/0109021]}.

\bibitem{Cardy:1991tv} J.L.~Cardy and D.C.~Lewellen,
  {\it Bulk and boundary operators in conformal field theory},
  Phys.\ Lett.\  B {\bf 259} (1991) 274.

\bibitem{Lewellen:1991tb} D.C.~Lewellen,
  {\it Sewing constraints for conformal field theories on surfaces with
       boundaries}, Nucl.\ Phys.\  B {\bf 372} (1992) 654.

\bibitem{Fjelstad:2006aw} J.~Fjelstad, J.~Fuchs, I.~Runkel and C.~Schweigert,
{\it Uniqueness of open/closed rational CFT with given algebra of open states},
{\tt hep-th/0612306}.

\bibitem{Cardy:1986gw} J.L.~Cardy,
{\it Effect of boundary conditions on the operator 
     content of two-dimensional conformally invariant theories},
Nucl.\ Phys.\  B {\bf 275} (1986) 200.
 
\bibitem{Shapere:1988zv} A.D.~Shapere and F.~Wilczek,
{\it Selfdual models with theta terms}, Nucl.\ Phys.\ B {\bf 320} (1989) 669. 

\bibitem{Asatani:1996jc} T.~Asatani, T.~Kuroki, Y.~Okawa, F.~Sugino
and T.~Yoneya, {\it T-duality transformation and universal structure of 
non-critical string field theory},
Phys.\ Rev.\  D {\bf 55} (1997) 5083 {\tt [hep-th/9607218]}.

\bibitem{Polchinski:1996na} J.~Polchinski,
  {\it Lectures on D-branes}, {\tt hep-th/9611050}.

\bibitem{Fuchs:2007fw} J.~Fuchs, C.~Schweigert and K.~Waldorf,
{\it Bi-branes: Target space geometry for world sheet topological defects},
{\tt hep-th/0703145}.

\bibitem{Kac:1987gg} V.G.~Kac and A.K.~Raina,
{\it Bombay Lectures on Highest Weight Representations of Infinite 
     Dimensional Lie Algebras}, Adv.\ Ser.\ Math.\ Phys.\ {\bf 2} (1987) 1.

\bibitem{DiFrancesco:1997nk} P.~Di Francesco, P.~Mathieu and D.~Senechal,
{\it Conformal Field Theory}, Springer 1997.

\bibitem{Fuchs:2004dz} J.~Fuchs, I.~Runkel and C.~Schweigert,
{\it TFT construction of RCFT correlators\ III: Simple currents},
Nucl.\ Phys.\ B {\bf 694} (2004) 277 {\tt [hep-th/0403157]}.

\bibitem{Fjelstad:2005ua} J.~Fjelstad, J.~Fuchs, I.~Runkel and
C.~Schweigert, {\it TFT construction of RCFT correlators\ V: Proof of modular
invariance and factorisation},
Theo.\ and Appl.\ of\ Cat.\ {\bf 16} (2006) 342 {\tt [hep-th/0503194]}.

\bibitem{Schellekens:1989am} A.N.~Schellekens and S.~Yankielowicz,
{\it Extended chiral algebras and modular invariant partition functions},
Nucl.\ Phys.\ B {\bf 327} (1989) 673.

\bibitem{Frohlich:2003hm} J.~Fr\"ohlich, J.~Fuchs, I.~Runkel and C.~Schweigert, 
{\it Correspondences of ribbon categories},
Adv.\ Math.\ {\bf 199} (2006) 192 {\tt [math.CT/0309465]}.
  
\bibitem{Recknagel:1998ih} A.~Recknagel and V.~Schomerus,
{\it Boundary deformation theory and moduli spaces of D-branes},
Nucl.\ Phys.\ B {\bf 545} (1999) 233 {\tt [hep-th/9811237]}.

\bibitem{Gaberdiel:2005sz} M.R.~Gaberdiel, A.O.~Klemm and I.~Runkel,
{\it Matrix model eigenvalue integrals and twist fields in the
su(2)-WZW model}, 
JHEP {\bf 0510} (2005) 107 {\tt [hep-th/0509040]}.

\bibitem{Creutzig:2006wk} T.~Creutzig, T.~Quella and V.~Schomerus,
{\it New boundary conditions for the $c \,{=}\,{-}2$ ghost system},
{\tt hep-th/0612040}.

\bibitem{Brunner:2000wx} I.~Brunner and V.~Schomerus,
{\it On superpotentials for D-branes in Gepner models},
JHEP {\bf 0010} (2000) 016 {\tt [hep-th/0008194]}.

\bibitem{Dijkgraaf:1989hb} R.~Dijkgraaf, C.~Vafa, E.P.~Verlinde and
H.L.~Verlinde, 
{\it The operator algebra of orbifold models},
Commun.\ Math.\ Phys.\  {\bf 123} (1989) 485.

\bibitem{Fuchs:2003id} J.~Fuchs, I.~Runkel and C.~Schweigert,
{\it TFT construction of RCFT correlators\ II: Unoriented world sheets},
Nucl.\ Phys.\  B {\bf 678} (2004) 511 {\tt  [hep-th/0306164]}.

\end{thebibliography}
\end{document}